\documentclass[10pt,conference]{IEEEtran} 
\usepackage[utf8]{inputenc}

\title{}
\author{wuyue16}

\usepackage[english]{babel}
\usepackage[utf8]{inputenc}
\usepackage{CJKutf8}

\usepackage[normalem]{ulem}

\usepackage{amsfonts,amsmath,amssymb,graphicx}
\usepackage{cuted}  
\usepackage[usenames,dvipsnames]{xcolor}
\usepackage{setspace}
\usepackage{hyperref}
\usepackage{amsmath}

\usepackage{pdfpages}

\newcommand{\be}{\begin{eqnarray}}
\newcommand{\ee}{\end{eqnarray}}

\newcommand{\E}{\mathcal{E}}

\usepackage{color,balance,xspace,verbatim,ifthen,engord}

\usepackage{amsmath,wasysym,amsthm,marvosym,stackengine}

\usepackage{booktabs,colortbl,diagbox,multirow,tabularx,tablefootnote}

\usepackage{caption}
\usepackage{subcaption}
\usepackage{graphicx,epsfig,epstopdf,wrapfig}
\graphicspath{{./figcam/}} 
\DeclareGraphicsExtensions{.pdf,.mps,.png,.jpg,.eps,.PNG,.JPG}
\usepackage{algorithm,algorithmic}

\usepackage{url}

\usepackage{tikz}
\usetikzlibrary{shapes.multipart}
\usetikzlibrary{calc}
\usetikzlibrary{arrows.meta}
\usepackage{relsize}

\newcommand*\circled[1]{\tikz[baseline=(char.base)]{\node[shape=circle,draw,inner sep=0.5pt] (char) {#1};}}

\newcommand{\nosection}[1]{\vspace{3pt}\noindent\textbf{#1}}

\usepackage{courier}

\usepackage{enumitem}

\usepackage[noadjust]{cite}

\usepackage{xcolor,soul}
\usepackage{float}
\usepackage{enumitem}
\usepackage[capitalize]{cleveref}

\newcommand{\code}[1]{\texttt{#1}}

\newcommand{\remove}[1]{}

\definecolor{lightgray}{gray}{0.6}
\definecolor{lightblue}{rgb}{0.9,0.9,1}
\definecolor{aqua}{rgb}{0.0, 1.0, 1.0}

\newcommand{\seqalgo}{Parity Blossom\xspace}
\newcommand{\paralgo}{Fusion Blossom\xspace}

\newcommand{\papertitle}{Fusion Blossom: Fast MWPM Decoders for QEC}
\newcommand{\personnel}{Yue Wu and Lin Zhong\\ Department of Computer Science, Yale University, New Haven, CT}

\makeatletter
\newcommand{\factlabel}[2]{%
   \protected@write \@auxout {}{\string \newlabel {#1}{{#2}{\thepage}{#2}{#1}{}} }%
   \hypertarget{#1}{#2.}
}
\makeatother
\newcounter{factcounter}
\newcommand{\definefact}[1]{{\factlabel{fact:#1}{\noindent\textit{Fact \thefactcounter}}}{\addtocounter{factcounter}{1}}}
\newcommand{\reffact}[1]{\ref{fact:#1}}

\makeatletter
\newcommand{\definelemma}[2]{%
   \protected@write \@auxout {}{\string \newlabel {lemma:#1}{{\textit{Lemma: #2}}{\thepage}{#2}{lemma:#1}{}} }%
   \hypertarget{lemma:#1}{\noindent\textit{Lemma: #2.}}
}
\makeatother
\newcommand{\reflemma}[1]{\ref{lemma:#1}}

\makeatletter
\newcommand{\nosectionlabel}[2]{%
   \protected@write \@auxout {}{\string \newlabel {#1}{{\textbf{#2}}{\thepage}{#2}{#1}{}} }%
   \hypertarget{#1}{\noindent\textbf{#2.}}
}
\makeatother

\makeatletter
\newcommand{\definitionlabel}[2]{%
   \protected@write \@auxout {}{\string \newlabel {#1}{{\textit{#2}}{\thepage}{#2}{#1}{}} }%
   \protected@write \@auxout {}{\string \newlabel {#1s}{{\textit{#2s}}{\thepage}{#2}{#1}{}} }%
   \hypertarget{#1}{\noindent\textit{Definition: #2.}}
}
\makeatother

\newcommand{\dns}{\!\!}  
\newcommand{\qns}{\!\!\!\!}  
\newcommand{\dqns}{\qns\qns}  
\newcommand{\tqns}{\qns\qns\qns}  
\newcommand{\qqns}{\qns\qns\qns\qns}  
\newcommand{\compactsetminus}{\!\setminus\!}
\newcommand{\spaciousland}{\;\land\;}

\newcommand{\ancestryminus}[2]{\mathcal{A}({#1 \compactsetminus #2})}
\newcommand{\circleminus}[2]{C({#1 \compactsetminus #2})}

\newcommand{\obstacles}{obstacles\xspace}

\makeatletter
\newcommand{\definetrim}[2]{%
  \define@key{Gin}{#1}[]{\setkeys{Gin}{trim=#2,clip}}%
}
\makeatother
\definetrim{evaluationtrim}{0 1.1cm 0 0}

\newcommand{\decodingprocedurebox}[1]{#1}

\definecolor{primalinterfacecolor}{RGB}{29,108,171}
\definecolor{dualinterfacecolor}{RGB}{208,35,36}
\definecolor{paritydecodercolor}{RGB}{227,237,238}

\newcommand{\theoremvaliddualvariables}[1]{%
\nosectionlabel{#1}{Theorem: Feasible Dual Variables}%
Solutions for the two disjoint sub-problems determine the values of the dual variables $y_S, S\in \mathcal{O}_1^*\cup \mathcal{O}_2^*$. These values plus setting $y_S$ to 0 for $S\in \mathcal{O}^*\setminus \mathcal{O}_1^*\cup \mathcal{O}_2^*$ constitute a feasible solution to the original dual problem.%
}

\newcommand{\theoremcoverfiniteoverlap}[1]{%
\nosectionlabel{#1}{Theorem: Node Cover Finite Overlap}%
Given two nodes $S_1$ and $S_2$, $\text{Cover}(S_1) \cap \text{Cover}(S_2)$ is a finite set.%
}

\newcommand{\theoremobstacledetection}[1]{%
\nosectionlabel{#1}{Theorem: Tight Edge Detection (Cover)}%
There exists a tight edge between two different nodes $S_1$ and $S_2$ if and only if Cover($S_1$) and Cover ($S_2$) overlap. That is,%
\begin{gather*}%
\exists e=(v_1,v_2) \in E, v_1\in S_1 \spaciousland v_2\in S_2 \spaciousland w_e = \dqns\sum_{S \in \mathcal{O}^* | e \in \delta(S)}\dqns y_S\\%
\Longleftrightarrow \text{Cover}(S_1) \cap \text{Cover}(S_2) \neq \varnothing%
\end{gather*}%
}

\newcommand{\theoremobstacledetectionoptimized}[1]{%
\nosectionlabel{#1}{Theorem: Tight Edge Detection (Pseudo-Cover)}%
There exists a tight edge between two different nodes $S_1$ and $S_2$ with $\Delta y_{S_1} + \Delta y_{S_2} > 0$ if and only if there exists two different nodes $S_3$ and $S_4$ with $\Delta y_{S_3} + \Delta y_{S_4} > 0$ whose Pseudo-Covers meet on a decoding graph edge. %
That is,%
\begin{gather*}%
\exists S_1, S_2, e=(v_1,v_2)\in E,\\%
v_1\in S_1, v_2\in S_2, \Delta y_{S_1} + \Delta y_{S_2} > 0, w_e = \dqns\sum_{S \in \mathcal{O}^* | e \in \delta(S)}\dqns y_S\\%
\Longleftrightarrow \qquad \exists S_3, S_4, e' = (v_3, v_4) \in E_M, \qquad\qquad\\%
v_3 \in \overline{\text{Cover}}(S_3), v_4 \in \overline{\text{Cover}}(S_4),\\%
\Delta y_{S_3} + \Delta y_{S_4} > 0, e' \subseteq \overline{\text{Cover}}(S_3) \cup \overline{\text{Cover}}(S_4)%
\end{gather*}%
}

\newcommand{\theoremnonnegativevertexdual}[1]{%
\nosectionlabel{#1}{Theorem: Non-negative Vertex Dual}%
Given the error probability for any independent error source $P(e)\leqslant 0.5$, dual variables $y_v$, $\forall v\in V$ are non-negative during the whole process of finding the solution by the blossom algorithm.%
}

\begin{document}

\title{\papertitle}
\author{\personnel}
\maketitle
\thispagestyle{plain}
\pagestyle{plain}

\newcommand{\enableappendix}{1}

\begin{abstract}

The Minimum-Weight Perfect Matching (MWPM) decoder is widely used in Quantum Error Correction (QEC) decoding.
Despite its high accuracy, existing implementations of the MWPM decoder cannot catch up with quantum hardware, e.g., 1 million measurements per second for superconducting qubits.
They suffer from a backlog of measurements that grows exponentially and as a result, cannot realize the power of quantum computation.
We design and implement a fast MWPM decoder, called \seqalgo, which reaches a time complexity almost proportional to the number of defect measurements.
We further design and implement a parallel version of \seqalgo called \paralgo.
Given a practical circuit-level noise of 0.1\%, \paralgo can decode a million measurement rounds per second up to a code distance of 33.
\paralgo also supports stream decoding mode that reaches a 0.7 ms decoding latency at code distance 21 regardless of the measurement rounds.

\end{abstract}

\section{Introduction}
\label{sec:introduction}
Quantum error correction (QEC) is essential for fault-tolerant quantum computing.
The decoder of QEC must be fast enough to avoid exponential backlog effect discussed by Terhal~\cite{terhal2015quantum}.
That is, it must process all the syndrome bits generated by the quantum hardware within a smaller period of time.
Such fast decoders are known as \emph{online} decoders.
Also, a decoder design should be \emph{scalable} to support large code distances in order to reach the desired logical error rate.

No scalable online MWPM decoders have been reported.
Fowler, Adam and Lloyd~\cite{fowler2012towardsta} reported an almost linear-time MWPM decoder without a publicly accessible implementation.
Higgott and Gidney recently reported an open-sourced, almost linear-time MWPM decoder at~\cite{higgott2023sparse}.
Since they are sequential algorithms, they will eventually fail to reach the throughput requirement at some large code distance.
Fowler also suggested an idea to parallelize the MWPM decoder~\cite{fowler2013minimum}, without providing any empirical data regarding its performance.

We design \paralgo as an online MWPM decoder that scales to arbitrarily large code distance $d$, using parallelization.
\paralgo is inspired by recently reported parallel realizations of the Union-Find (UF) decoder~\cite{das2022afs,skoric2022parallel,tan2022scalable,liyanage2023scalable} and by the relationship between the UF decoder and the MWPM decoder revealed in~\cite{wu2022interpretation}.
\paralgo drastically speeds up QEC decoding to sub-microsecond per measurement round by using parallel CPU cores.
Taking a rotated surface code of 0.1\% circuit-level noise on a 64-core CPU as an example, it can decode up to $d = 33$ with throughput of one million rounds per second using batch decoding.
Using stream decoding, it achieves a constant 0.7 ms average latency at $d = 21$ regardless of the number of measurement rounds.
To the best of our knowledge, \paralgo is the first publicly available parallel MWPM decoder~\cite{fusion-blossom}, implemented in Rust with Python binding~\cite{fusion-blossom-py}.

The key ideas of \paralgo are two. First, it recursively divides a decoding problem into two sub-problems that can be solved independently and efficiently fuses their solutions, according to a tree structure computed offline. Second, it leverages a fast sequential MWPM decoder called \seqalgo, which implements a novel variant of the blossom algorithm.
\seqalgo leverages the property of the \textit{syndrome graph}~\cite{wu2022interpretation} where the MWPM problem is defined: the syndrome graph is constructed from a much sparser graph called \emph{decoding graph}.
\seqalgo works on the decoding graph to solve the MWPM problem for the syndrome graph.

In an impressive parallel work, Higgott and Gidney~\cite{higgott2023sparse} present Sparse Blossom, an implementation of blossom algorithm that shares the key idea of \seqalgo: identifying tight edges using the decoding graph, and the same mathematical foundation. Sparse Blossom features several novel optimizations that are not used by \seqalgo. This paper presents the following contributions that complement those by Sparse Blossom.
\begin{itemize}
    \item The mathematical foundation behind Sparse Blossom and \seqalgo (\ref{sec:geometric-interpretation}).
    \item \paralgo, a parallel MWPM decoder that could be based on either Sparse Blossom or \seqalgo (\ref{sec:parallel}).
    \item A unified framework for implementing matching-based decoders including novel mathematically grounded optimizations (\ref{sec:implementation}).
\end{itemize}

We evaluate \seqalgo and \paralgo in \S\ref{sec:evaluation} and discuss related work in~\S\ref{sec:related}. Our implementation is open-source and available at~\cite{fusion-blossom}.

\ifnum \enableappendix=0
Due to space limitations, we have omitted proofs of the theorems presented in this paper.
However, all the results are rigorously established and the interested reader can find the proofs in the full version of this work.
\fi

\section{Background}\label{sec:background}
We first define the necessary data structures for decoding a surface code. More information can be found in~\cite{wu2022interpretation}.

\subsection{Quantum Error Correction (QEC) Codes}
\label{sec:qec}
We aim at decoding codes that can be represented by a data structure called \emph{model graph}. Such codes include many of the topological codes~\cite{fowler2012topological}.
Such a code consists of data qubits and stabilizers.  Data qubits store the quantum information while stabilizers allow errors in data qubits to be observed classically: an error in a data qubit will impact the measurement outcome of the adjacent stabilizers that are designed to detect this type of error. 

\begin{figure*}[ht]
    \renewcommand*\thesubfigure{(\arabic{subfigure})}  
    \centering
    \begin{subfigure}{.16\linewidth}
        \centering
        \color{lightgray}\decodingprocedurebox{\includegraphics[width=0.8\linewidth]{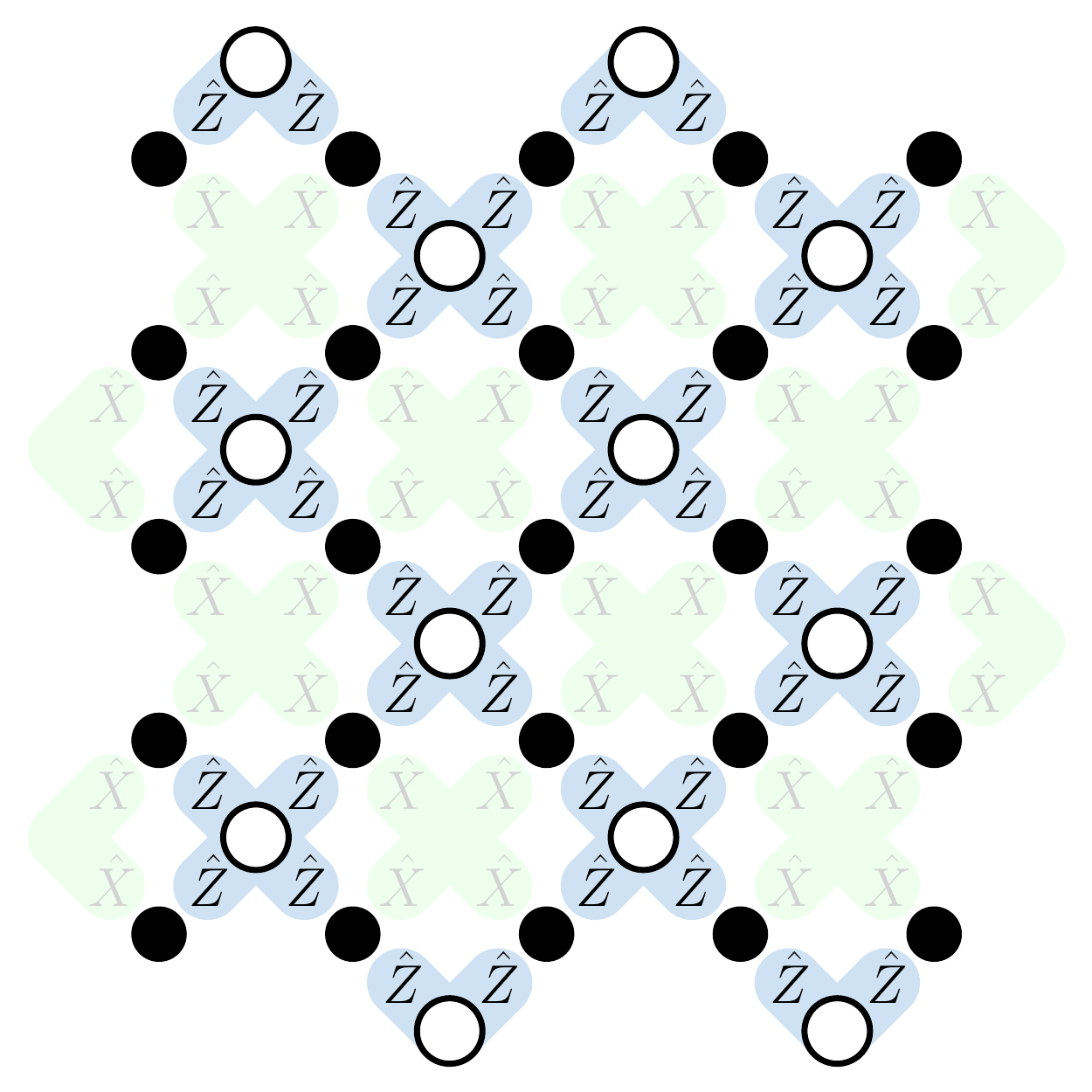}}
        \caption{Surface Code}
        \label{fig:surface-code}
    \end{subfigure}
    \begin{subfigure}{.16\linewidth}
        \centering
        \color{lightgray}\decodingprocedurebox{\includegraphics[width=0.8\linewidth]{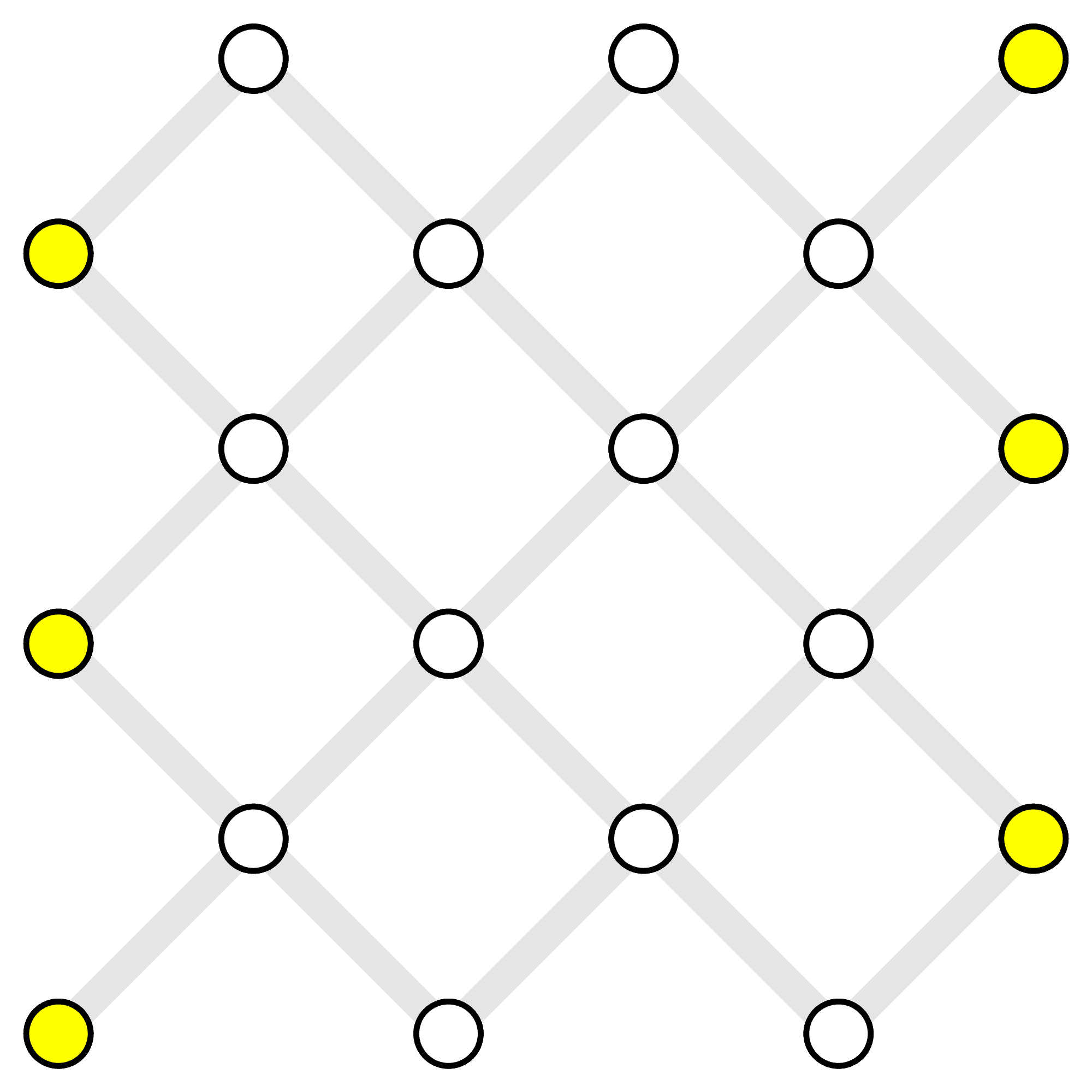}}
        \caption{Model Graph}
        \label{fig:model-graph}
    \end{subfigure}
    \begin{subfigure}{.16\linewidth}
        \centering
        \color{lightgray}\decodingprocedurebox{\includegraphics[width=0.8\linewidth]{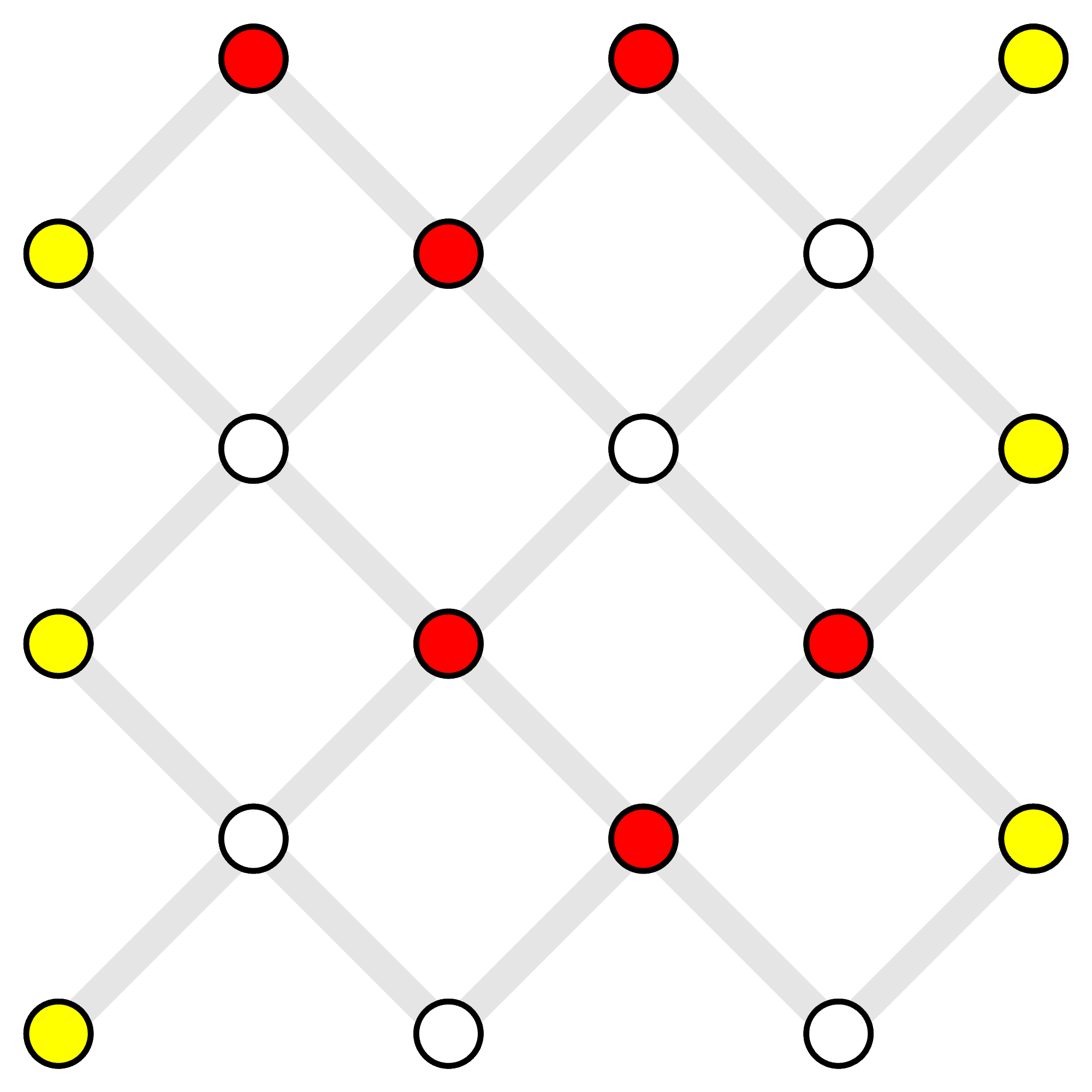}}
        \caption{Decoding Graph}
        \label{fig:decoding-graph}
    \end{subfigure}
    \begin{subfigure}{.16\linewidth}
        \centering
        \color{lightgray}\decodingprocedurebox{\includegraphics[width=0.8\linewidth]{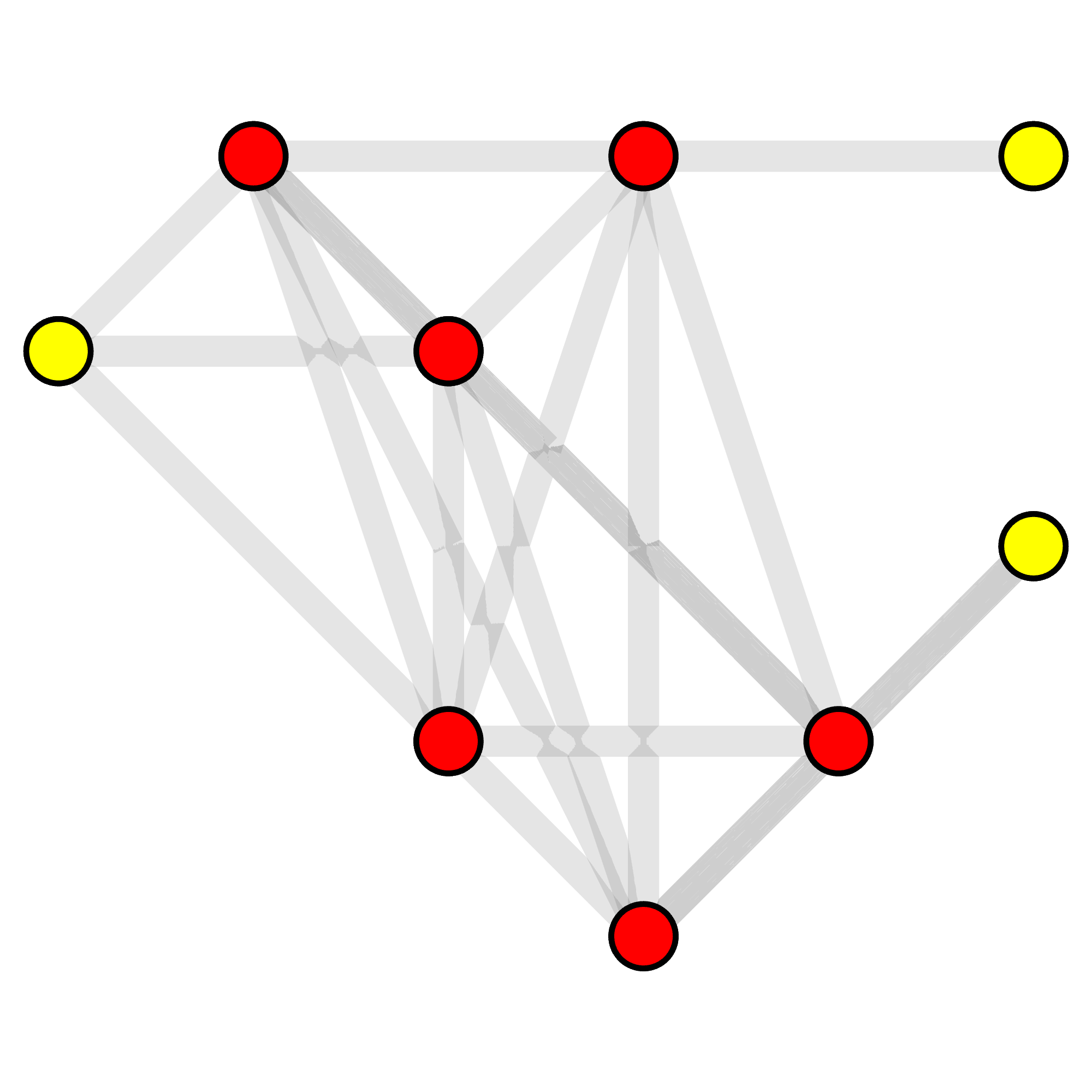}}
        \caption{Syndrome Graph}
        \label{fig:syndrome-graph}
    \end{subfigure}
    \begin{subfigure}{.16\linewidth}
        \centering
        \color{lightgray}\decodingprocedurebox{\includegraphics[width=0.8\linewidth]{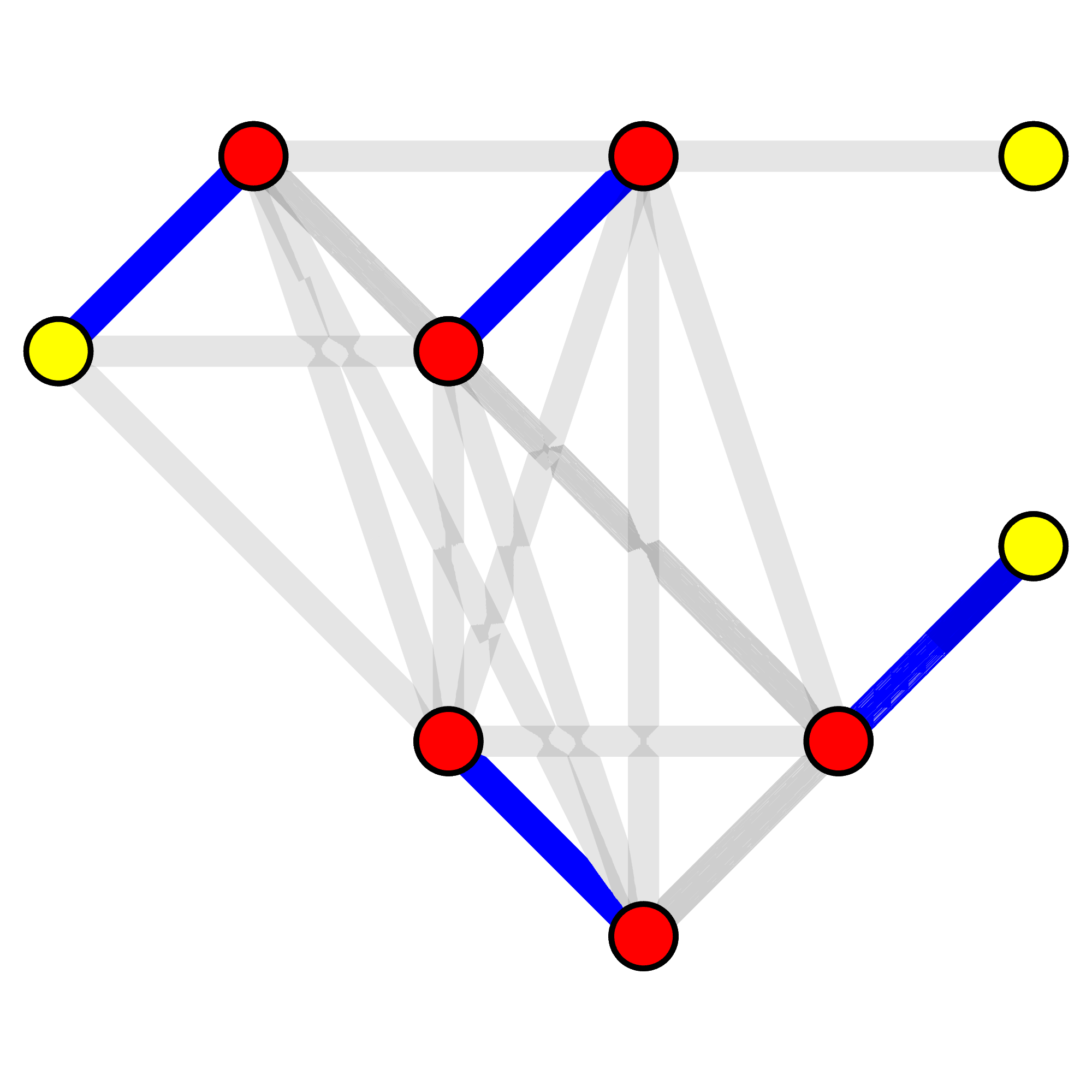}}
        \caption{MWPM}
        \label{fig:syndrome-graph-result}
    \end{subfigure}
    \begin{subfigure}{.16\linewidth}
        \centering
        \color{lightgray}\decodingprocedurebox{\includegraphics[width=0.8\linewidth]{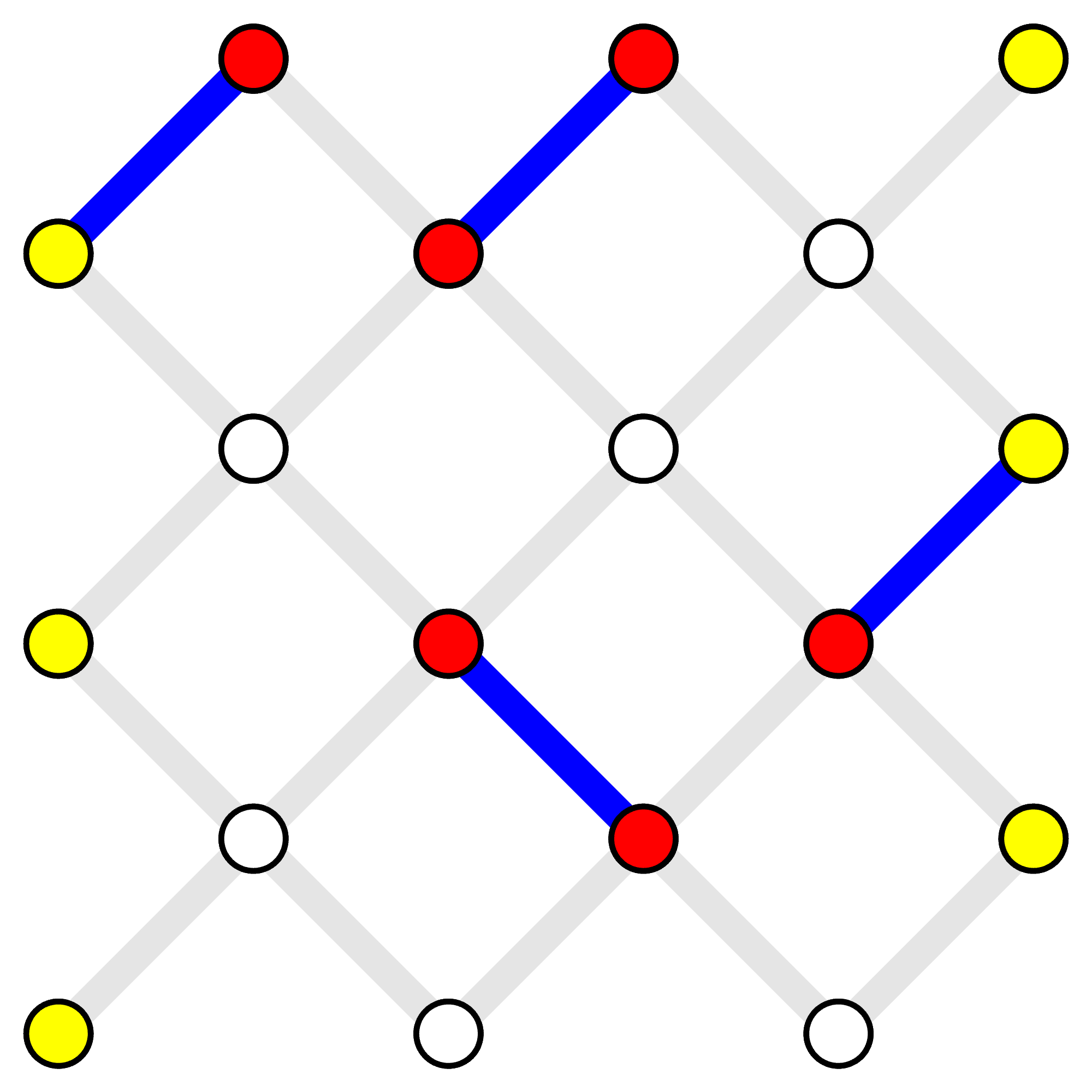}}
        \caption{Most-Likely Error}
        \label{fig:decoding-graph-result}
    \end{subfigure}
    \caption{The procedure of an MWPM decoder. (2) Create model graph given the code and noise model. Yellow vertices are virtual boundaries. (3) Create decoding graph given model graph and the observed syndrome. The white (red) vertices corresponds to normal (defect) measurement result. (3) Solve an MWPM on the syndrome graph. Blue edges are selected in the MWPM. (4) Translate MWPM into a subset of edges in the decoding graph, which corresponds to a most-likely error.}
    \label{fig:existing-mwpm-decoder}
\end{figure*}

\paragraph{Model Graph}
Following~\cite{wu2022interpretation}, we represent such a QEC code with a \emph{model graph} $G_M = (V_M, E_M)$. A vertex $v \in V_M$ corresponds to a stabilizer measurement result.
Each edge $e \in E_M$ corresponds to an independent error source and connects with two vertices that correspond to the measurement outcomes of stabilizers adjacent to this error source. 
We add a \textit{virtual vertex} to an edge if the corresponding error source only connects with a single vertex. 
This results in a two-dimensional graph as show in \cref{fig:model-graph}.
The model graph is sparse because $|E_M|$ is $O(|V_M|)$. 
Edges in the model graph are weighted. The weight of an edge can be computed from the error model of the corresponding error source $P(e)$  as 
$w_e = \log\left(\frac{1 - P(e)}{P(e)}\right)$.

The model graph can be generalized to be three-dimensional to account for erroneous stabilizer measurement shown in \cref{fig:batch-vs-stream-decoding}. The third dimension comes from multiple rounds of measurement, with each round represented by the two-dimensional model graph as shown in \cref{fig:model-graph}. 
Vertices corresponding to the measurement outcomes of the same stabilizer in two consecutive rounds are connected with a new edge, which represents the potential measurement error. 

\paragraph{Error Pattern \& Syndromes}
When an independent error source in a code experiences an error, it will ``flip'' the measurement outcome of the two adjacent stabilizers. Because a stabilizer is adjacent to multiple independent error sources, its measurement outcome is determined by the parity of the number of erroneous sources: Only if an odd number of adjacent sources experience error, the stabilizer will have a defect measurement outcome.

Because $E_M$ denote the set of independent error sources in the code, $\E\subseteq E_M$ denotes the subset that experience an error, or \emph{error pattern}.
$P(\mathcal{E}), \forall \E\subseteq E_M$ indicates the probability that $\mathcal{E}$ happens. It is the \emph{error model} of the code and can be obtained by characterizing the quantum hardware.

One can compute the error model from the error model for each independent error source, $P(e)$, as below:
\begin{equation}
\small
\begin{split}
P(\mathcal{E}) 
    = \prod_{e \in \mathcal{E}} P(e) \prod_{f \in E_M\setminus\E} (1-P(f))
    \propto \prod_{e \in \mathcal{E}} \frac{P(e)}{(1-P(e))}
\end{split}
\end{equation}

Given an error pattern $\E$, $S(\E)$ denotes the set of vertices that correspond to the defect measurement outcomes in the decoding graph and it is known as the \emph{syndrome} of $\E$.

Given the syndrome $\mathcal{S}$ and the model graph, a decoder seeks to find an error pattern that produces $\mathcal{S}$.
The \emph{decoding graph} is the model graph with the syndrome $\mathcal{S}$ marked, as shown in \cref{fig:decoding-graph}. The Union-Find decoder~\cite{delfosse2021almost} uses the decoding graph~\cite{wu2022interpretation} to find an error pattern that can produce the syndrome. 

\paragraph{Most-Likely Error Decoder}

A Most-Likely Error (MLE) decoder tries to find the most likely error pattern that generates the syndrome $S$.
\[ \arg\max_{\mathcal{E} | S(\mathcal{E}) = \mathcal{S}} P(\mathcal{E}) = \arg\max_{\mathcal{E} | S(\mathcal{E}) = \mathcal{S}} \prod_{e \in \mathcal{E}} \frac{P(e)}{1 - P(e)} \]

The MLE decoding problem then becomes a problem for the decoding graph: find a subset of edges $\E \subseteq E_M$ that generates the observed syndrome $S(\mathcal{E}) = \mathcal{S}$ while \textit{maximizing} $P(\E) = \prod_{e \in \mathcal{E}} \frac{P(e)}{(1-P(e))}$.
Note that we use $\E$ as error pattern and subset of edges interchangeably because they represent the same thing.
Since it's more common to define the summation of weights in graph problems, we can equivalently translate the problem into \textit{minimizing} $W(\E) = \sum_{e \in \E} w_e$.

$$\arg\min_{\E | S(\E) = \mathcal{S}} \sum_{e \in \E} w_e$$

\paragraph{MWPM Decoder}\label{ssec:mwpm-decoder}

The Minimum-Weight Perfect Matching (MWPM) decoder is an exact MLE decoder when the error model can be precisely represented by a model graph.
Unlike the Union-Find decoder,  the MWPM decoder uses the \emph{syndrome graph}, $G(V,E)$, which is generated from the decoding graph by creating an edge between any two defect vertices and removing normal vertices $v \in V_M \setminus \mathcal{S}$ (and their incident edges). 
That is, $V=\mathcal{S}$ and $E=\{(u,v)|\exists u,v\in \mathcal{S}\}$.
The weight of an edge in the syndrome graph is calculated as that of a minimum-weight path between them.  
As its name suggests, the MWPM decoder finds an MLE error pattern by finding a minimum-weight perfect matching for the syndrome graph.
We illustrate the workflow of the MWPM decoder in \cref{fig:existing-mwpm-decoder}.

Because the fastest known algorithm to solve the MWPM problem for a general graph is the blossom algorithm~\cite{edmonds1973matching}, most implementations of the MWPM decoder use off-the-shelf MWPM libraries such as Kolmogorov's blossom V library~\cite{kolmogorov2009blossom} and the Lemon library by Dezs{\H{o}} \textit{et al}~\cite{dezsHo2011lemon}.
These implementations must go through all the stages in \cref{fig:existing-mwpm-decoder}.
Because the syndrome graph is complete, i.e, $|E| = |V|^2$, even the fastest implementations known~\cite{micali1980v,vazirani2012simplification} have a time complexity of $O(\sqrt{|V|} |E|) = O(|V|^{2.5})$, scaling faster than the number of defect stabilizer measurements $|V|$.

The key idea behind \seqalgo is that it removes the stage of building the syndrome graph. As a result, \seqalgo reaches an average runtime of almost $O(|V|)$ given sufficiently low error rate.
In doing so, unlike the blossom algorithm, \seqalgo does not work for general graphs but decoding graphs representing syndromes of QEC codes.
We derive this insight from our prior work~\cite{wu2022interpretation}, which shows that the UF decoder can be considered as an approximation of the MWPM decoder. 
Like the UF decoder, \seqalgo uses the decoding graph.

\subsection{Blossom Algorithm in General}\label{ssec:blossom-in-general}

We next describe the blossom algorithm~\cite{edmonds1973matching}.
We elide details that are irrelevant to our contributions.
The blossom algorithm formulate the MWPM problem as an integer linear-programming (ILP) problem.
Given any graph $G = (V, E)$ and edge weights $w_e, \forall e \in E$, the MWPM problem solves a perfect matching $x_e, \forall e \in E$ with minimum total weight $\sum_{e\in E} w_e x_e $.
A solution is represented by $x_e, \forall e \in E$, with all selected edges $x_e = 1$ and others $x_e = 0$.
A perfect matching requires that for every vertex $v \in V$, there is a unique edge with $x_e = 1$ incident to $v$, and all other incident edges have $x_e = 0$.
There is no constraint on the incident edges for a virtual vertex.

The blossom algorithm solves the above ILP problem by first relaxing the integer constraint, becoming a linear-programming (LP) problem.
It then adds some more constraints to the LP problem so that all optimal ILP solutions are optimal LP solutions~\cite{cook1999computing}.
It solves the following LP problem.

\begin{align}\label{eq:primal-lp}
\min\;\;\quad \sum_{e \in E} w_e &x_e & \tag{1}\\
\text{subject to}\quad \sum_{e \in \delta(v)} x_e &= 1        &\forall v \in V \tag{1a}\label{eq:primal-lp-constraint-a}\\
                       \sum_{e \in \delta(S)} x_e &\geqslant 1    &\forall S \in \mathcal{O} \tag{1b}\label{eq:primal-lp-constraint-b}\\
                        x_e &\geqslant 0                            & \forall e \in E \tag{1c}\label{eq:primal-lp-constraint-c}
\end{align}
where $\mathcal{O} = \{ S | S \subseteq V \land |S| > 1\land |S| = 1 \mod 2 \}$  and $\delta(S) = \{ e | e = (u, v) \in E \land ((u \in S \land v \notin S) \lor (u \notin S \land v \in S)) \}$. $e \in \delta(S)$ is called a \emph{hair} of $S$ and has one and only one incident vertex inside $S$.

The blossom algorithm creatively exploits the \textit{dual} formulation of the same problem.

\begin{align}\label{eq:dual-lp}
\max\quad \sum_{v \in V} y_v + \sum_{S \in \mathcal{O}} y_S \quad\quad & & \tag{2}\\
\text{subject to}\quad w_e - \sum_{v \in e} y_v - \dqns \sum_{S \in \mathcal{O} | e \in \delta(S)} \dqns y_S &\geqslant 0  &\forall e \in E \tag{2a}\label{eq:dual-lp-constraint-a}\\
                       y_S &\geqslant 0    &\forall S \in \mathcal{O} \tag{2b}\label{eq:dual-lp-constraint-b}
\end{align}

\vspace{1ex}\definitionlabel{def:tight}{Tight Edge} For edge $e\in E$, we say it is tight when $w_e=\sum_{S \in \mathcal{O}^* | e \in \delta(S)} y_S$, where  $\mathcal{O}^* = \{ S | S \subseteq V \land |S| = 1 \mod 2 \}$.

If an edge $e$ is not tight, its primal $x_e$ must be zero, thanks to the Complementary Slackness theorem. This means that the final solution to the primal problem only includes tight edges.

\vspace{1ex}\definitionlabel{def:blossom}{Blossom} In the blossom algorithm, $S\in\mathcal{O}$ is a blossom if and only if $y_S>0$~\cite{kolmogorov2009blossom}. Blossoms are defined inductively as below.

\begin{enumerate}[topsep=0.1em, leftmargin=*, label=\textit{\arabic*.}]
    \item An odd number of vertices connected in a circle by tight edges form a blossom.
    \item An odd number of vertices or blossoms connected in a circle by tight edges form a blossom.
\end{enumerate}

\vspace{1ex}\definitionlabel{def:node}{Node} A blossom or vertex that is not the child of any other blossom is called a node. A node also includes an odd number of vertices. We denote it with the set of its vertices.

\subsubsection{Blossom algorithm}

The blossom algorithm organizes tight edges and nodes in \emph{alternating trees} and matched pairs.
The matching solution includes alternating edges in alternating trees and those between matched pairs.
A matched pair represents an MWPM solution for the vertices included by the two nodes.

The blossom algorithm starts with an empty matching solution that is feasible to the dual problem and evolves it toward a feasible solution to the primal problem while maintaining the dual feasibility.
It terminates when there is no alternating tree: all nodes are in matched pairs. 

The primal phase seeks to increase the number of edges in the matching solution: it does so by updating the alternative trees. It computes a vector $\Delta \vec{y}$ of which an element $\Delta y_S\in \{0,+1,-1\}$ is the update for $y_S$ of node $S$.  $\Delta \vec{y}$ is essentially the \emph{direction of update} of the dual variables corresponding to nodes. 
Because \seqalgo does not innovate in the primal phase, we refer readers to~\cite{kolmogorov2009blossom} for details of the primal phase.

\vspace{1ex}\definitionlabel{def:direction}{Direction} $\Delta \vec{y}$, the direction of updating $\vec{y}$.

\vspace{1ex}The dual phase seeks to update the dual variables $y_S$ along the direction $\Delta \vec{y}$ computed in the primal phase. In doing so, it must maintain the dual feasibility. That is, it ensures constraints (\ref{eq:dual-lp-constraint-a}) and (\ref{eq:dual-lp-constraint-b}) are always true.

\vspace{1ex}\definitionlabel{def:obstacle}{Obstacle} An obstacle is a dual constraint (\ref{eq:dual-lp-constraint-a} or \ref{eq:dual-lp-constraint-b}) that updating a dual variable according to $\Delta \vec{y}$ may violate.

\vspace{1ex}A key job of the dual phase is to detect obstacles and stop before it violates any of the dual constraints.  When the dual phase detects an obstacle, it stops after reporting the detected obstacles to the primal phase. 
Existing implementations of the blossom algorithm detect obstacles using the syndrome graph. A key idea of \seqalgo and Sparse Blossom~\cite{higgott2023sparse} is to do it using the decoding graph.

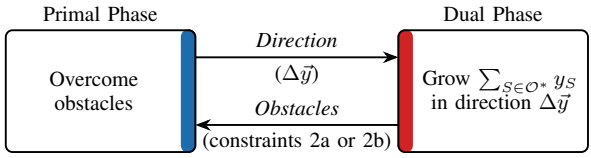
\begin{figure}[!t]
    \centering
    \scalebox{0.9}{
        \begin{tikzpicture}%
\path[primalinterfacecolor,thick,draw,fill=primalinterfacecolor,rounded corners=1mm] (2.5999999999999996,0.0) rectangle (2.8,1.8);%
\path[thick,draw,rounded corners=1mm] (0.0,0.0) rectangle (2.8,1.8);%
\node[color=black,align=center,font=\small] at (1.2999999999999998,0.9) {Overcome\\\obstacles};%
\path[dualinterfacecolor,thick,draw,fill=dualinterfacecolor,rounded corners=1mm] (5.799999999999999,0.0) rectangle (5.999999999999999,1.8);%
\path[thick,draw,rounded corners=1mm] (5.799999999999999,0.0) rectangle (8.6,1.8);%
\node[color=black,align=center,font=\small] at (7.299999999999999,0.9) {Grow $\sum_{S \in \mathcal{O}^*} y_S$\\in direction  $\Delta\vec{y}$};%
\node[color=black,align=center,font=\small] at (1.4,2.05) {Primal Phase};%
\node[color=black,align=center,font=\small] at (7.199999999999999,2.05) {Dual Phase};%
\path[thick,-{Stealth},draw] (2.8,1.4) -- (5.799999999999999,1.4);%
\node[color=black,align=center,font=\small] at (4.3,1.65) {\ref{def:direction}};%
\node[color=black,align=center,font=\small] at (4.3,1.15) {($\Delta \vec{y}$)};%
\path[thick,-{Stealth},draw] (5.799999999999999,0.4) -- (2.8,0.4);%
\node[color=black,align=center,font=\small] at (4.3,0.65) {\ref{def:obstacles}};%
\node[color=black,align=center,font=\small] at (4.3,0.15000000000000002) {(constraints \ref{eq:dual-lp-constraint-a} or \ref{eq:dual-lp-constraint-b})};%
\end{tikzpicture}
    }
    \caption{High-level structure of the blossom algorithm}
    \label{fig:primal-dual-interface}
\end{figure}

\subsection{Blossom Algorithm in QEC}
\label{sec:blossom-qec}

We show some special properties of the blossom algorithm when it solves the MWPM problem for a syndrome graph.

\vspace{1ex}\theoremnonnegativevertexdual{theorem:non-negative-vertex-dual}

\vspace{1ex}With \ref{theorem:non-negative-vertex-dual}, we can simplify the LP problem for QEC decoding as follows.
We define a set that includes both blossoms and single vertices $\mathcal{O}^* = \{ S | S \subseteq V, |S| = 1 \mod 2 \}$.
\begin{align}\label{eq:qec-dual-lp}
\small
\max\quad \sum_{S \in \mathcal{O}^*} y_S \quad\quad & & & \tag{3}\\
\text{subject to}\quad w_e - \dqns \sum_{S \in \mathcal{O}^* | e \in \delta(S)} \dqns y_S &\geqslant 0  &\forall e &\in E \tag{3a}\label{eq:qec-dual-lp-constraint-a}\\
                       y_S &\geqslant 0    &\forall S &\in \mathcal{O}^* \tag{3b}\label{eq:qec-dual-lp-constraint-b}
\end{align}

With the above simplification, the blossom algorithm can be simplified in description so that a single vertex can be treated as a blossom. As a result, the inductive defintion of blossoms is reduced to:
\begin{enumerate}[topsep=0.1em, leftmargin=*, label=\textit{\arabic*.}]
    \item A single vertex is a blossom.
    \item An odd number of  blossoms connected in a circle by tight edges form a blossom.
\end{enumerate}

As a result, we will use blossom to refer to both vertices and proper blossoms when discussing the blossom algorithm in the rest of the paper.

\section{Geometric Interpretation}\label{sec:geometric-interpretation}

Before describing our decoder designs, we provide their mathematical foundation, which is based on a novel geometric interpretation of  the blossom algorithm working on the decoding graph, inspired by~\cite{wu2022interpretation}.
By linking the notion of blossom with a geometric object (Cover) on the decoding graph, we prove key theorems used by our design.
Because this interpretation relies on the non-negative weights and non-negative dual variables given in Eq.~\ref{eq:qec-dual-lp-constraint-b}, it is not applicable to the blossom algorithm working on general graphs.

\subsection{Geometry of Decoding Graph}
The geometric interpretation is based on viewing an edge $e = (u, v) \in E_M$ as a straight, continuous line of length $w_e\geqslant 0$. This line consists of points, which do not include $u$ and $v$. 
We do consider vertices as points; we say $u$ and $v$ are incident to $e$, not on $e$, i.e., $u,v\notin e$. 
Under the above interpretation, the entire decoding graph comprises of points.

When $w_e>0$, $\forall p\in e$, $p$ partitions the line into two segments $(u,p)$, of length $w_{(u,p)}$, and $(p,v)$, of length $w_{(p,v)}$. We have $w_{(u,p)}> 0$, $w_{(p,v)}> 0$, and $w_{(u,p)}+w_{(p,v)}=w_e$.
 We can conveniently view that $p\in e$ breaks $e$ into two edges, each with its own weight. 
We call these edges \emph{segment edges} when it is necessary to distinguish them from $e\in E_M$.
Similarly, two points on $e$ would break it into three segment edges whose weights can be similarly determined and are positive.
With this, we can extend the notion of path to two arbitrary points $p$ and $q$ as the set of edges connecting them, some of which are segment edges.

When $w_e=0$, we call it a zero edge and $e=\varnothing$. Zero edges are necessary for decoding erasure errors~\cite{wu2022erasure,teoh2022dual}. Note that segment edges by definition always have a positive weight. 

\vspace{1ex}\definitionlabel{def:distance}{Distance} We define the \emph{distance} between two vertices $u, v \in V_M$ on the decoding graph as the weight of a minimum-weight path between them, noted as $\text{Dist}(u, v)$.
This definition of distance can be readily extended for two arbitrary points $p$ and $q$ of the decoding graph, according to the path definition above.

\vspace{1ex}A point $r$ on a minimum-weight path between $p$ and $q$ partitions the path into two paths, one between $p$ and $r$ and the other between $r$ and $q$. They are also the minimum-weight paths between $p$ and $r$ and between $r$ and $q$, respectively.
And we have $\text{Dist}(p,r)+\text{Dist}(r,q)=\text{Dist}(p,q)$. 

We note that minimum-weight paths between vertices in the decoding graph are related to the edges in the syndrome graph: the edge weight between two vertices in the syndrome graph is the same as the weight of a minimum-weight path between the corresponding vertices in the decoding graph. 
The notions of point and distance are key for understanding how the blossom algorithm can be adapted to work on the decoding graph. 

\vspace{1ex}\definitionlabel{def:circle}{Circle}
A circle of radius of $d$ around $v \in V_M$, $C(v,d)$,  is defined as the set of points whose distance from $v$ is no greater than $d$. That is,  $C(v,d) = \{ p | \text{Dist}(p, v) \leqslant d \}$.

A Circle consists of boundary and inside. The boundary of $C(v,d)$ is simply $\{ p | \text{Dist}(p, v)=d \}$. Likewise, the inside $C(v,d)$ is simply $\{ p | \text{Dist}(p, v)<d \}$.

\subsection{Blossom on Decoding Graph}

We relate dual variables in the blossom algorithm to the geometric objects of the decoding graph.

\subsubsection{Geometric Notions}
The inductive definition of \ref{def:blossom} allows a tree representation of a blossom: The blossom is the root; the children of the root are also blossoms; a child can also have its own children and so on. We call the set of vertices and blossoms represented by the (grand)children of this tree the \emph{descendants} of the root blossom.

\vspace{1ex}\definitionlabel{def:blossom-progeny}{Progeny} Given a blossom $S$, $\mathcal{D}(S)$ is the set that includes $S$ itself and all its descendants. We call it the \emph{progeny} of $S$.
Furthermore, we define
\begin{itemize}
    \item $\mathcal{D}_v(S)=\{ D | D \in \mathcal{D}(S) \land v \in D \}$ consists of the members of $S$'s progeny that include vertex $v$.
    \item $\mathcal{D}_{u \setminus v}(S)=\{ D | D \in \mathcal{D}(S) \land u \in D \land v \notin D \}$ is the set of progeny members of $S$ that includes $u$ but not $v$.
\end{itemize}

\vspace{1ex}\definitionlabel{def:blossom-ancestry}{Ancestry} Given a vertex $v\in V$ in the syndrome graph, its \emph{ancestry} $\mathcal{A}(v)$ is the set of all blossoms that include $v$. Let $\ancestryminus{u}{v}$ denote the subset of $u$'s Ancestry whose members do not include vertex $v$.

\vspace{1ex}
\definitionlabel{def:cover}{Cover} 
Given a blossom $S$, it covers the set of points defined by the union of circles centered at $\forall v\in S$ with $d=\sum_{D \in \mathcal{D}_v(S)} y_D$.  That is, 
\[\text{Cover}(S)=\cup_{v\in S} C(v,\sum_{D \in \mathcal{D}_v(S)} y_D).\]

That is, Cover$(S)$ consists of Circles around $\forall v\in S$. 
The \emph{boundary} of a Cover consists of points of the Cover that are not inside any of its Circles. 
Because a Circle consists of a finite number of edges, $\text{Cover}(S)$ also consists of a finite number of edges. 
For a blossom of a single vertex $v$, its Cover is simply $\text{Cover}(v)=C(v,y_v)$.

We emphasize that blossoms are defined on the syndrome graph while their $\text{Cover}$s are defined on the decoding graph. As a result, the notion of Cover is an important bridge between the decoding and syndrome graphs.

\subsubsection{Obstacle Detection on Decoding Graph}
Because the dual phase detects obstacles based on the syndrome graph, we must find a way to do so on the decoding graph. The key insight and theoretical result of this work is the next theorem, which show exactly how to do it.

First of all, we note that detecting obstacles from dual constraints \ref{eq:dual-lp-constraint-b} is independent from the choice of syndrome vs. decoding graphs. Therefore, we only need to focus on those from dual constraints \ref{eq:dual-lp-constraint-a}.
Second, because obstacles only occur on edges between different \ref{def:nodes}, it only needs to watch them to detect obstacles. Formally, we have

\vspace{1ex}\theoremobstacledetection{theorem:obstacle-detection}

An obstacle of \ref{eq:dual-lp-constraint-a} is detected if such a tight edge exists and $\Delta y_{S_1}+\Delta y_{S_2}>0$. That is, it can be detected by examining Covers of nodes on the decoding graphs.

\subsection{\seqalgo}\label{ssec:seq-key-idea}
The key idea of \seqalgo, as well as Sparse Blossom~\cite{higgott2023sparse}, is to detect obstacles using the decoding graph, leveraging the result of \ref{theorem:obstacle-detection}.
Therefore, \seqalgo, like Sparse Blossom, uses the existing design of the primal phase, e.g, that of Blossom V~\cite{kolmogorov2009blossom}.
Only in the dual phase, they eschew the use of the syndrome graph.
We will describe our implementation of \seqalgo in \S\ref{sec:implementation}.

Using the decoding graph to detect obstacles is more advantageous than the syndrome graph given a low physical error rate $p \ll 1$.
As explained in \S\ref{sec:qec}, generating the syndrome graph itself already takes quadratic time $O(|V|^2)$.
On the decoding graph, however, large \ref{def:covers} are exponentially unlikely with its size, so the average time complexity scales with roughly $O(|V|)$.
Note that when $|V|$ is small or when $p$ is large, it might be faster to use the syndrome graph.

\section{\paralgo}\label{sec:parallel}

\begin{figure*}[!t]
    \renewcommand*\thesubfigure{(\arabic{subfigure})}  
    \centering
    \begin{subfigure}{.16\textwidth}
        \centering
        \color{lightgray}\includegraphics[width=0.8\linewidth]{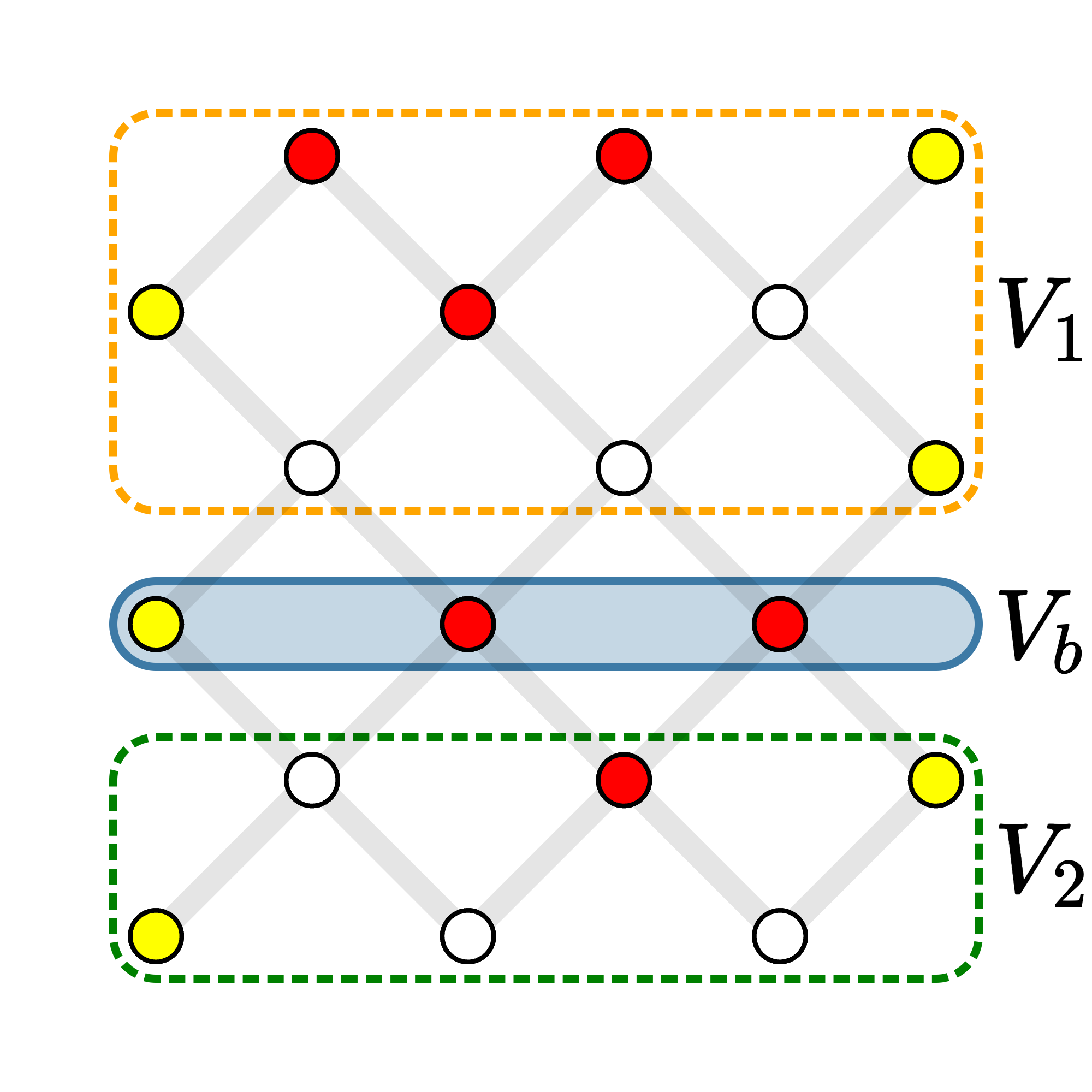}
        \caption{Partition}
        \label{fig:fusion-temporary-boundary}
    \end{subfigure}
    \begin{subfigure}{.16\textwidth}
        \centering
        \color{lightgray}\includegraphics[width=0.8\linewidth]{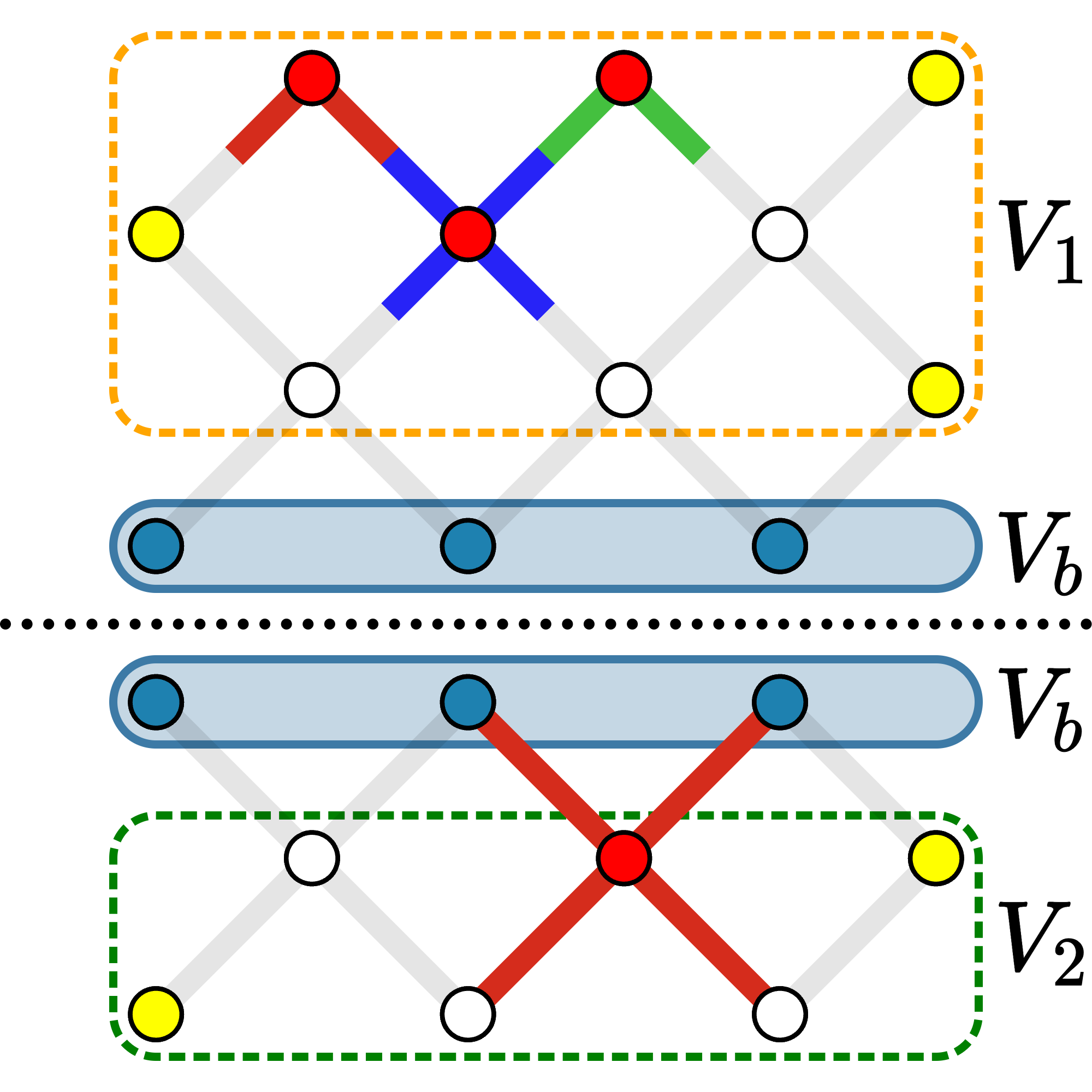}
        \caption{Individual Step 1}
        \label{fig:fusion-solve-individual-step-1}
    \end{subfigure}
    \begin{subfigure}{.16\textwidth}
        \centering
        \color{lightgray}\includegraphics[width=0.8\linewidth]{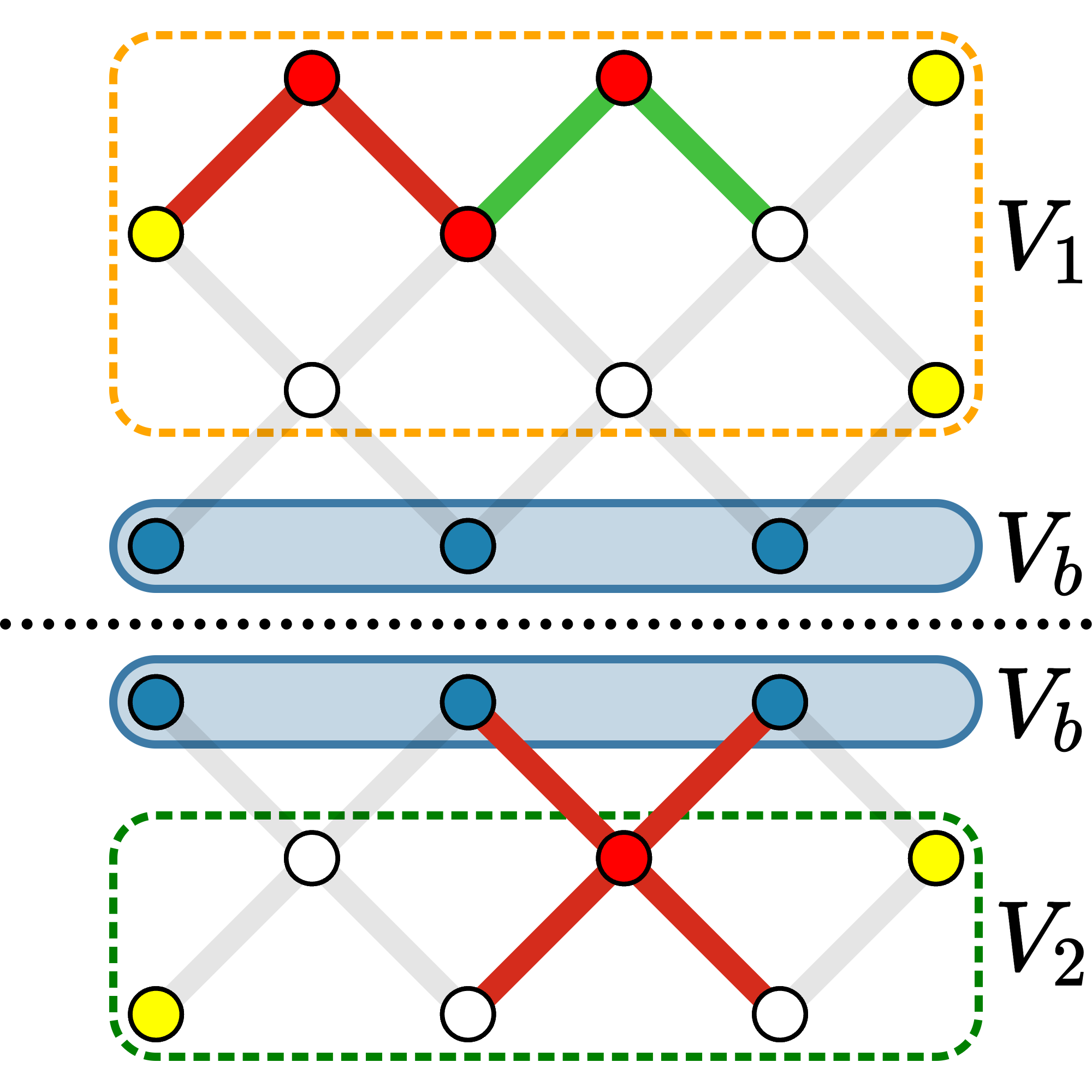}
        \caption{Individual Step 2}
        \label{fig:fusion-solve-individual-step-2}
    \end{subfigure}
    \begin{subfigure}{.16\textwidth}
        \centering
        \color{lightgray}\includegraphics[width=0.8\linewidth]{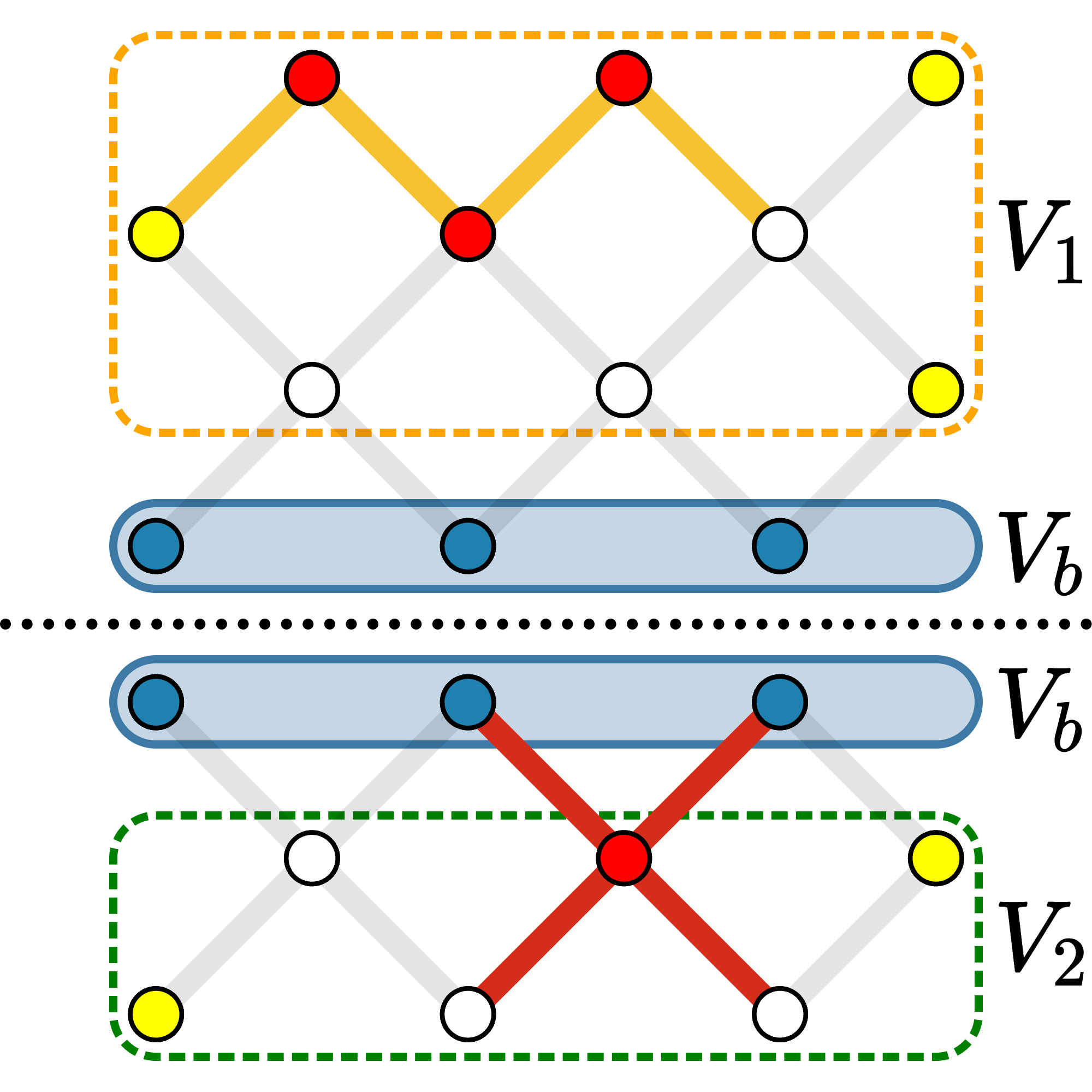}
        \caption{Individual Step 3}
        \label{fig:fusion-solve-individual-step-3}
    \end{subfigure}
    \begin{subfigure}{.16\textwidth}
        \centering
        \color{lightgray}\includegraphics[width=0.8\linewidth]{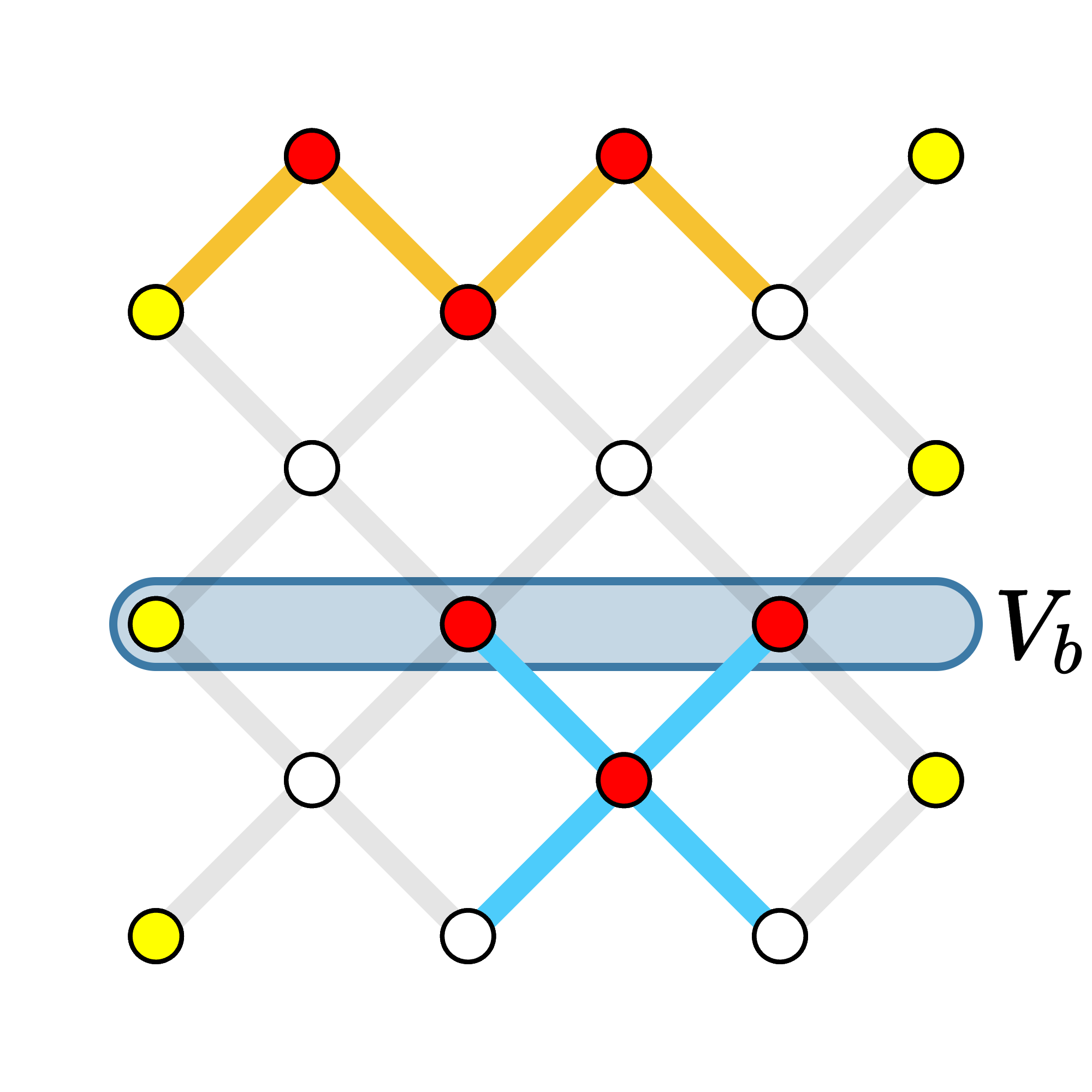}
        \caption{Recover $V_b$}
        \label{fig:fusion-recover}
    \end{subfigure}
    \begin{subfigure}{.16\textwidth}
        \centering
        \color{lightgray}\includegraphics[width=0.8\linewidth]{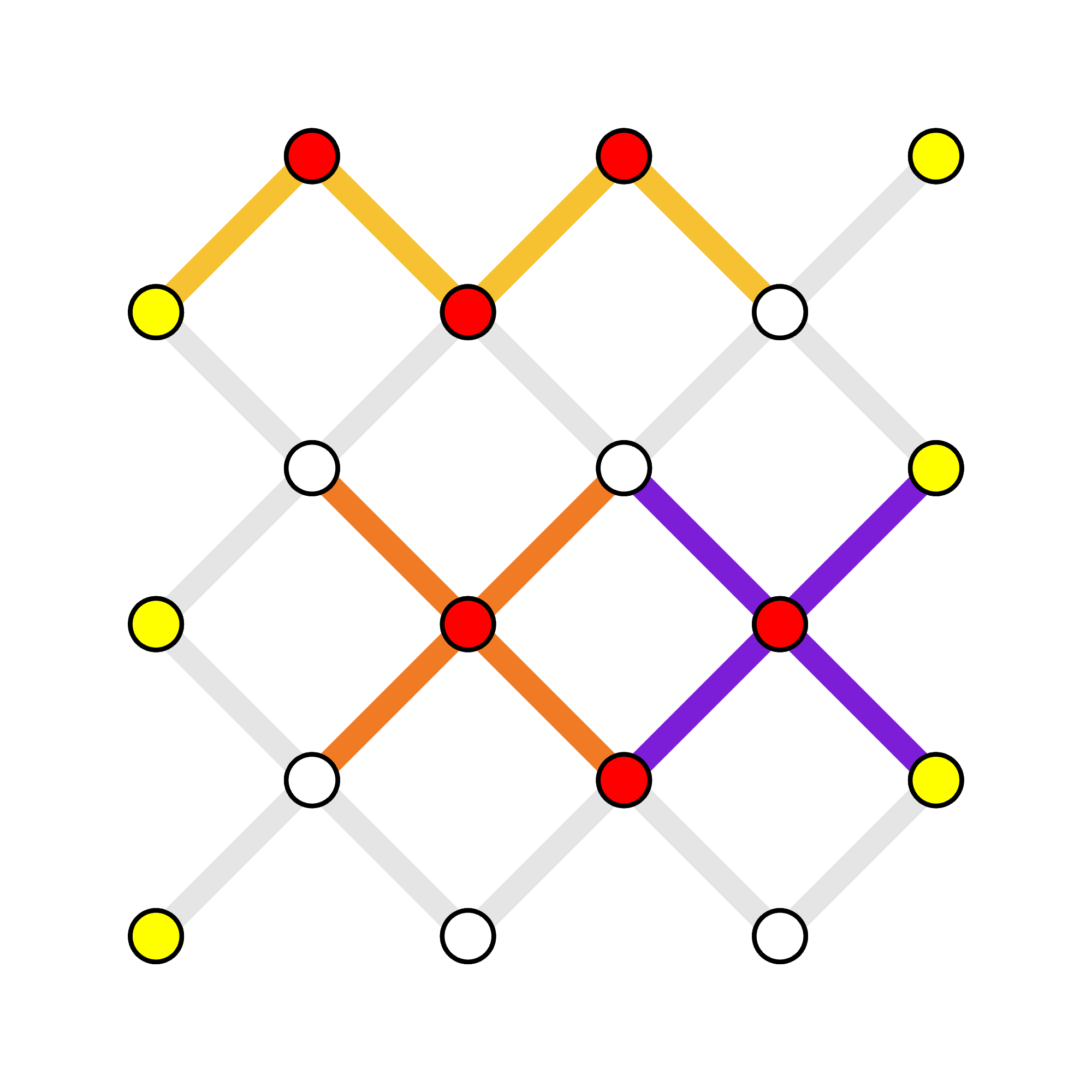}
        \caption{Fusion Completed}
        \label{fig:fusion-solve-all}
    \end{subfigure}
    \caption{Fusion Blossom example. The two sub-problems solves their local MWPM (2-4) individually, in parallel. The fusion operation first (5) recovers the temporary boundary vertices and then (6) evolves the intermediate state to a global MWPM.}
    \label{fig:fusion-example}
\end{figure*}

We next describe a parallel algorithm of solving the MWPM problem for QEC, called \emph{\paralgo}.
\paralgo recursively divides a decoding problem into sub-problems that can be solved independently and then recursively ``fuses'' their solutions to produce the solution to the original problem. 
We represent this recursive division/fusion as a full binary tree, called a \textit{fusion tree}.
Every leaf in the fusion tree invokes an MWPM solver, while other nodes fuse the solutions from their two children, also leveraging the MWPM solver.
In our implementation, the MWPM solver is \seqalgo.

We next provide the mathematical formulation of division and fusion in \S\ref{ssec:lp-decomposition} and \S\ref{ssec:fusion-operation}, respectively. We discuss how \paralgo can make tradeoffs between decoding time and latency in \S\ref{ssec:fusion-tree}.

\subsection{Division}\label{ssec:lp-decomposition}

As illustrated by \cref{fig:fusion-temporary-boundary}, a carefully selected set of vertices $V_b \subset V_M$, e.g., a minimum vertex cut~\cite{west2001vertexcut} of the decoding graph, can divide a decoding graph into two disjoint graphs that include $V_1$ and $V_2$, respectively. The only requirement of $V_b$ is that, there is no edge in the decoding graph that connects vertices from both $V_1$ and $V_2$.

With $V_b$, we can create two sub-problems, one working on the subgraph covering vertices $V_1 \cup V_b$ and the other that covering $V_2 \cup V_b$, as illustrated by \cref{fig:fusion-solve-individual-step-1,fig:fusion-solve-individual-step-2,fig:fusion-solve-individual-step-3}.
Each of the sub-problems treats a vertex from $V_b$ as a virtual vertex that can be matched arbitrary times. 
This effectively relaxes the parity constraints on vertices $V_b$ in the sub-problems, which will be tightened later by fusion.
For the $i$-th sub-problem, $i\in \{1,2\}$, the primal and dual formulations as in Eq.~\ref{eq:primal-lp} and \ref{eq:qec-dual-lp}, respectively, have $E$ and $\mathcal{O}^*$ as follows.
\begin{align*}
    E_i &= \{ e | e = (u, v) \in E \spaciousland u, v \in V_i \cup V_b \} \\
    \mathcal{O}_i^* &= \{ S | S \in \mathcal{O}^* \spaciousland S \subseteq V_i \}
\end{align*}

The process of division stops when the subproblem is adequately small for invoking the MWPM solver directly. We call such subproblems \emph{leaf problems} and the corresponding subgraphs \emph{leaf partitions}.

\subsection{Fusion}\label{ssec:fusion-operation}

After the sub-problems are solved independently, 
their solutions form an intermediate state for the original problem
in terms of the values of the primal and dual variables. The fusion operation invokes the MWPM solver to find a solution to the original problem starting with this intermediate state.

\paragraph{Correctness}
We next show that the intermediate state is indeed a valid state for the blossom algorithm. For the primal variables, we remove the matchings to the temporary boundary vertices $V_b$.
Those matched pairs break into alternating trees and search for new matchings.
Except for those, the matchings within $V_1$ or $V_2$ are preserved.
We also create an alternating tree for each defect vertex in $V_b$.
We simply keep the existing dual variables, as shown in \cref{fig:fusion-solve-individual-step-3} to \cref{fig:fusion-recover}, given

\vspace{1ex}\theoremvaliddualvariables{theorem:valid-dual-variables}

\paragraph{Speed}
We estimate the average time complexity of fusion operation to be no worse than $O(p|V_b|)$ where $p \ll 1$ is the physical error rate. 
Note the expected number of defect vertices in $V_b$ is also $O(p|V_b|)$.
Our estimate is based on two intuitions. First, the fusion operation only needs to break about the same number of matched pairs from the sub-problem solution to match the defect vertices in $V_b$. Second, due to the objective of minimum weight, it is more likely to find matches for these vertices close to $V_b$.
We note this estimate is independent of the size of the sub-problems $|V_1|$ and $|V_2|$.
We confirm this independence empirically in \S\ref{ssec:evaluation-fusion-time}.

\subsection{Schedule Design: Leaf Partitions and Fusion Tree}\label{ssec:fusion-tree}

\begin{figure}[!t]
    \centering
    \includegraphics[width=0.98\linewidth]{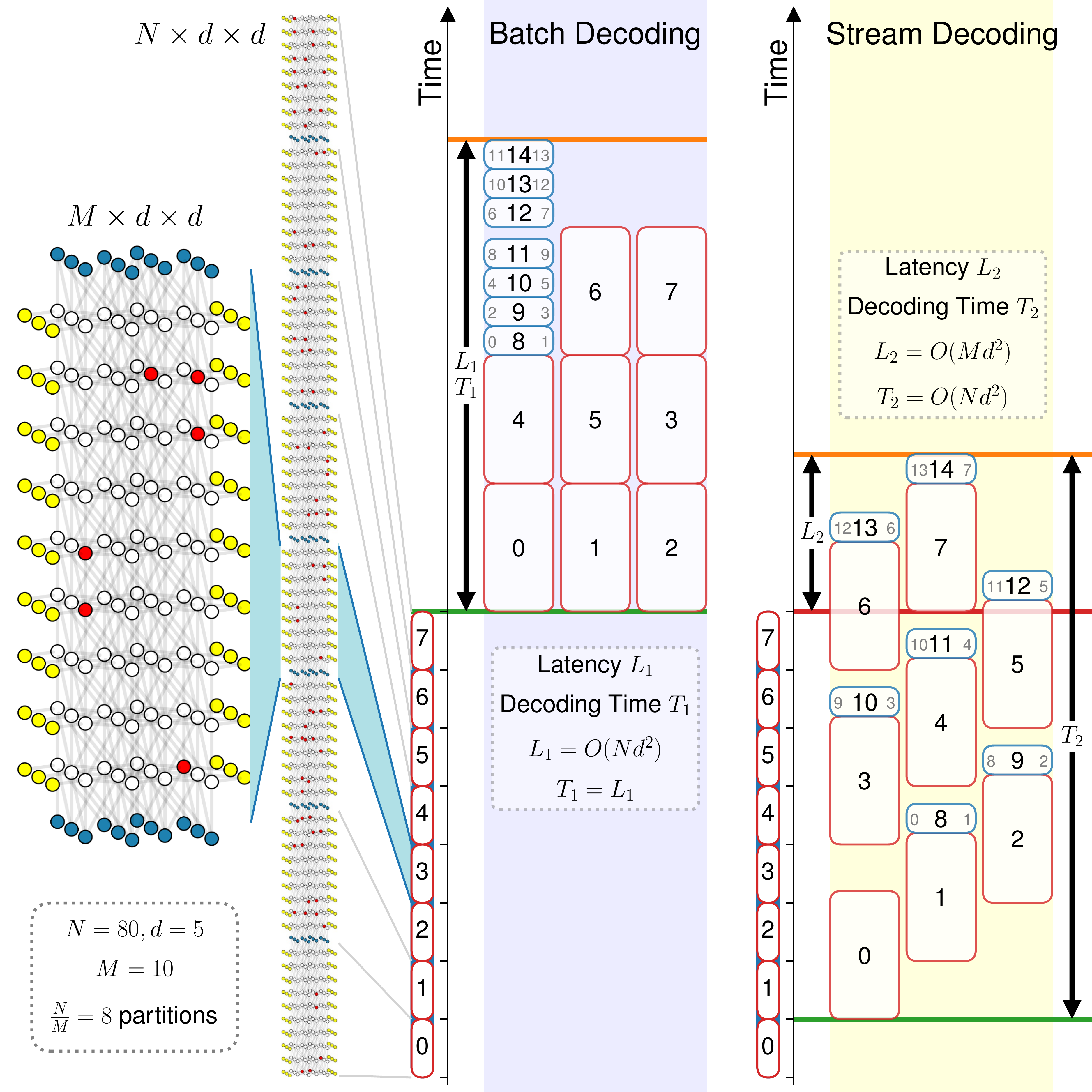}
    \caption{A 3D decoding graph (left), its Batch (center) and Stream (right) Decoding using three CPU cores. The Batch decoding starts when all $N=80$ rounds of measurements are ready; it fuses solutions according to the Balanced tree in~\cref{fig:fusion-tree}.
    The Stream decoding starts decoding whenever $M=10$ rounds of measurements for a leaf node are ready; it fuses solutions according to the Linear tree in~\cref{fig:fusion-tree}.}
    \label{fig:batch-vs-stream-decoding}
\end{figure}

When the leaf partitions are properly chosen, 
there can be multiple ways to fuse their solutions, allowing different tradeoffs between decoding time and latency. 
In this case, the fusion tree defines the space for scheduling leaf and fusion operations.
One particularly relevant case is illustrated in \cref{fig:batch-vs-stream-decoding}, where the decoding graph is a stream of measurement rounds, each a two-dimensional graph.
In this case, a leaf partition is simply a subgraph that consists of $M$ consecutive measurement rounds.

\begin{figure}[!t]
    \includegraphics[width=0.98\linewidth]{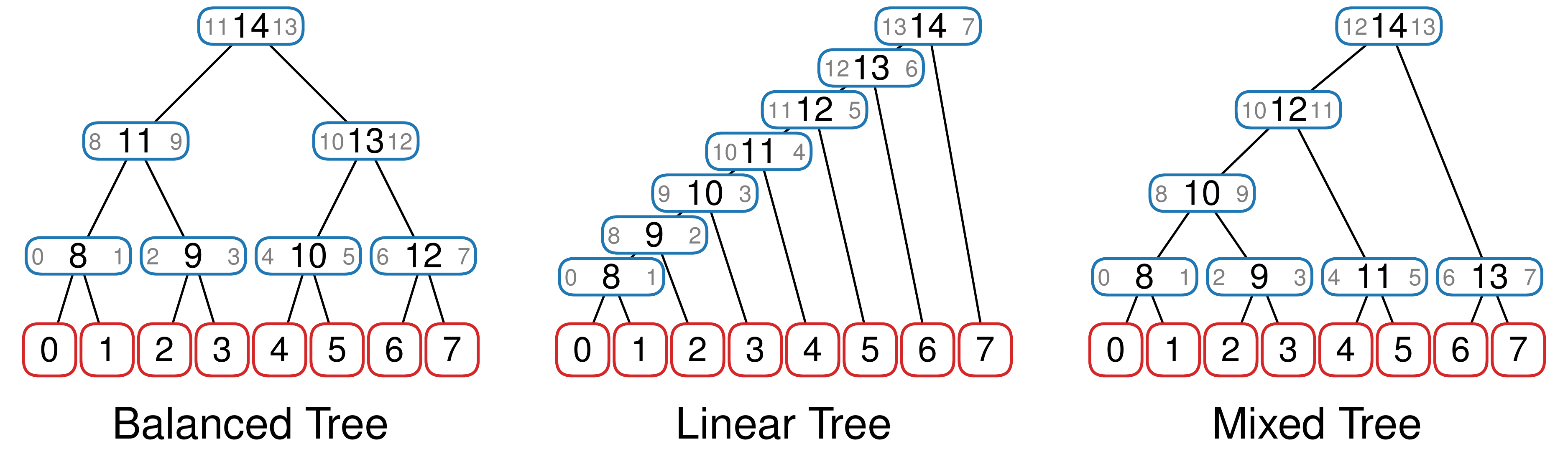}
    \caption{Fusion Trees. The leaves need to be fused recursively into a single root, which represents a global MWPM. Because a parent depends on the children, the paths between leaves and the root determine the decoding time and latency.}
    \label{fig:fusion-tree}
\end{figure}

When the measurement rounds of a leaf partition become available, it invokes the MWPM solver to produce a solution.
In an online system, as the measurement rounds stream in, the leaf partitions finish one by one.
However, there are many ways in which their solutions can be fused, with three examples shown in \cref{fig:fusion-tree}, each making a different trade-off between decoding time and latency. As illustrated in Fig.~\ref{fig:batch-vs-stream-decoding}, we define
\begin{itemize}
\item Decoding Time: $T$, the time from when decoding starts to when it finishes.
\item Latency: $L$, the time from when all measurements are ready to when decoding finishes.
\item Measurement Rounds: $N$, the number of rounds of stabilizer measurements.
\item Leaf Partition Size: $M$, the number of measurement rounds in each leaf partition.
\end{itemize}

We note the throughput of the system is related to decoding time as $N/T$: how many rounds it can decode per unit time.

\nosection{Batch Decoding.}
Prior studies generally assumed that the syndrome of all $N$ rounds of measurement is available at the time of decoding,
which is known as batch decoding (see \cref{fig:batch-vs-stream-decoding} (center)).
For batch decoding, the decoding latency and time are the same, i.e., $L=T$, and the decoding time is determined by the longest path from a leaf to the root given enough parallel resources. Therefore, the \textit{balanced tree}, as shown in \cref{fig:fusion-tree} (left), is preferable since its longest path (from leaf to root) is the shortest.

\nosection{Stream Decoding.}
In contrast to batch decoding, stream decoding starts as soon as enough rounds of measurement are ready for a leaf node (see \cref{fig:batch-vs-stream-decoding} (right)).
As a result, the decoding latency can be substantially shorter than the decoding time, i.e., $L<T$.
More importantly, to determine the decoding latency, one can no longer simply consider paths between leaves and the root but must add the time when the rounds of measurement for a leaf is ready (assuming those for the first leaf is ready at time zero).
To minimize the decoding latency, one must balance the path length plus the ready time for all paths, allowing a shorter path for a later leaf.
For example, when there are enough parallel resources that are fast enough, the path between the last leaf and the root determines the decoding latency. In this case, the \emph{linear tree} (\cref{fig:fusion-tree} (center)) is preferable.

With the balanced and linear trees as the two extreme cases in mind, we can create a continuum of trees between them called \textit{mixed trees}.  Given the parallel resources and decoding setup, e.g., $M$ and $N$, one must examine this continuum to find the tree that achieves the shortest decoding latency.
To construct a mixed tree, one selects a height in the balanced tree, keeps balanced sub-trees below the height but constructs a linear tree above it. The higher this height, the smaller path difference between earlier and later leaves. For the balanced and linear trees, this mix height is root and leaf, respectively. \cref{fig:fusion-tree} (right) shows the mixed tree with the mix height of one.
In our latency evaluation (\S\ref{ssec:evaluation-latency}), we use the mixed tree that minimizes the decoding latency.

We note that the mix height can be determined dynamically: the decoder can start with a balanced tree and switch to a linear tree to optimize the performance of the system.

\section{Implementation}
\label{sec:implementation}

We next describe our implementation of \seqalgo and \paralgo, including major ideas for optimizations.
We implemented these algorithms in Rust with 12k lines of code.

\subsection{Unified Framework for Matching Decoders}
\label{sec:unified}
\begin{figure}[!t]
    \centering
    \scalebox{0.9}{
        \begin{tikzpicture}%
\path[primalinterfacecolor,thick,draw,fill=primalinterfacecolor,rounded corners=1mm] (2.3,2.0) rectangle (2.5,3.0);%
\path[thick,draw,rounded corners=1mm] (0.19999999999999996,2.0) rectangle (2.5,3.0);%
\node[color=black,align=center,font=\small] at (1.2499999999999998,2.5) {\hyperref[ssec:impl-primal-serial]{\code{Standard}}};%
\path[primalinterfacecolor,thick,draw,fill=primalinterfacecolor,rounded corners=1mm] (2.3,0.5) rectangle (2.5,1.5);%
\path[thick,draw,rounded corners=1mm] (0.19999999999999996,0.5) rectangle (2.5,1.5);%
\node[color=black,align=center,font=\small] at (1.2499999999999998,1.0) {\code{Union-Find}};%
\path[dualinterfacecolor,thick,draw,fill=dualinterfacecolor,rounded corners=1mm] (6.1,2.0) rectangle (6.3,3.0);%
\path[thick,draw,rounded corners=1mm] (6.1,2.0) rectangle (8.4,3.0);%
\node[color=black,align=center,font=\small] at (7.35,2.5) {\code{Standard}};%
\path[dualinterfacecolor,thick,draw,fill=dualinterfacecolor,rounded corners=1mm] (6.1,0.5) rectangle (6.3,1.5);%
\path[thick,draw,rounded corners=1mm] (6.1,0.5) rectangle (8.4,1.5);%
\node[color=black,align=center,font=\small] at (7.35,1.0) {\hyperref[ssec:impl-dual-serial]{\code{Parity}}};%
\node[color=black,align=center,font=\small] at (1.2499999999999998,3.6000000000000005) {Primal Modules};%
\path[primalinterfacecolor,thick,draw,fill=primalinterfacecolor,rounded corners=1mm] (2.4999999999999996,3.2000000000000006) rectangle (2.6999999999999997,4.000000000000001);%
\path[thick,dotted,draw,rounded corners=1mm] (0.0,0.3) rectangle (2.6999999999999997,4.2);%
\node[color=black,align=center,font=\small] at (7.35,3.6000000000000005) {Dual Modules};%
\path[dualinterfacecolor,thick,draw,fill=dualinterfacecolor,rounded corners=1mm] (5.9,3.2000000000000006) rectangle (6.1000000000000005,4.000000000000001);%
\path[thick,dotted,draw,rounded corners=1mm] (5.9,0.3) rectangle (8.6,4.2);%
\path[thick,-{Stealth},draw] (2.7,3.8000000000000007) -- (5.8999999999999995,3.8000000000000007);%
\node[color=black,align=center,font=\small] at (4.3,4.000000000000001) {\ref{def:direction}};%
\path[thick,-{Stealth},draw] (5.8999999999999995,3.4000000000000004) -- (2.7,3.4000000000000004);%
\node[color=black,align=center,font=\small] at (4.3,3.6000000000000005) {\ref{def:obstacles}};%
\path[densely dashed,-,draw] (3.1999999999999997,2.5) -- (2.4999999999999996,2.5);%
\path[densely dashed,-,draw] (5.4,2.5) -- (6.1,2.5);%
\path[densely dashed,-,thick,primalinterfacecolor,draw] (3.3,1.75) -- (2.4999999999999996,2.2);%
\path[densely dashed,-,thick,dualinterfacecolor,draw] (5.3,1.75) -- (6.1,1.3);%
\path[densely dashed,-,draw] (3.5,1.0) -- (2.4999999999999996,1.0);%
\path[densely dashed,-,draw] (5.1,1.0) -- (6.1,1.0);%
\path[thick,draw,fill=paritydecodercolor] (4.3,1.75) ellipse[x radius=1.2,y radius=0.3];%
\node[color=black,align=center,font=\footnotesize] at (4.3,2.5) {traditional\\MWPM Decoder};%
\node[color=black,align=center,font=\footnotesize] at (4.3,1.75) {\hyperref[ssec:impl-serial]{Parity Blossom}};%
\node[color=black,align=center,font=\footnotesize] at (4.3,1.0) {(weighted)\\UF Decoder};%
\end{tikzpicture}
    }
    \caption{Unified Framework for Matching Decoders.}
    \label{fig:matching-decoders}
\end{figure}
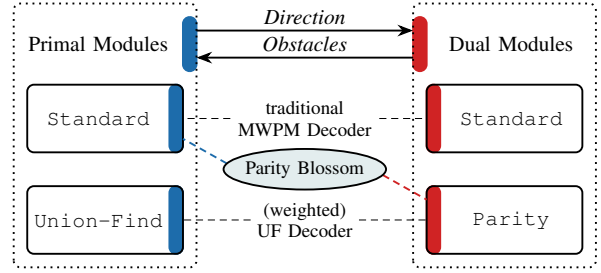

A key idea behind our implementation is to use a unified framework for the blossom algorithm and its variants, including \seqalgo, Union-Find~\cite{delfosse2021almost}, and more, as illustrated by \cref{fig:matching-decoders}.
As the blossom algorithm iterates between the primal and dual phases as shown in \cref{fig:primal-dual-interface}, our unified framework implements them in modules with a narrow, well-defined interface with each other, marked as red and blue in \cref{fig:primal-dual-interface,fig:matching-decoders}.
The interface allows any primal module to work with any dual module.

This framework serves three purposes.
(\textit{i}) First, it shows how these variants and the blossom algorithm are related.
(\textit{ii}) Second, it allows code reuse between their implementations.
For example, \seqalgo optimizes the dual module (\code{Parity}) to work on the decoding graph instead of the syndrome graph (\S\ref{ssec:impl-dual-serial}).
It uses the same primal module (\code{Standard}) as the original blossom algorithm, with slight modification to support virtual vertices (\S\ref{ssec:impl-primal-serial}).
For another example, the UF decoder and \seqalgo share the same \code{Parity} dual module. 
The UF decoder uses its own primal module (\code{Union-Find}) that computes the direction approximately, compared to the blossom algorithm~\cite{wu2022interpretation}.
(\textit{iii}) Third, this framework reveals the existence of previously unknown variants that can achieve different trade-offs between accuracy and speed when applied to QEC.
It also allows such new variants to be easily implemented.
For example, one can design a new primal module based on the \code{Standard} primal module, which sets a limit to the size of alternating trees.
Once an alternating tree reaches the size limit, the module treats it as an invalid cluster like the \code{Union-Find} primal module.
When the size limit is infinite, the decoder is identical to \seqalgo; when the limit is zero, it is identical to the UF decoder.
As a result, by adjusting the size limit, we can produce a continuum of decoders between the UF decoder and \seqalgo, making different tradeoffs between decoding accuracy and speed.

\subsection{\seqalgo}\label{ssec:impl-serial}

As shown in \cref{fig:matching-decoders}, \seqalgo uses the \code{Parity} Dual Module and the \code{Standard} Primal Module.

\subsubsection{\emph{\code{Parity}} Dual Module}\label{ssec:impl-dual-serial}
Because \seqalgo uses the decoding graph to detect \ref{def:obstacles}, given \ref{theorem:obstacle-detection}, the \code{Parity} dual module must efficiently track the Covers of nodes.

Our first implementation idea comes from the UF decoder~\cite{delfosse2020linear}: we maintain the boundary edges for each Cover.
This is efficient and sufficient for updating Covers because when dual variables are adjusted according to $\Delta \vec{y}$, the boundary edges of their Covers change.

Our second idea removes implementation complication resulting from the fact that some vertices from the decoding graph may belong to multiple Covers. 
Zero edges, representing erasure errors~\cite{wu2022erasure,teoh2022dual}, specifically contribute to this complication because their vertices can belong to many Covers. 
To ensure that a vertex belongs to at most one ``Cover'', our idea is to use Pseudo-Covers, derived from Covers as follows. 

\vspace{1ex}\definitionlabel{def:pseudo-cover}{Pseudo-Cover}
All Covers with a single vertex are Pseudo-Cover. That is, when the blossom algorithm starts, all the Covers on the decoding graph are Pseudo-Covers, each of them with a single defect vertex.
(\textit{i}) At the beginning of a dual phase, for a node with $\Delta y_S < 0$, its Pseudo-Cover is derived by removing all boundary non-defect vertices. 
(\textit{ii}) The Pseudo-Cover for a node $S$ with $\Delta y_S > 0$ is derived by modifying how its Cover grows.
When adding a vertex to a growing Pseudo-Cover, the growing stops if the vertex is already inside another Pseudo-Cover.
We denote the Pseudo-Cover of Cover$(S)$ with $\overline{\text{Cover}}(S)$. 

\vspace{1ex} Since a vertex belongs to at most one $\overline{\text{Cover}}$, its memory usage is constant.
Also, since the incident vertices of an edge $e = (u, v)$ each belongs to at most one $\overline{\text{Cover}}$, there are at most two covered segment edges on $e$.
That is, the memory usage of an edge is also constant.
The next theorem says that Pseudo-Covers can also be used to detect tight edges.

\vspace{1ex}\theoremobstacledetectionoptimized{theorem:obstacle-detection-optimized}

Using this theorem, the \code{Parity} dual module checks whether such $S_3$ and $S_4$ exists for each decoding graph edge and reports all detected obstacles to the primal module.
We note that the above theorem differs from \ref{theorem:obstacle-detection} in a fundamental way: it does not detect all obstacles but at most one for each decoding graph edge.

We have not implemented an important optimization used by 
Sparse Blossom~\cite{higgott2023sparse} and Blossom V~\cite{kolmogorov2009blossom}: they sort and process edges based on when they will become tight using priority queues. Instead, our dual module implementation enumerate all edges, like the UF decoder~\cite{delfosse2020linear}.

\subsubsection{\emph{\code{Standard}} Primal Module}\label{ssec:impl-primal-serial}
We base the \code{Standard} primal module on the implementation of blossom V~\cite{kolmogorov2009blossom} with three optimizations. We note these optimizations can be generally useful beyond \seqalgo.

First, it supports virtual vertices. Prior works, using the blossom V library, had to emulate them, which is inefficient.

Second, each time the dual module is invoked, it prepares Pseudo Covers based on the directions of their blossoms, which incurs significant overhead. To amortize this overhead, our \code{Standard} primal module handle all reported \obstacles each time it is invoked, instead of returning to the dual module after handling one.

Third, we grow all alternating trees simultaneously like the UF decoder and Sparse Blossom~\cite{higgott2023sparse}.
It corresponds to ``the multiple-tree approach with fixed $\delta$'' reported in~\cite{kolmogorov2009blossom} and applied by the blossom V library to only 5\% of the nodes.
As shown by the authors of Sparse Blossom, this approach explores fewer edges on average and thus is faster.

\subsection{\paralgo}\label{ssec:impl-parallel-modules}
We implement \paralgo with \seqalgo as the MWPM solver, and using the Rayon parallel programming library~\cite{rayon-rs}. 
A manager thread reads the fusion tree from leaves up.
It creates a job for each node in the fusion tree when the jobs for the children nodes have returned. 
The manager thread inserts new jobs into a queue from which a group of worker threads remove jobs and complete them. 
A job invokes \seqalgo implemented in the unified framework as the MWPM solver. 

During the fusion operation, the MWPM solver is invoked to evolve the intermediate state to an optimum for the fused problem.
A na\"ive implementation would construct the internal data structures for solving the fused problem from the output of solving the children sub-problems, leading to excessive memory copying. 
Instead, our implementation allows the MWPM solver invoked by a parent to reuse, i.e., operate directly on, the internal data structures of the MWPM solver invoked by its child. 
As the system progresses from a leaf toward the root in the fusion tree, it maintains the internal data structures for the MWPM solver invoked by each node of the fusion tree, with those of a parent including those of its children.

Organizing the internal data structures hierarchically based on the fusion tree as described above brings an additional opportunity to optimize the MWPM solver when invoked by a parent for the fusion operation. 
Since the fusion operation only changes the \ref{def:pseudo-covers} in a small region around the boundary vertices $V_b$, the MWPM solver ideally should only work on data structures related to this small region.
We achieve this by allowing the MWPM solver invoked by the parent to ``invoke'' the MWPM solver of its child to evolve the child's internal data structures.
This is possible because the child's internal data structures are maintained in place. 

\subsection{Other Optimizations}

To improve performance, we implement an optional feature that leverages \code{unsafe} Rust to bypass safety checks when it is safe to do so at the algorithm level.
By enabling the optional \code{dangerous\_pointer} feature, it results in a speedup of approximately 2x compared to the standard build.

Since the initialization time scales with $|V|$ yet the decoding time scales with $p|V|$, it is more practical to reset the decoder between simulation shots rather than creating a new one.
Our optimization achieves a reset time of $O(1)$.
The key idea is to avoid enumerating all the decoding graph edges to reset the Pseudo Covers stored on them.
In implementing this idea, the decoder keeps a global timestamp and each edge has its own timestamp, all initialized to 0.
The global timestamp advances by 1 on a global reset.
An edge is invalid if its timestamp does not match the global one.
Only when an invalid edge is being accessed, it is reset and its timestamp is updated to the global timestamp.

\section{Evaluation}\label{sec:evaluation}

We evaluate our implementations of \seqalgo and \paralgo with both macro and micro benchmarks.
The evaluation answers the following questions.

\begin{itemize}
    \item Correctness: Are they exact MWPM decoders?
    \item Throughput: How many rounds of measurement can be decoded per unit time?
    \item Latency: How long does it take from when the last round of measurement arrives to when decoding finishes? 
    \item Scalability: How throughput changes with code distance?
\end{itemize}

We verify the correctness of our implementations by comparing against the blossom V library~\cite{kolmogorov2009blossom} over millions of randomized test cases with tractable code distances up to $19$. We focus on throughput, latency, and scalability in the rest of this section.

\subsection{Setup}\label{ssec:setup}
\subsubsection{Noise Model}
We use the circuit-level noise model~\cite{landahl2011fault} with a physical error rate of $0.1\%$.
We use a rotated surface code shown in \cref{fig:surface-code}.
It has $n = d^2$ data qubits and $(d^2-1)/2$ Z (X) stabilizers.
Given a syndrome of $N$ noisy rounds of measurement, the Z (X) decoding graph has $(N+1)(d^2-1)/2$ ordinary vertices and $(N+1)(d+1)$ virtual vertices, a total of $(N+1)(d+1)^2/2$.
Since the X and Z decoding graphs can be decoded independently, we only use the Z decoding graph for evaluation.
For simplicity, we use format $N \times d \times d$ to represent the code.

\subsubsection{Measurement}
We evaluate the decoding speed on a Linux server with dual Intel Xeon Platinum 8375C CPUs, a total of 64 cores, each supporting two hyper-threads.
The server is an M6i instance from AWS (Amazon Web Services).
Our results do not include the initialization time, during which the one-time, expensive memory allocation is performed.
Once initialized, the decoder works on 100 simulation shots consecutively. Between two shots, the decoder is reset with a constant overhead, which is included in the result. Each shot by default includes $10^5$ rounds of measurement.

\subsubsection{Baseline}\label{sssec:evaluation-baseline}
For Sparse Blossom, we use the authors' own implementation through its Python binding~\cite{pymatchingv2} with the same setup as the above, with batch optimization enabled. 
For the traditional MWPM decoder, we use the blossom V library~\cite{kolmogorov2009blossom} with the following optimizations.
It pre-computes a complete graph of $V_M$ offline to reduce the runtime overhead of constructing the syndrome graph.
It also eliminates edges in the complete graph if they have higher weight than the two vertices matching to virtual boundary respectively, because these edges would never be selected in an MWPM.

\subsubsection{Metrics}
Given the decoding time $T$ and measurement rounds $N$, we define
\begin{itemize}
    \item Throughput: $N/T$, decoded rounds per unit time.
    \item Decoding time per measurement round: $T/N$, the inverse of throughput. Since this is easier to compare with the measurement cycle of a quantum hardware, we use it in lieu of throughput in the figures.
\end{itemize}

\begin{figure}[!t]
    \renewcommand*\thesubfigure{(\alph{subfigure})}  
    \centering
    \begin{subfigure}{.49\linewidth}
        \centering
        \includegraphics[width=\linewidth,evaluationtrim]{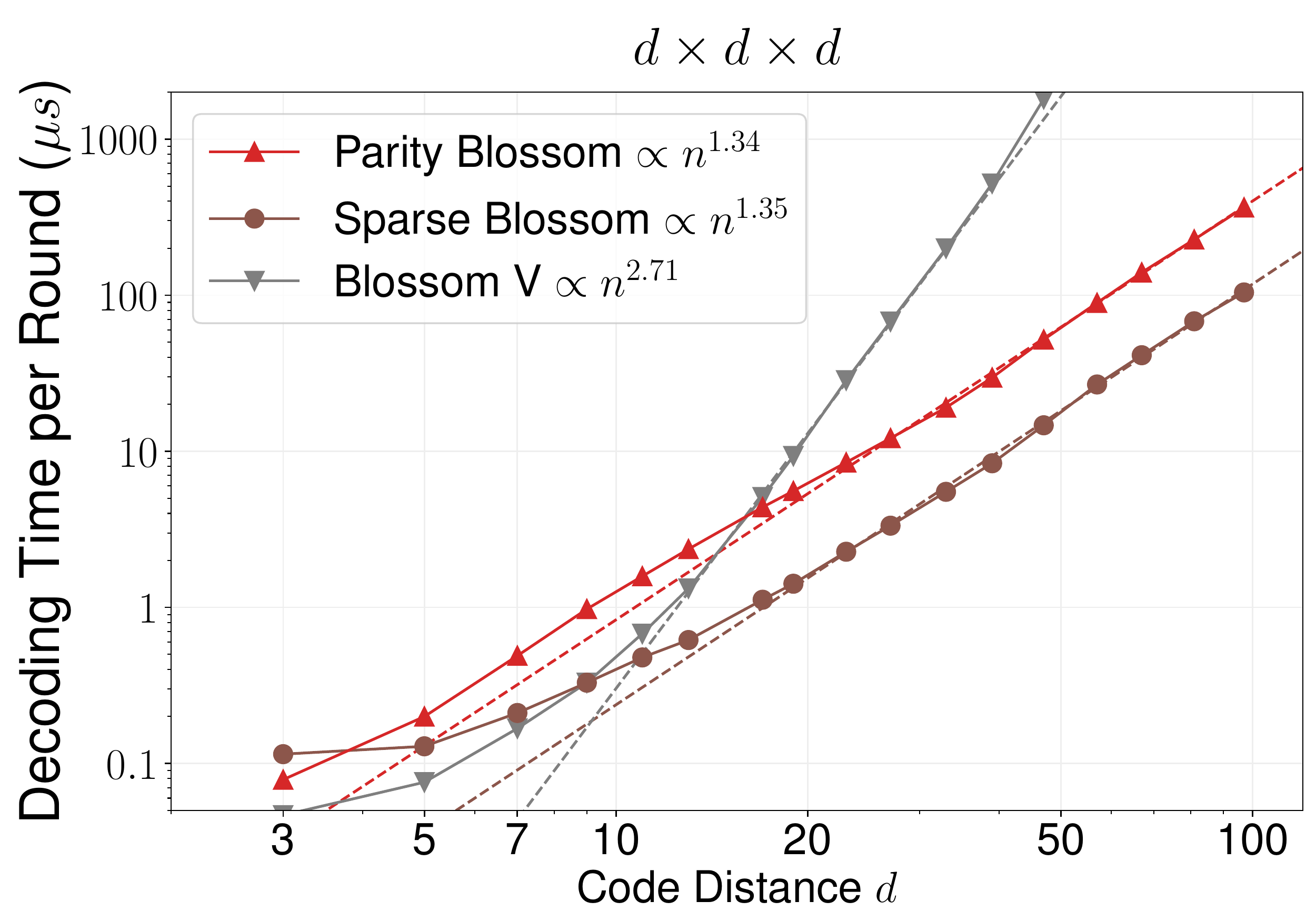}
        \caption{Code Distance $d$}
        \label{fig:serial-decoding-time-d}
    \end{subfigure}
    \begin{subfigure}{.49\linewidth}
        \centering
        \includegraphics[width=\linewidth,evaluationtrim]{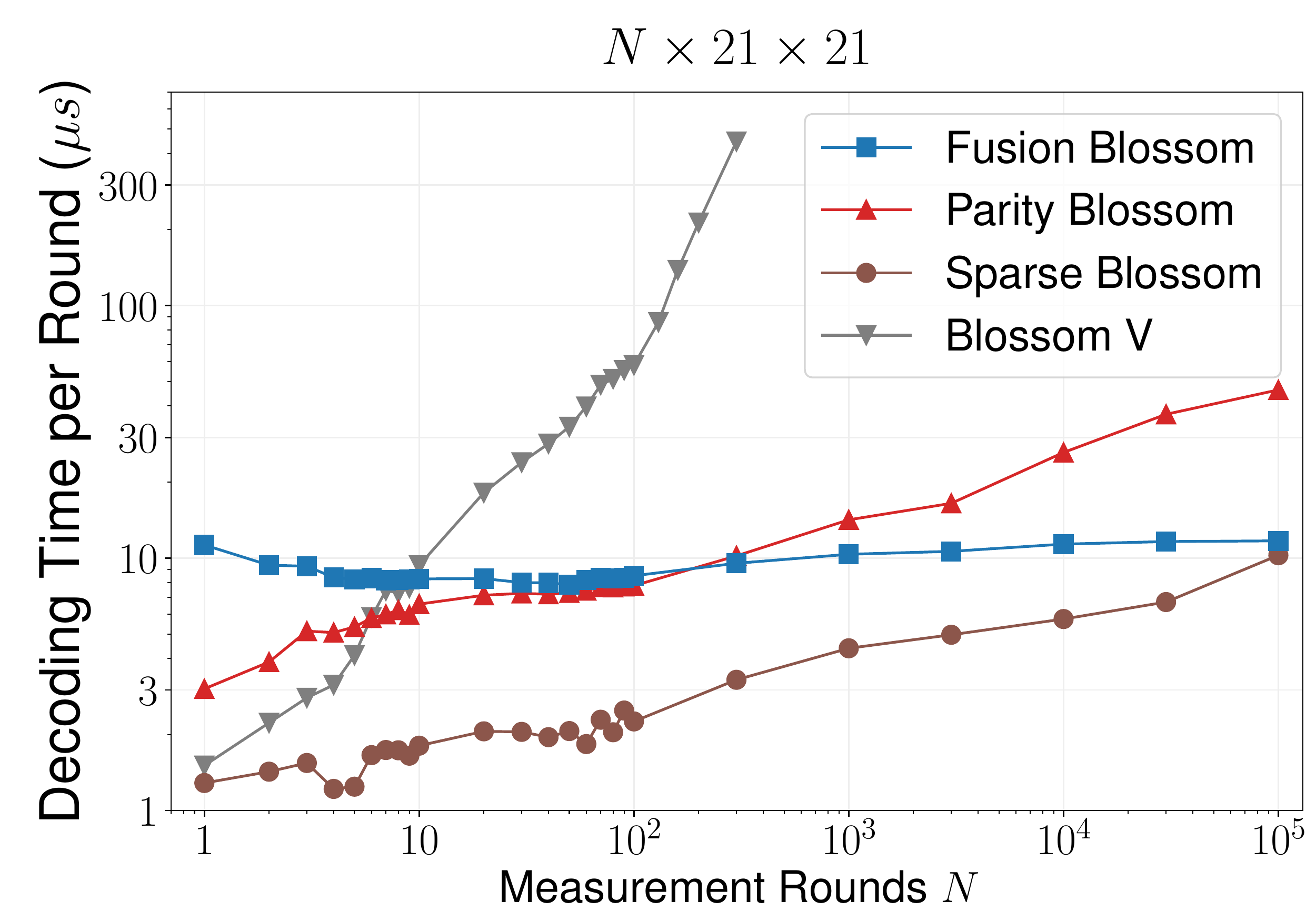}
        \caption{Measurement Rounds $N$}
        \label{fig:serial-decoding-time-T}
    \end{subfigure}
    \caption{Decoding time with a single thread (lower the better).}
    \label{fig:serial-decoding-time}
    
    \begin{subfigure}{.49\linewidth}
        \centering
        \includegraphics[width=\linewidth,evaluationtrim]{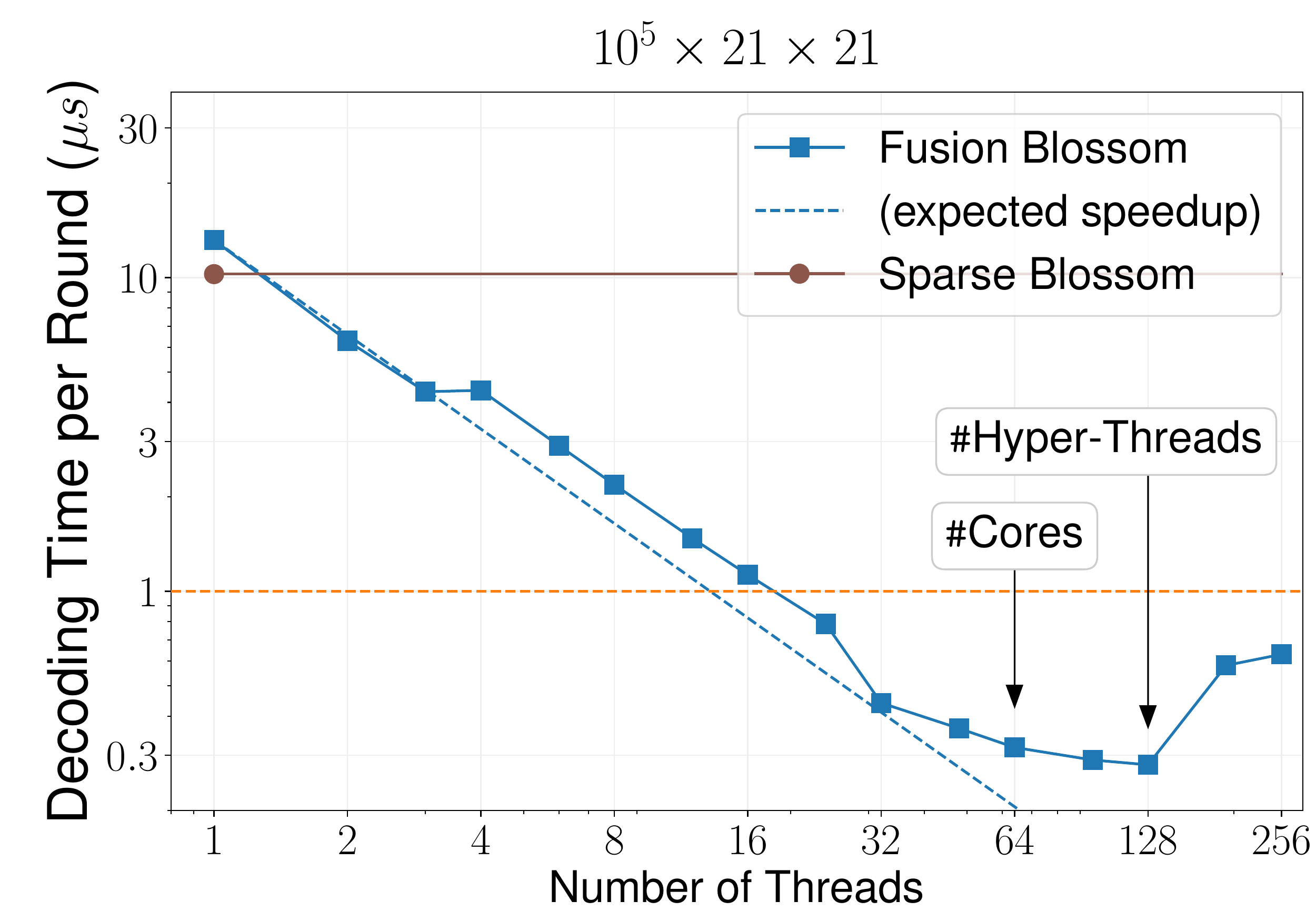}
        \caption{Number of Threads}
        \label{fig:parallel-decoding-time-threads}
    \end{subfigure}
    \begin{subfigure}{.49\linewidth}
        \centering
        \includegraphics[width=\linewidth,evaluationtrim]{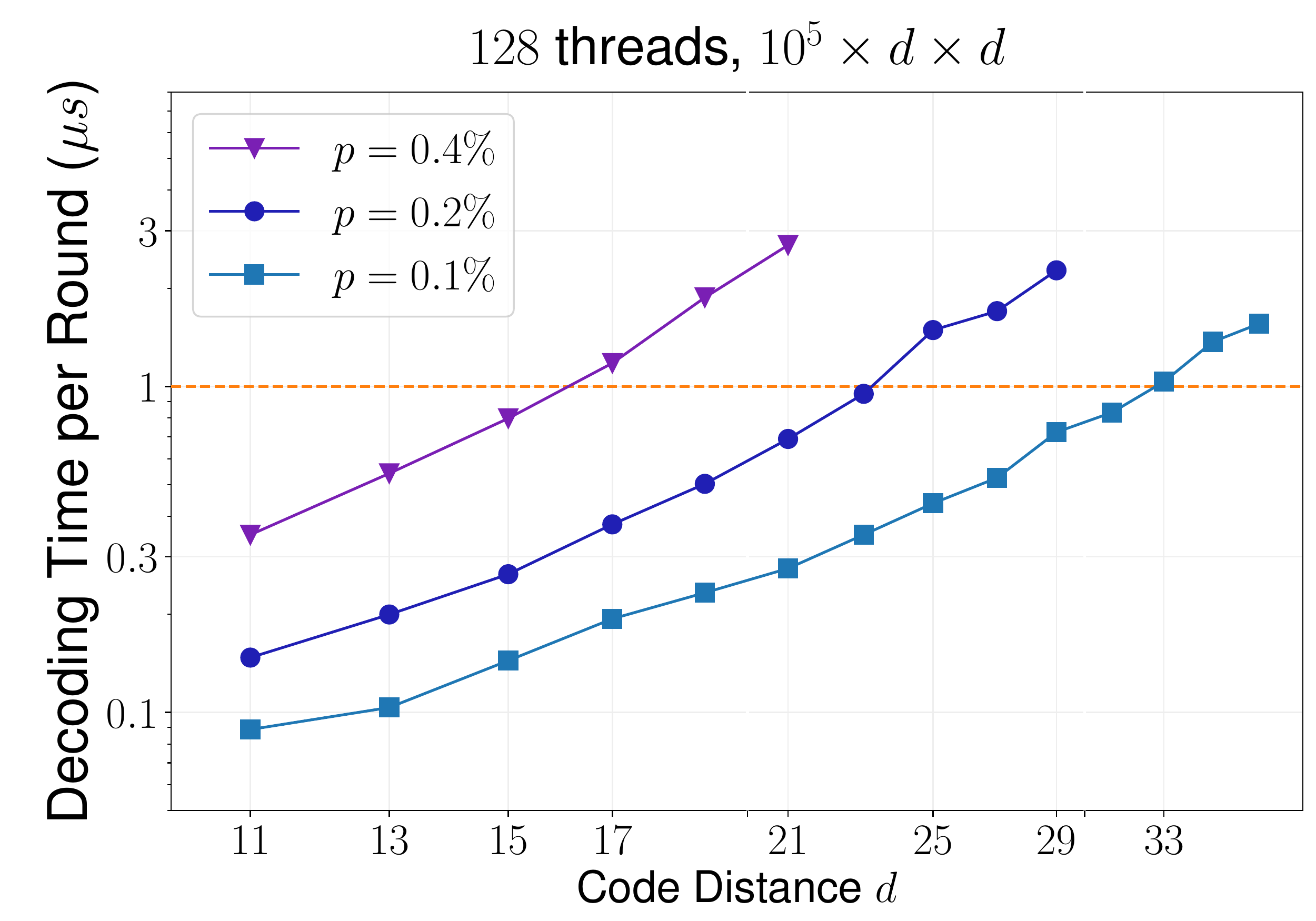}
        \caption{Code Distance $d$}
        \label{fig:parallel-decoding-time-d}
    \end{subfigure}
    \caption{Parallel decoding time (lower the better).}
    \label{fig:parallel-decoding-time}
\end{figure}

\begin{figure*}[!htb]
    \centering
    \begin{minipage}{.228\linewidth}
        \renewcommand*\thesubfigure{}
        \centering
        \begin{subfigure}{.99\linewidth}
            \centering
            \includegraphics[width=\linewidth,evaluationtrim]{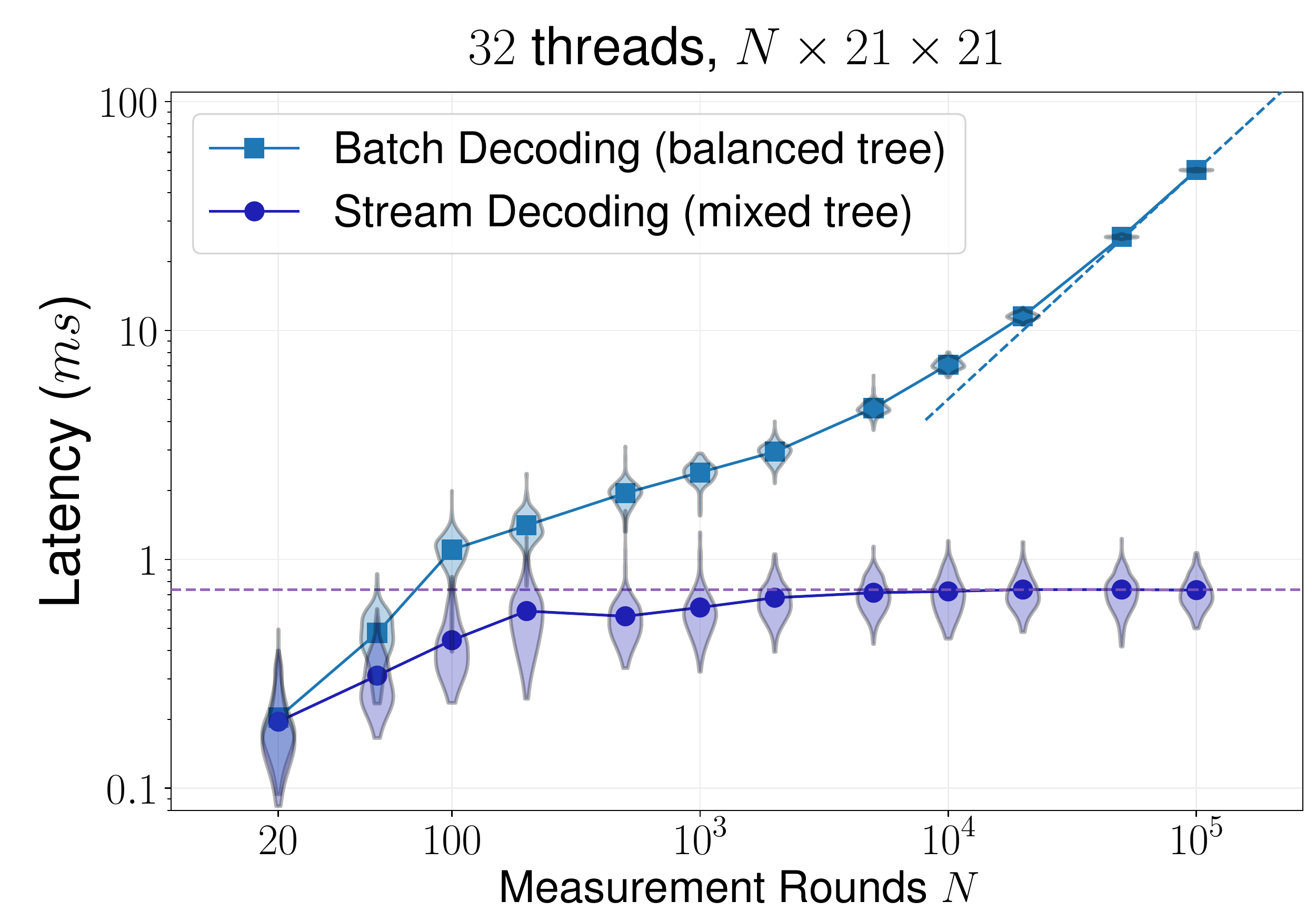}
            \caption{Measurement Rounds $N$}
            \label{fig:evaluation-latency}
        \end{subfigure}
        \caption{Latency}
        \label{fig:latency}
    \end{minipage}%
    \hspace{3.5ex}
    \begin{minipage}{0.735\linewidth}
        \renewcommand*\thesubfigure{(\alph{subfigure})}  
        \centering
        \begin{subfigure}{.31\linewidth}
            \centering
            \includegraphics[width=\linewidth,evaluationtrim]{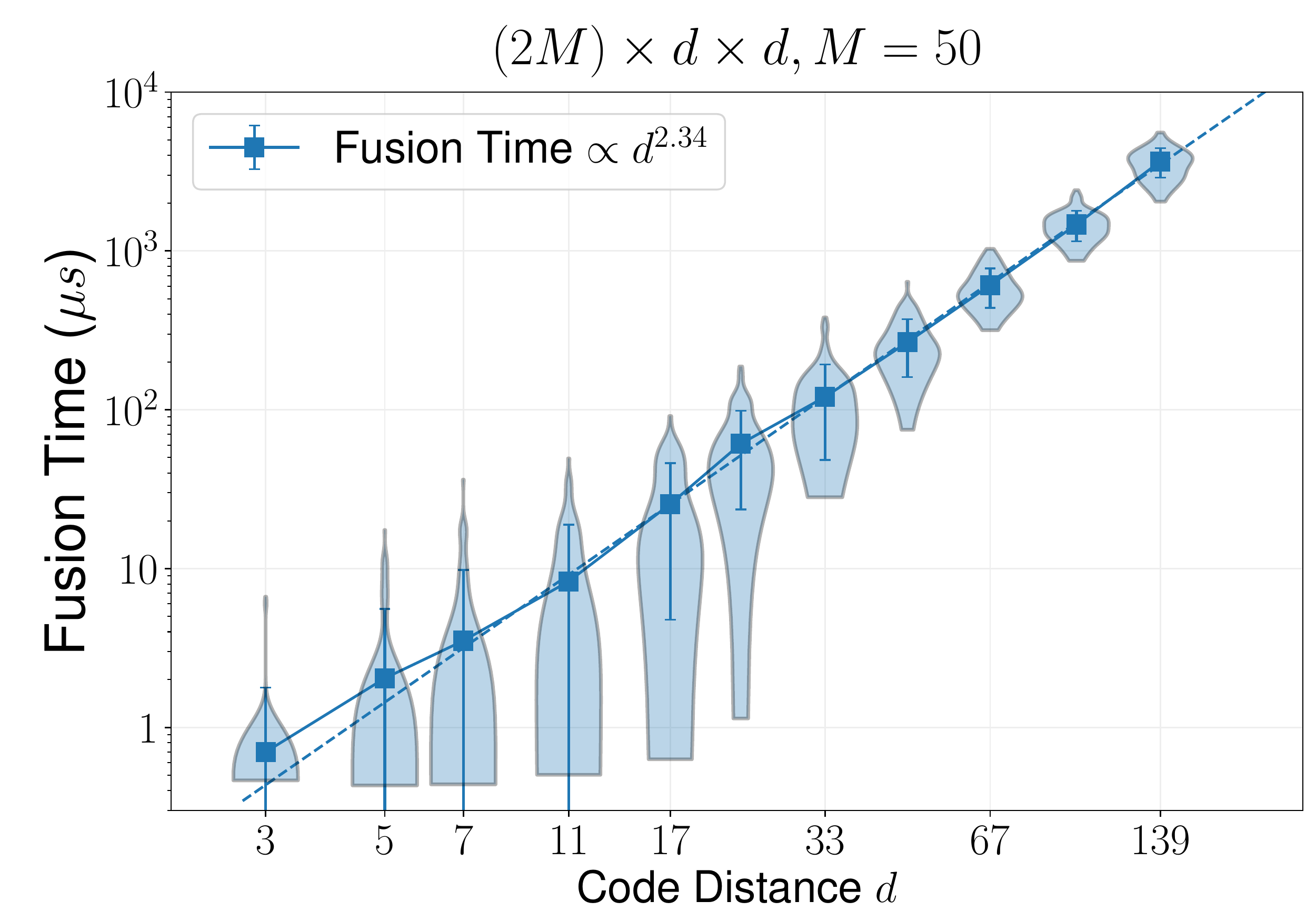}
            \caption{Code Distance $d$}
            \label{fig:fusion-time-d}
        \end{subfigure}
        \begin{subfigure}{.31\linewidth}
            \centering
            \includegraphics[width=\linewidth,evaluationtrim]{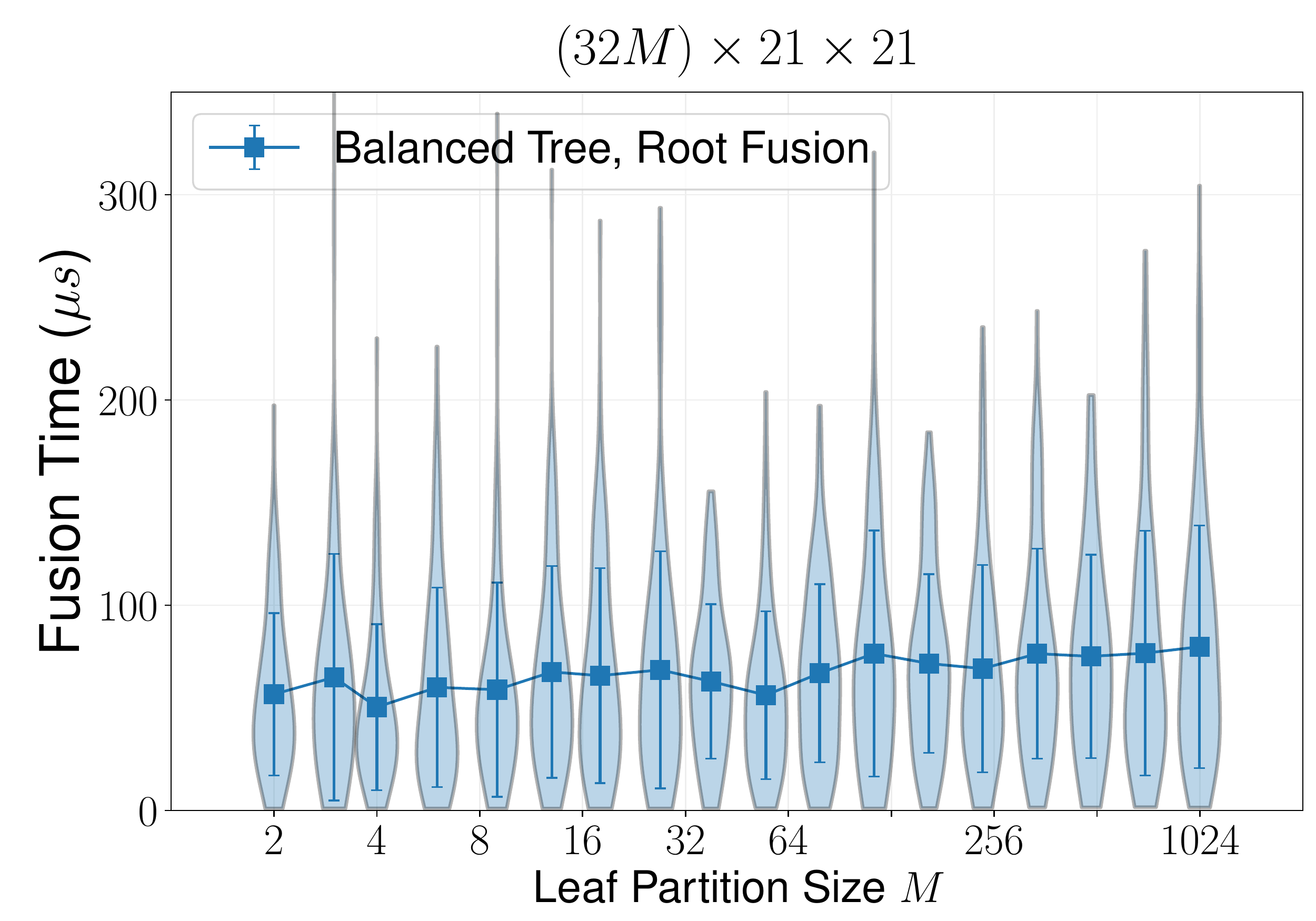}
            \caption{Leaf Partition Size $M$}
            \label{fig:fusion-time-delta-T}
        \end{subfigure}
        \begin{subfigure}{.31\linewidth}
            \centering
            \includegraphics[width=\linewidth,evaluationtrim]{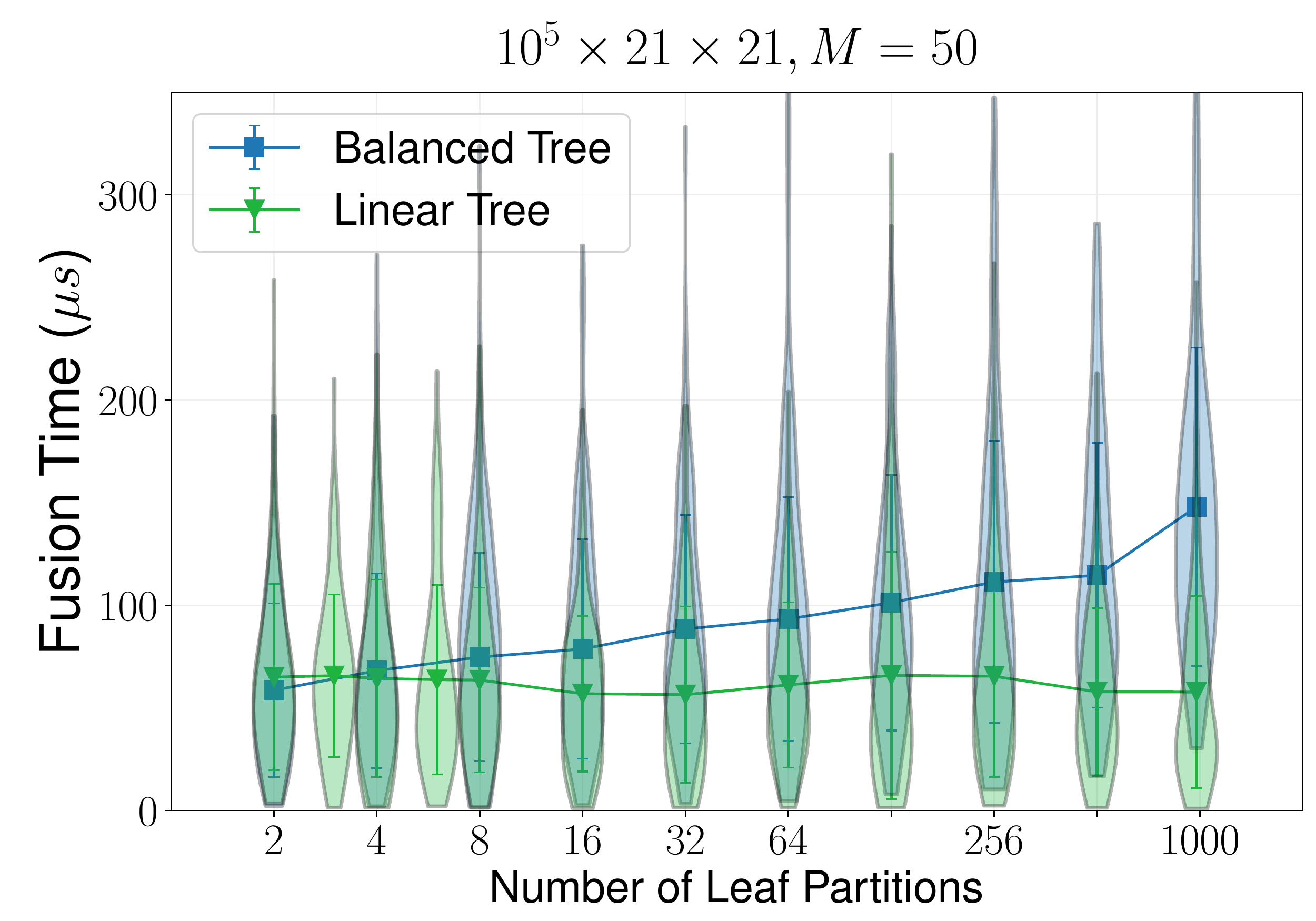}
            \caption{Number of Leaf Partitions}
            \label{fig:fusion-time-children-count}
        \end{subfigure}
        \caption{Fusion time with a single thread (lower the better).}
        \label{fig:evaluations-micro-benchmark}
    \end{minipage}
\end{figure*}

\subsection{Results}
\label{ssec:result}

\subsubsection{Throughput}\label{ssec:evaluation-throughput}
We use batch decoding in \cref{fig:batch-vs-stream-decoding} for all throughput evaluations, assuming the syndrome is ready when the decoding begins. Instead of throughput, we report data in its inverse, i.e., decoding time per round ($T/N$).

We first show the advantage of using the decoding graph over the syndrome graph, confirming the findings also reported by Higgott and Gidney in~\cite{higgott2023sparse}.
We benchmark the throughput on a single thread.
As shown in \cref{fig:serial-decoding-time-d}, the decoding time of \seqalgo and Sparse Blossom scales almost linearly with the number of qubits $n = d^2$, which is the theoretically lower bound.
In contrast, the traditional MWPM decoder based on the blossom V library scales poorly with the number of qubits $n$.
Not surprisingly, \seqalgo is roughly 4x slower than Sparse Blossom in this case, because we have not incorporated some important optimizations (\S\ref{ssec:impl-dual-serial}).

Second, when the number of rounds $N$ grows in \cref{fig:serial-decoding-time-T}, both \seqalgo and Sparse Blossom see decoding time per round increases~\cite{higgott2023sparse}, due to increasing pressure on the memory hierarchy.
Surprisingly, that of \paralgo remains steady as $N$ grows and beats that of \seqalgo at large $N \geqslant 10^3$, despite that \paralgo is not supposed to enjoy any algorithmic advantage over \seqalgo with a single thread.
This is because \paralgo divides the problem equally into small ones and solving a small problem enjoys better cache locality.
On the other hand, a small $M$ incurs more fusion operations and more overhead. Therefore, we empirically find the optimal $M = 100$ and use it as the default leaf partition size.

Not surprisingly, \paralgo beats all serial MWPM decoders when more threads are available.
As shown in \cref{fig:parallel-decoding-time-threads}, the throughput of \paralgo increases almost linearly with the number of threads, until the maximum number of hyper-threads (128) supported by the processor.
At that point, it reaches the minimum decoding time per round (0.3~us).
Using all 128 hyper-threads in the processor, \paralgo can decode up to $d=33$ at $p=0.1\%$ with less than 1~us decoding time per round, as shown in~\cref{fig:parallel-decoding-time-d}.

\subsubsection{Latency}\label{ssec:evaluation-latency}
We observe a constant latency regardless of the measurement rounds $N$ in the stream decoding (\cref{fig:batch-vs-stream-decoding}), compared to the linearly growing latency in the batch decoding.
We emulate the stabilizer measurement cycle of 1~us, which is similar to that of state-of-the-art superconducting quantum hardware~\cite{google2023suppressing}.
We use a mixed fusion tree in which each balanced subtree at the mix height has 50 leaves. Each leaf deals with $M = 20$ rounds of measurement.
As shown in \cref{fig:evaluation-latency}, the average latency is roughly 0.7~ms regardless of the number of measurement rounds $N$.
For the batch decoding, as predicted in \cref{fig:batch-vs-stream-decoding}, the latency scales linearly with $N$.

\subsubsection{Fusion Time}\label{ssec:evaluation-fusion-time}
Given a low physical error rate, a fusion operation only changes a small region around the boundary vertices on average.
Thus, the fusion time should only increase with $|V_b| = O(d^2)$, as confirmed by \cref{fig:fusion-time-d}, but not $M$, the number of rounds, as confirmed by  \cref{fig:fusion-time-delta-T}.

Moreover, because a fusion operation may recursively invoke the children's MWPM solvers until leaf partitions are reached, the structure of the corresponding subtree of the fusion tree impacts fusion time. \cref{fig:fusion-time-children-count} show this with fusion time for both balanced and linear trees where the X axis is the number of leaf partitions in the subtree.

Interestingly, the mean of fusion time of the balanced tree increases with the number of leaf partitions while that of the linear tree largely remains constant.
This, again, is because a fusion operation only changes a small region around the boundary vertices on average. As a result, the operation is most likely to involve two leaf partitions next to each other and increasingly unlikely to involve partitions that are farther away from each other. In the balanced tree, the operation must travel through the entire height of tree to reach any two leaf partitions, even if they are next to each other. In contrast, in the linear tree, the operation is exponentially less likely to travel one level down the tree. For the example in \cref{fig:fusion-time-children-count} (center), fusion operation 14 is exponentially less likely to involve lower numbered leaf paritioned.

\subsubsection{Scalability}
\label{ssec:scalability}
Finally, we show that given enough ($\Omega(d^{2.68})$) parallel resources, e.g., cores, \paralgo can meet the throughput requirement by any code distance $d$ when the physical error rate $p$ is well below the threshold $p_{\text{th}}$, using both analysis and empirical data.

\newcommand{\scaleddecode}{2.68}
\newcommand{\scaledfusion}{2.34}
Let $K$ denote the number of threads, each handling a $(N/K) \times d \times d$ partition on its own core.
Note that the threads may reside in different machines, accessing shared memory via network with a constant-factor slowdown, e.g., using shared-memory rack-scale distributed systems like~\cite{lee2021mind}.
The decoding time of each thread scales with $O(d^{\scaleddecode}N/K)$,  according to \cref{fig:serial-decoding-time-d}.
The solutions from the $K$ concurrent threads can be fused with a balanced tree, with $O(d^{\scaledfusion}\log K)$ time, according to \cref{fig:fusion-time-d}.
The decoding time per round has a complexity of $O(d^{\scaleddecode}/K + d^{\scaledfusion}\log K/N)$.
Thus, given a lower bound of $K = \Omega(d^{\scaleddecode})$ and $N = \Omega(d^{\scaledfusion}\log K)$, the decoding time per round will be bounded.
This analysis assumes $N$ polynomially grows with $d$. This is reasonable because the lifetime of a logical qubit scales exponentially with $d$, i.e., $\max N \propto (p_{\text{th}}/p)^{(d+1)/2}$~\cite{fowler2012towardsta}.

Note the scaling factors of $O(d^{\scaleddecode})$ and $O(d^{\scaledfusion})$ are empirically derived from code distances up to $100$ (\cref{fig:serial-decoding-time-d,fig:fusion-time-d}), which corresponds to about $10^4$ physical qubits for each logical qubit, orders of magnitude higher than what is considered to be practical in the near future.

We note that $K =\Omega(d^{\scaleddecode})$ does not mean $d^{\scaleddecode}$ threads (or cores) are necessary.
When estimating how many cores are needed for a large $d$, one can empirically derive the number for a small $d$ and then extrapolate based on the scaling of $d^{\scaleddecode}$.
For example, to estimate how many cores are necessary to decode $d=51$ with $p=0.1\%$ using the setup in \cref{ssec:setup}, one can pick a data point in \cref{fig:parallel-decoding-time-threads} where $d=21$ roughly needs 20 cores to meat the throughput requirement.
We can estimate roughly $20 \times (51/21)^{\scaleddecode}$ or 216 cores are necessary for $d=51$.

\section{Related Work}
\label{sec:related}

As mentioned in \S\ref{sec:introduction}, Sparse Blossom is a contemporary work closely related to \seqalgo, sharing the key idea of solving the MWPM problem using the decoding graph. We provide that first rigorous mathematical foundation for this idea and contribute new implementation optimizations.  

Related to \paralgo, Fowler~\cite{fowler2013minimum} presented a parallel design of the MWPM decoder.
It partitions the qubits to parallel decoding units of customized hardware.
Each decoding unit handles a sufficiently large number of qubits so that the inter-unit communication is relatively rare.
Paradoxically, its success requires both a large number of decoding units (for lower decoding time) and a large number of qubits in each unit (for lower communication overhead).
To our best knowledge, perhaps not surprisingly, no implementation or empirical data has been reported for this design.
\paralgo, on the other hand, eliminates communications between the partitions and only synchronizes them during the fusion operations.
This minimizes the need for communication and is scalable.

There is a literature that seeks to parallelize solving the MWPM problem for general graphs. 
This literature, however, does not exploit the special structure of the QEC decoding problem as we do. As a result, its results have larger time complexity than \seqalgo and \paralgo when applied to QEC decoding.
For example, Peterson and Karalekas~\cite{peterson2022distributed} designed and implemented a distributed MWPM algorithm with $O(|V|^4)$ time complexity. 

Recently there is a growing interest in approximate algorithms for QEC decoding that sacrifice decoding accuracy to gain speed, e.g., parallelization with parallel-window technique~\cite{skoric2022parallel,tan2022scalable,bombin2023modular} and fast decoders with cryogenic chips~\cite{holmes2020nisq+,ueno2021qecool,ueno2022qulatis,ravi2023better}.
Perhaps the most relevant is the (weighted) Union-Find (UF) decoder~\cite{delfosse2020linear,huang2020fault} for which various design~\cite{das2022afs} and implementation~\cite{liyanage2023scalable} have been reported. 
The key idea of \seqalgo draws inspiration from how the UF decoder approximates the MWPM decoder~\cite{wu2022interpretation}.

\section*{Acknowledgments}
This work was supported in part by Yale University and NSF MRI Award \#2216030. The authors are grateful for the insightful discussion with Shruti Puri.

\bibliographystyle{IEEEtran}
\bibliography{ref}{}

\ifnum \enableappendix=1
\clearpage
\appendices
\section{Facts \& Lemmas}

\noindent \textit{a) Relevant facts about \textbf{the decoding problem}.}

\vspace{1ex}\definefact{non-negative-weights} 
Edge weights are non-negative in a decoding graph.
This is because $w_e =\log\frac{(1 - P_e)}{P_e}$ given reasonable physical error rate of $P_e \leqslant 50\%$, $w_e \ge 0$. Consequently, edge weights in a syndrome graph are also non-negative.

We note that even if $P_e > 50\%$, this error is equivalent to an always-happening error plus another error with probability $1-P_e < 50\%$. We can first apply this error to the syndrome by flipping the incident vertices and then decode with positive weighted edge $w_e =\log\frac{P_e}{(1 - P_e)}$.
 
\vspace{1ex}\definefact{triangular-relationship} The triangular relationship holds for the weights in a syndrome graph $G(V,E)$. That is, for any vertices $u, a, b \in V$, $w_{(a,b)} \leqslant w_{(a,u)} + w_{(b,u)}$. This is because syndrome graph weights are constructed from minimum-weight paths in the decoding graph.

\vspace{2ex}\noindent \textit{b) Relevant facts about \textbf{the blossom algorithm}, also see \S\ref{ssec:blossom-in-general}.}

\vspace{1ex}\definefact{dual-feasible} The blossom algorithm starts with some feasible dual variables and maintains the feasibility throughout the algorithm.
We assume the initial solution is $y_S = 0$, $\forall S \in \mathcal{O}^*$ given the non-negative weights $w_e \ge 0, \forall e \in E$ (\reffact{non-negative-weights}).

\vspace{1ex}\definefact{shrink-node-sandwich} When a dual variable decreases by $\Delta$, it must be a ``$-$'' node in an alternating tree, and there exists two ``$+$'' nodes in the tree that connects this ``$-$'' node with tight edges. The dual variables of these ``$+$'' nodes must increase by $\Delta$.

\vspace{2ex}\noindent \textit{c) Relevant facts \& lemmas about \textbf{the blossoms}.}

\vspace{1ex}\definefact{root-blossom-non-overlapping} Given two different nodes $S_1$ and $S_2$, $S_1 \cap S_2= \varnothing$ and $\mathcal{D}(S_1) \cap \mathcal{D}(S_2) = \varnothing$.
That is, different nodes do not share any vertices, or share any descendants.

\vspace{1ex}\definelemma{root-uniqueness}{Root Uniqueness} Any blossom $S$ has a unique node whose \ref{def:blossom-progeny} includes $S$, denoted as Root$(S)$. 

\vspace{1ex}\definelemma{ancestry}{Ancestry} Similar to the definitions in \ref{def:blossom-progeny},
\begin{itemize}
    \item $\mathcal{A}(v) = \mathcal{D}_v(\text{Root}(v))$.
    \item $\ancestryminus{u}{v} = \mathcal{D}_{u \setminus v}(\text{Root}(u))$.
\end{itemize}

\vspace{1ex}\definelemma{distinct-root-ancestry}{Distinct-Root Ancestry} If $\text{Root}(u) \neq \text{Root}(v)$, then $\ancestryminus{u}{v} = \mathcal{A}(u)$.

\vspace{1ex}\definefact{blossom-complete} Blossom algorithm maintains all blossoms that have positive dual variables.
\begin{align*}
    &\{S|S \in \mathcal{O}^* \land u \in S \land v \notin S \land y_S \neq 0\} \subseteq \ancestryminus{u}{v} \\
    &\qquad\qquad \subseteq \{S|S \in \mathcal{O}^* \land u \in S \land v \notin S\} \\
    &\{S|S \in \mathcal{O}^* \land e \in \delta(S) \land y_S \neq 0\} \subseteq \ancestryminus{u}{v} \cup \ancestryminus{v}{u} \\
    &\qquad\qquad \subseteq  \{S|S \in \mathcal{O}^* \land e \in \delta(S)\},\ \text{where}\ e = (u, v)
\end{align*}

\vspace{1ex}\definefact{blossom-summation} Given \reffact{blossom-complete} and $\ancestryminus{v_1}{v_2} \cap \ancestryminus{v_2}{v_1} = \varnothing$ we can simplify the dual variable summation,

\begin{align*}
    &\sum_{\substack{S \in \mathcal{O}^* \\ v_1 \in S \land v_2 \notin S}}\qns\;y_S = \dns\qns\sum_{\substack{S \in \mathcal{O}^* \\ v_1 \in S \land v_2 \notin S \land y_S \neq 0}}\dqns\;y_S = \qns\sum_{A \in \ancestryminus{v_1}{v_2}}\qns\dns y_A \\
    &\sum_{\substack{S \in \mathcal{O}^* \\ (v_1, v_2) \in \delta(S)}}\qns y_S = \dns\qns\sum_{\substack{S \in \mathcal{O}^* \\ (v_1, v_2) \in \delta(S) \land y_S \neq 0}}\dqns y_S = \qns\sum_{A \in \ancestryminus{v_1}{v_2}}\qns\dns y_A + \qns\sum_{A \in \ancestryminus{v_2}{v_1}}\qns\dns y_A
\end{align*}

\section{LP Simplification}

\theoremnonnegativevertexdual{theorem:non-negative-vertex-dual-with-proof}

\begin{proof}
Given \reffact{non-negative-weights} and \reffact{dual-feasible}, the initial state satisfies that $y_u \geqslant 0, \forall u \in V$.
Whenever a vertex dual variable $y_u$ decreases $\Delta y_u < 0$, it must be a node.
According to \reffact{shrink-node-sandwich}, there must exists two other nodes $S_a$ and $S_b$ increasing.
Since there are tight edges between a node and $u$, we can assume edges $(a, u)$ and $(b, u)$ are the constraints of tight edges from the two nodes.
Obviously $(u, a), (u, b) \in \delta(u)$, i.e. $y_u$ contributes to the slack of both edges $(u, a)$ and $(u, b)$.
According to \reffact{root-blossom-non-overlapping}, there are no other node containing $u$ since $\{ u \}$ is a node, and any non-zero dual variable that includes $a$ must be child node of $S_a$.
According to \reffact{shrink-node-sandwich}, there are two tight constraints.
\begin{align}
    y_u + y_a + \dqns\sum_{S \in \mathcal{O} | (u, a) \in \delta(S)}\dqns y_S = w_{(u, a)}\label{eq:tight1} \\
    y_u + y_b + \dqns\sum_{S \in \mathcal{O} | (u, b) \in \delta(S)}\dqns y_S = w_{(u, b)}\label{eq:tight2}
\end{align}

According to \reffact{dual-feasible} and \reffact{shrink-node-sandwich}, there are three constraints for the update amount $\Delta$ between vertices $a$, $b$ and $u$.
\begin{align}
    y_u - \Delta + y_a + \dqns\sum_{S \in \mathcal{O} | (u, a) \in \delta(S)}\dqns y_S + \Delta \leqslant w_{(u, a)}\label{eq:con1} \\
    y_u - \Delta + y_b + \dqns\sum_{S \in \mathcal{O} | (u, b) \in \delta(S)}\dqns y_S + \Delta \leqslant w_{(u, b)}\label{eq:con2} \\
    y_a + \dqns\sum_{S \in \mathcal{O} | (u, a) \in \delta(S)}\dqns y_S + \Delta + y_b + \dqns\sum_{S \in \mathcal{O} | (u, b) \in \delta(S)}\dqns y_S + \Delta \leqslant w_{(a, b)}\label{eq:con3}
\end{align}

Constraints (\ref{eq:con1}) and (\ref{eq:con2}) are automatically satisfied with any $\Delta$ value, but constraint \ref{eq:con3} requires that
$$
\Delta \leqslant \left( w_{(a, b)} - y_a - \dqns\sum_{S \in \mathcal{O} | (u, a) \in \delta(S)}\dqns y_S - y_b - \dqns\sum_{S \in \mathcal{O} | (u, b) \in \delta(S)}\dqns y_S \right) / 2
$$

Given $w_{(a,b)} \leqslant w_{(u,a)} + w_{(u,b)}$ (\reffact{triangular-relationship}) and \cref{eq:tight1,eq:tight2},
$$
\Delta \leqslant y_u
$$

That means in the next stage $y'_u = y_u - \Delta \ge 0$.
Since blossom algorithm starts with $y_u \geqslant 0$ and there is no chance of decreasing any $y_u$ below zero at any point of the algorithm, we can conclude that $y_u \geqslant 0$ stands throughout the algorithm.
\end{proof}

\section{Proofs of Decoding Graph}
\label{sec:decoding-geometry}

\vspace{1ex}\definelemma{edge-max-point}{Edge Max Point} 
For a (segment) edge $e=(s,t)$ of weight $w$ and a point $p\notin e$,
\begin{align*}  
\max_{q\in e} \text{Dist}(p,q) &= \min(\text{Dist}(p,s),\text{Dist}(p,t))+ \\ 
                               &w/2+|\text{Dist}(p,s)-\text{Dist}(p,t)|/2
\end{align*}

\begin{proof}
By the definition of \ref{def:distance}, we have $|\text{Dist}(p,t)-\text{Dist}(p,s)|\leqslant w$. We can denote
$$|\text{Dist}(p,t)-\text{Dist}(p,s)|=w-\delta$$
where $0\leqslant \delta \leqslant w$. 
Without loss of generality, we assume $\text{Dist}(p,s)\leqslant \text{Dist}(p,t)$.  $\forall q\in e$, let $w_1=\text{Dist}(q,s)$. We have 
\begin{align*}
    \text{Dist}(p,q)&=\min(\text{Dist}(p,s)+w_1,\text{Dist}(p,t)+w-w_1)\\
                    &\tqns=\min(\text{Dist}(p,s)+w_1,(\text{Dist}(p,s)+w-\delta)+w-w_1)\\
                    &\tqns=\text{Dist}(p,s) + \min(w_1,2w-\delta-w_1)
\end{align*}
Thus, we have 
\begin{align*}
    \max_{q\in e}\text{Dist}(p,q)&=\text{Dist}(p,s)+w-\delta/2\\
    &= \min(\text{Dist}(p,s),\text{Dist}(p,t))+w-\delta/2
\end{align*}
\end{proof}

\vspace{1ex}\definelemma{edge-min-point}{Edge Min Point} For a (segment) edge $e=(s,t)$ of weight $w$ and a point $p\notin e$, 
\begin{align*}
\min_{q\in e} \text{Dist}(p,q) = \min(\text{Dist}(p,s),\text{Dist}(p,t))
\end{align*}

\vspace{1ex}\definelemma{edge-max-min}{Edge Max-Min Bound} For a (segment) edge $e$ of weight $w$ and a point $p\notin e$,
\[ \max_{q \in e} \text{Dist}(p, q) - \min_{q \in e} \text{Dist}(p, q) \geqslant w/2. \]

\vspace{1ex}\definelemma{circle-edges-with-proof}{Circle-Edges} A circle $C(v, d)$ on the decoding graph is a union of a finite number of vertices and (segment) edges with their incident points.

\begin{proof}
There are a finite number of vertices $V_M$.
Given an edge $e=(s,t)$ of weight $w$, if $w=0$, the edge is empty $e = \varnothing$;
If $w>0$, there are four situations regarding its relationship:
\begin{itemize}
    \item if $\forall p\in e$, $\text{Dist}(p,v)>d$, $e$ is outside the circle.
    \item if $\forall p\in e$, $\text{Dist}(p,v)\leqslant d$, $e \cup \{ s, t \}$ is inside the circle.
    \item if $\exists r\in e$, $\text{Dist}(s,v)\leqslant d$, and $\exists t\in e$, $\text{Dist}(t,v)>d$, 
    there must be one or two points that are of distance $d$ to $v$. When there is a single point, it partitions $e$ into two segments: one inside the circle and the other outside.
    \item When there are two points that are of distance $d$ to $v$, they partition $e$ into three segments: the middle segment lies outside the circle while the other two inside. 
\end{itemize}
With the above, we can conclude that a circle covers a finite number of vertices and (segment) edges with their incident points. That is, a circle is a union of a finite number of sets each represented by an closed intervals of (segment) edge. 
\end{proof}

\section{Proofs of Blossom}
\label{sec:blossom-geometry}

The blossom algorithm solves the MWPM problem for the \emph{syndrome} graph.
Our \seqalgo algorithm solves the same problem but work on the decoding graph. Therefore, to prove that \seqalgo solves the same problem, we relate dual variables in the blossom algorithm to the geometric objects of the decoding graph.

We imagine the syndrome graph is overlaid over its decoding graph. As $V\subseteq V_M$, a vertex in the syndrome graph is aligned with its correspondent in the decoding graph, as shown in \cref{fig:overlay-decoding-syndrome-graph}.

\begin{figure}[!htb]
    \centering
    \includegraphics[width=0.6\linewidth]{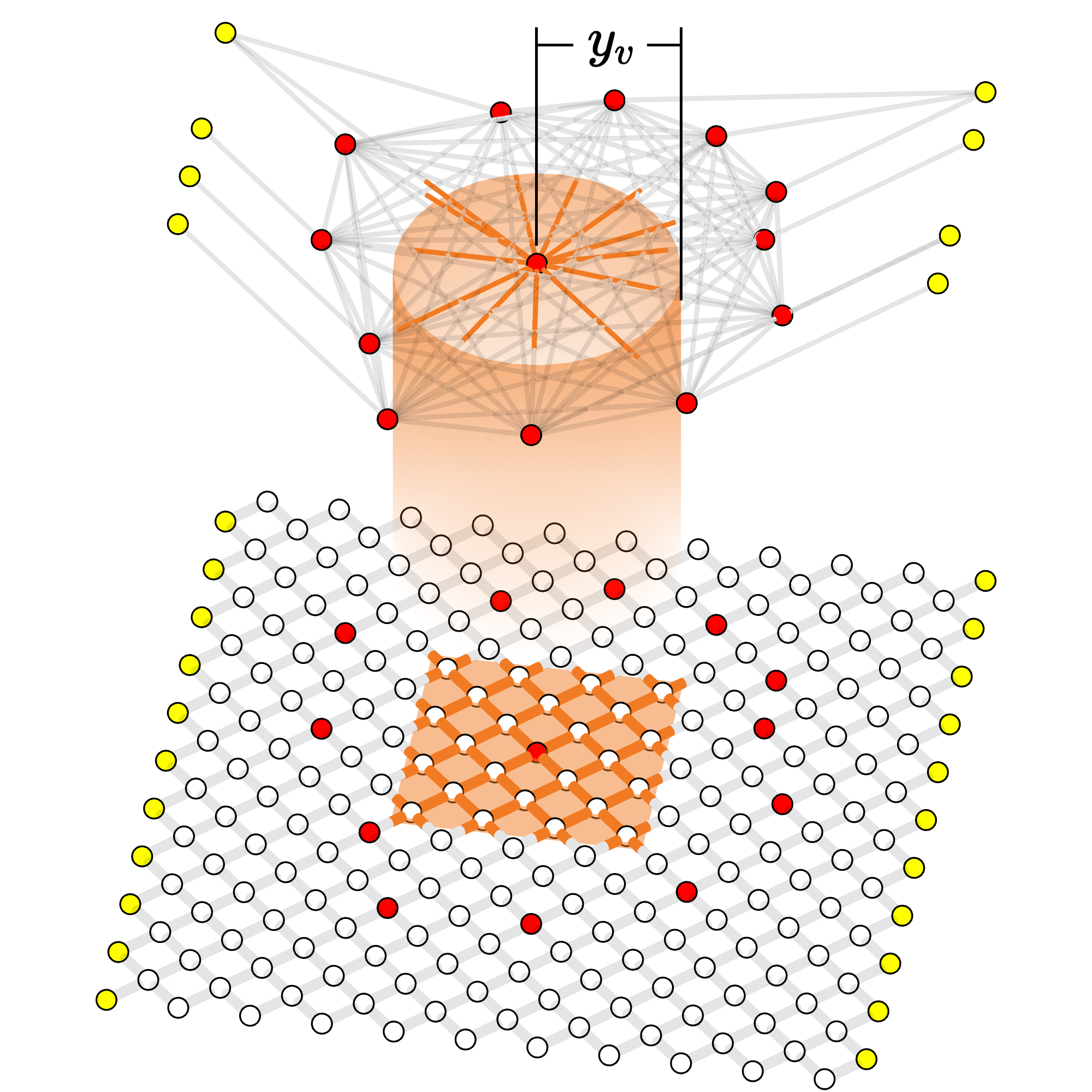}
    \caption{Overlaying syndrome graph on top of its decoding graph. The center vertex is $v$. The dual variable $y_v$ on the syndrome graph corresponds to a \ref{def:circle} of the same radius on the decoding graph. In the decoding graph, the circle $C(v, y_v)$ consists of all the orange (segment) edges. Despite its square appearance, it is actually a ``circle'' in Manhattan geometry.}
    \label{fig:overlay-decoding-syndrome-graph}
\end{figure}

We define
\begin{align*}
    C(v) &= C(v, \sum_{A \in \mathcal{A}(v)} y_A) \\
    \circleminus{u}{v} &= C(u, \sum_{A \in \mathcal{A}({u \setminus v})} y_A)
\end{align*}

\vspace{1ex}\definelemma{root-cover}{Root Cover} If $S$ is a node, i.e., $S = \text{Root}(S)$, \[\text{Cover}(S)=\cup_{v\in S} C(v).\]

\vspace{1ex}\definelemma{distinct-root-circle}{Distinct-Root Circle} If $\text{Root}(u) \neq \text{Root}(v)$, then $\circleminus{u}{v} = C(u)$.

\vspace{1ex}\definelemma{tight-edge}{Tight Edge}
An edge $(v_1, v_2)$ in the syndrome graph becomes tight if and only if $\circleminus{v_1}{v_2}$ and $\circleminus{v_2}{v_1}$ on the decoding graph overlap. That is,%
\begin{align*}%
w_e = \dqns\sum_{S \in \mathcal{O}^* | e \in \delta(S)}\dqns y_S\ \ \Longleftrightarrow\ \  \circleminus{v_1}{v_2} \cap \circleminus{v_2}{v_1} \neq \varnothing%
\end{align*}%

\begin{proof}
\nosection{Sufficiency.} 
Assume on the decoding graph
$\exists p \in C(v_1, \sum_{A \in \mathcal{A}({v_1 \setminus v_2})} y_A) \cap C(v_2, \sum_{A \in \mathcal{A}({v_2 \setminus v_1})} y_A)$. 
By the definition of Circle,
\begin{align*}
    \text{Dist}(v_1, p) \leqslant \sum_{A \in \mathcal{A}({v_1 \setminus v_2})} y_A \\
    \text{Dist}(v_2, p) \leqslant \sum_{A \in \mathcal{A}({v_2 \setminus v_1})} y_A
\end{align*}

According to the triangular relationship, the weight of the edge between $v_1$ and $v_2$ in the syndrome graph has $w_e = \text{Dist}(v_1, v_2) \leqslant \text{Dist}(v_1, p) + \text{Dist}(v_2, p)$. This and \reffact{blossom-summation} have
\begin{align*}
    w_e \leqslant \sum_{S \in \mathcal{O}^* | e \in \delta(S)} y_S
\end{align*}

Since the edge slackness constraints (\ref{eq:qec-dual-lp-constraint-b}) also says $\geqslant$ of the above inequality, we have
\begin{align*}
    w_e = \sum_{S \in \mathcal{O}^* | e \in \delta(S)} y_S
\end{align*}

\nosection{Necessity.}\label{theorem:tight-edge-detection-necessity-proof}
Suppose an edge $e = (v_1, v_2)$ is tight, there exists a minimum-weight path from $v_1$ to $v_2$ in the decoding graph that consists of edges $e_1=(v_1,a), e_2, \cdots, e_n=(b,v_2)$ where
\begin{align*}
    \sum_{i=1}^{n} w_{e_i} = w_e= \sum_{S \in \mathcal{O}^* | e \in \delta(S)} y_S
\end{align*}

Since Dist$(v_1,v)$ is continuous and monotonic as $v$ moves from $v_1$ to $v_2$ along the minimum-weight path,  $\exists p \in e_j$ that split the edge $e_j$ into two parts weighted $w_1$ and $w_2$, where

\begin{align*}
    \text{path}\ (v_1, p)&:\quad \sum_{i=1}^{j-1} w_{e_i} + w_1 &\tqns= \sum_{S \in \mathcal{O}^* | v_1 \in S \land v_2 \notin S}\tqns y_S \\
    \text{path}\ (p, v_2)&:\quad w_2 + \sum_{i=j+1}^{n} w_{e_i} &\tqns= \sum_{S \in \mathcal{O}^* | v_1 \notin S \land v_2 \in S}\tqns y_S
\end{align*}

Given the distance definition and \reffact{blossom-summation},

\begin{align*}
    \text{Dist}(v_1, p) &\leqslant \sum_{S \in \mathcal{O}^* | v_1 \in S \land v_2 \notin S}\tqns y_S &\dqns= \sum_{A \in \mathcal{A}({v_1 \setminus v_2})} y_A \\
    \text{Dist}(v_2, p) &\leqslant \sum_{S \in \mathcal{O}^* | v_1 \notin S \land v_2 \in S}\tqns y_S &\dqns= \sum_{A \in \mathcal{A}({v_2 \setminus v_1})} y_A
\end{align*}

That is, $\exists p$ belongs to both $C({v_1 \compactsetminus v_2})$ and $C({v_2 \compactsetminus v_1})$.
\end{proof}

\theoremcoverfiniteoverlap{theorem:cover-finite-overlap-with-proof}

\begin{proof}
Because $S_1$ and $S_2$ are nodes, 

\begin{align*}
    \text{Cover}(S_1) \cap \text{Cover}(S_2) = \left(\dns\qns\underset{\quad v_1\in S_1}{\cup}\dns\qns\ C(v_1)\right) \cap \left(\dns\qns\underset{\quad v_2\in S_2}{\cup}\dns\qns\ C(v_2)\right).
\end{align*}

Suppose $|\text{Cover}(S_1) \cap \text{Cover}(S_2)|$ is infinite, there must exist two vertices $v_1 \in S_1$ and $v_2 \in S_2$ such that $|C(v_1) \cap C(v_2)|$ is infinite.

Since a circles includes a finite number of edges, there must exists a (segment) edge $f$ of a nonzero weight $w > 0$ with
$$f \subseteq C(v_1) \cap C(v_2).$$

By the definition of Circle, we have 
\begin{align*}
    \forall p\in f,\ &\text{Dist}(v_1, p) \leqslant \sum_{A \in \mathcal{A}(v_1)} y_A~~\text{and} \\
                    &\text{Dist}(v_2, p) \leqslant \sum_{A \in \mathcal{A}(v_2)} y_A
\end{align*}

Moreover, with \reflemma{edge-max-min},
\begin{gather*}
    \exists p\in f, \text{Dist}(v_2, p) \leqslant \sum_{A \in \mathcal{A}(v_2)} y_A - w/2
\end{gather*}

Given the triangular inequality of the distance function $\text{Dist}(v_1, v_2) \le \text{Dist}(v_1, p) + \text{Dist}(v_2, p)$,
\begin{align*}
    \text{Dist}(v_1, v_2) + w/2 \leqslant \sum_{A \in \mathcal{A}(v_1)} y_A + \sum_{A \in \mathcal{A}(v_2)} y_A.
\end{align*}

With \reffact{blossom-summation} and \reflemma{distinct-root-ancestry},
\begin{align*}
    \sum_{A \in \mathcal{A}(v_1)} y_A + \sum_{A \in \mathcal{A}(v_2)} y_A
    = \sum_{S \in \mathcal{O}^* | e \in \delta(S)}\!\!\!\!\!\!\! y_S
\end{align*}

The syndrome graph edge $e = (v_1, v_2)$ has a weight $w_e = \text{Dist}(v_1, v_2)$, thus
\begin{align*}
    w_e + w/2 \leqslant \dqns\sum_{S \in \mathcal{O}^* | e \in \delta(S)}\dqns y_S
    \quad\implies\quad w_e < \dqns\sum_{S \in \mathcal{O}^* | e \in \delta(S)}\dqns y_S
\end{align*}

The above violates the edge slackness constraints (\ref{eq:qec-dual-lp-constraint-b}). As the result, the theorem must be true.
\end{proof}

\theoremobstacledetection{theorem:obstacle-detection-with-proof}

\begin{proof}
Since $S_1$ and $S_2$ are different nodes, we have $C(v_1 \setminus v_2) = C(v_1)$ and $C(v_2 \setminus v_1) = C(v_2)$ per \reflemma{distinct-root-circle}.

\nosection{Necessity.}
With \reflemma{tight-edge},
\begin{align*}
&w_e = \qqns\sum_{S \in \mathcal{O}^* | (v_1, v_2) \in \delta(S)}\qqns y_S\\
&\Longrightarrow  C(v_1 \compactsetminus v_2) \cap C(v_2 \compactsetminus v_1) \neq \varnothing\\
&\Longrightarrow  C(v_1) \cap C(v_2) \neq \varnothing\\
&\Longrightarrow  \text{Cover}(S_1) \cap \text{Cover}(S_2) \neq \varnothing
\end{align*}

The last step is true because the \ref{def:cover} of a node consists of circles of all its vertices, per \reflemma{root-cover}.
\begin{align*}
\text{Cover}(S) = \cup_{v \in S} C(v)
\end{align*}

\nosection{Sufficiency.}
\begin{align*}
&\text{Cover}(S_1) \cap \text{Cover}(S_2) \neq \varnothing\\
&\Longrightarrow  \exists v_1\in S_1, v_2\in S_2, C(v_1) \cap C(v_2) \neq \varnothing\\ 
&\Longrightarrow  C(v_1 \compactsetminus v_2) \cap C(v_2 \compactsetminus v_1) \neq \varnothing\\
&\Longrightarrow   w_{(v_1,v_2)} = \qqns\sum_{S \in \mathcal{O}^* | (v_1, v_2) \in \delta(S)}\qqns y_S
\end{align*}

\end{proof}

\section{Proofs of Fusion Blossom}
\label{sec:fusion-blossom-theorems}

$V_b \subset V_M$ divides a decoding graph $G(E_M, V_M)$ into two disjoint graphs that include $V_1$ and $V_2$, respectively if there is no edge in the decoding graph that connects vertices from both $V_1$ and $V_2$. As a result, it divides the MWPM problem into two disjoint sub-problems. For the $i$-th sub-problem, $i\in \{1,2\}$, the primal and dual formulations as in Eq.~\ref{eq:primal-lp} and \ref{eq:qec-dual-lp}, respectively, have $E$ and $\mathcal{O}^*$ as follows.
\begin{align*}
    E_i &= \{ e | e = (u, v) \in E \spaciousland u, v \in V_i \cup V_b \} \\
    \mathcal{O}_i^* &= \{ S | S \in \mathcal{O}^* \spaciousland S \subseteq V_i \}
\end{align*}

\theoremvaliddualvariables{theorem:valid-dual-variables-with-proof}

\begin{proof}\label{theorem:valid-dual-variables-proof}
Apparently $\mathcal{O}_1^* \cap \mathcal{O}_2^* = \varnothing$ and $\mathcal{O}_1^* \cup \mathcal{O}_2^* \subseteq \mathcal{O}^*$.
Likewise,  $E_1\cup E_2 \subseteq E$.

The MWPM solutions for the sub-problems must satisfy the following by definition
\begin{align*}
    \sum_{S \in \mathcal{O}_1^* | e \in \delta(S)} \qns y_S &\leqslant w_e,\ \forall e \in E_1 \\
    \sum_{S \in \mathcal{O}_2^* | e \in \delta(S)} \qns y_S &\leqslant w_e,\ \forall e \in E_2
\end{align*}

Additionally, since any vertex in $V_b$ does not appear in any blossom of $\mathcal{O}_1^*$ or $\mathcal{O}_2^*$,
\begin{align*}
    &e \notin \delta(S),\ \forall S \in \mathcal{O}_1^*,\ \forall e \in E_2 \\
    &e \notin \delta(S),\ \forall S \in \mathcal{O}_2^*,\ \forall e \in E_1
\end{align*}

These four can be combined as

\[  \sum_{S \in \mathcal{O}_1^*\cup \mathcal{O}_2^* | e \in \delta(S)} \dqns y_S \leqslant w_e,\ \forall e \in E_1\cup E_2  \]

Because  $y_S$ is 0 for $S\in \mathcal{O}^*\setminus (\mathcal{O}_1^* \cup \mathcal{O}_2^*)$, we can further simplify the above as 

\[  \sum_{S \in \mathcal{O}^*| e \in \delta(S)} \qns y_S \leqslant w_e,\ \forall e \in E_1\cup E_2  \]

We only need to prove the constraint is also met for $\forall e\in E\setminus (E_1\cup E_2)$, which can be noted as $e=(v_1,v_2)$ where $v_1 \in V_1$ and $v_2 \in V_2$. $e$ corresponds to a minimum-weight path between $v_1$ and $v_2$ in the original decoding graph.
Since $V_b$ divides the decoding graph into two disjoint parts, this path must go through $\exists v_b \in V_b$.
Since a minimum-weight path go through $v_1, v_b, v_2$, we have $w_{(v_1, v_2)} = w_{(v_1, v_b)} + w_{(v_2, v_b)}$.
Because $(v_1,v_b)\in E_1$ and $(v_2,v_b)\in E_2$, we have 
\begin{align*}
    \sum_{S \in \mathcal{O}_1^* | (v_1, v_2) \in \delta(S)} \dqns y_S \leqslant w_{(v_1, v_b)} \\
    \sum_{S \in \mathcal{O}_2^* | (v_1, v_2) \in \delta(S)} \dqns y_S \leqslant w_{(v_2, v_b)}
\end{align*}

Again because  $y_S$ is 0 for $S\in \mathcal{O}^* \setminus (\mathcal{O}_1^* \cup \mathcal{O}_2^*)$, we have
\begin{align*}
    \sum_{S \in \mathcal{O}^* | (v_1, v_2) \in \delta(S)} \dqns y_S \leqslant w_{(v_1, v_b)} + w_{(v_2, v_b)} = w_{(v_1, v_2)}
\end{align*}

That is,
\begin{align*}
    \sum_{S \in \mathcal{O}^* | e \in \delta(S)} \qns y_S \leqslant  w_e,\ \forall e\in E\setminus (E_1\cup E_2)
\end{align*}

Overall, the dual variables are feasible in (\ref{eq:qec-dual-lp-constraint-a}) and (\ref{eq:qec-dual-lp-constraint-b})
\begin{align*}
    w_e - \dqns \sum_{S \in \mathcal{O}^* | e \in \delta(S)} \dqns y_S &\geqslant 0,\ \forall e \in E\\
    y_S &\geqslant 0,\ \forall S \in \mathcal{O}^*
\end{align*}

\end{proof}

\section{Proofs of \code{Parity} Dual Module}\label{ap:proof-parity-dual}

We optimize the algorithm by tracking \ref{def:pseudo-covers} (\S\ref{ssec:impl-dual-serial}) instead of \ref{def:covers}.
Pseudo-Covers are similar to Covers, but differ by only a few boundary vertices.
The primary complication introduced by Pseudo-Cover stems from the presence of zero edges, as all the vertices connected by zero edges can belong to an arbitrary number of Covers.
Additional complications arise when \paralgo partitions the decoding graph, as some of the zero edges may not be known to a Pseudo-Cover when it is constructed from a sub-problem.
Therefore, it is natural to question whether Pseudo-Covers are equally effective as Covers in detecting \ref{def:obstacles}.

Here we prove the correctness of using Pseudo-Covers, as stated in \ref{theorem:obstacle-detection-optimized-with-proof}.
We start with a few definitions and lemmas exclusively used in the proof of this theorem.
We will use the example in \cref{fig:pseudo-cover-island} throughout.
Pseudo-Covers have the following properties.
\begin{gather*}
    \overline{\text{Cover}}(S) \subseteq \text{Cover}(S)\\
     \text{Cover}(S) \setminus \overline{\text{Cover}}(S) \subseteq V_M\\
    \overline{\text{Cover}}(S_1) \cap \overline{\text{Cover}}(S_2) \cap V_M = \varnothing,\ \forall S_1 \neq S_2
\end{gather*}

\vspace{1ex}\definelemma{no-free-zero-vertex}{No Free Zero Edge Vertex} Given a zero edge $e=(u,v)\in E_M$, if there is a node $S$ such that $u\in\overline{\text{Cover}}(S)$, there is a node $T$ such that $v\in\overline{\text{Cover}}(T)$.

\begin{figure}[!tb]
    \renewcommand*\thesubfigure{(\arabic{subfigure})}  
    \centering
    \begin{subfigure}{.49\linewidth}
        \centering
        \includegraphics[width=0.8\linewidth]{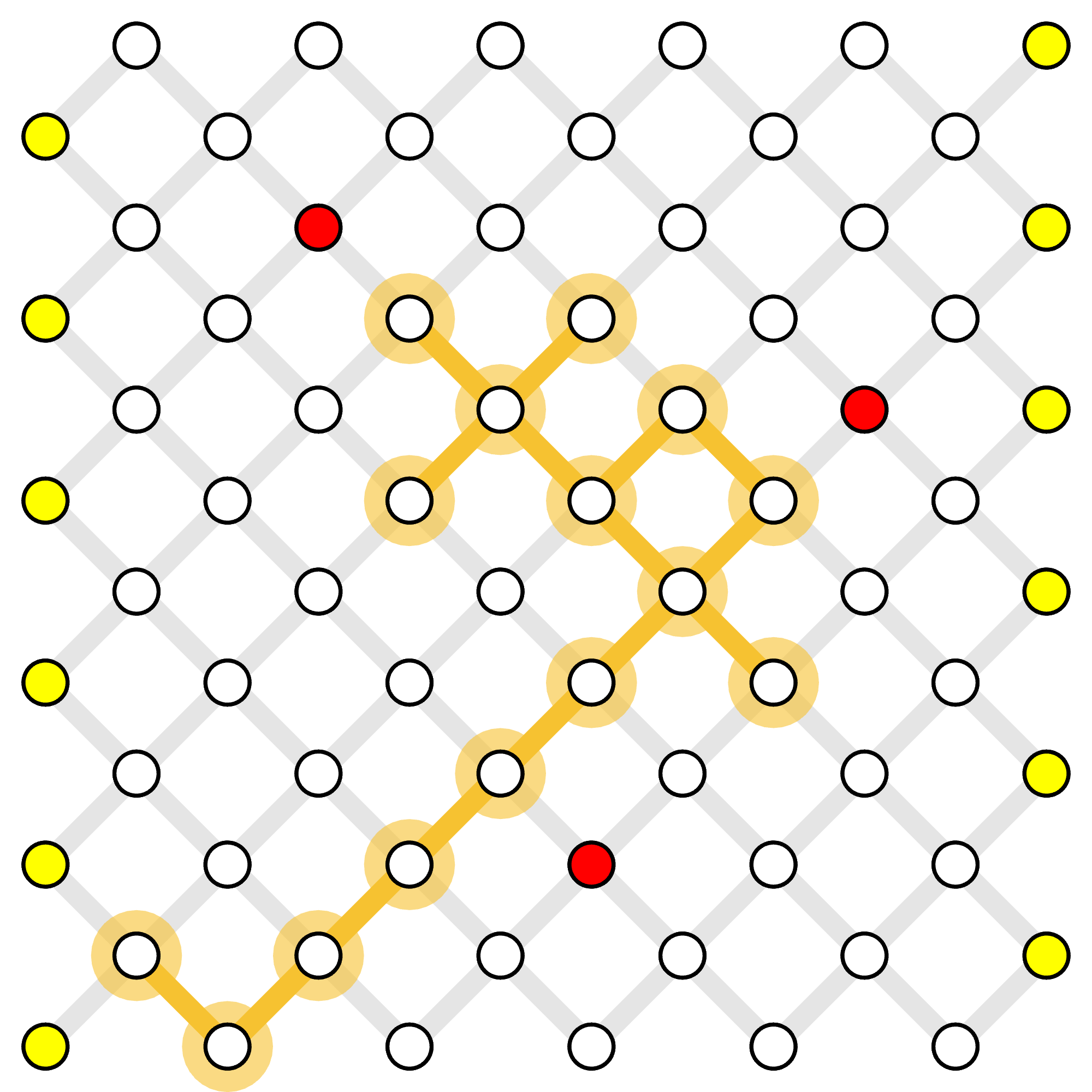}
        \caption{Island}
        \label{fig:pseudo-cover-island-island}
    \end{subfigure}
    \begin{subfigure}{.49\linewidth}
        \centering
        \includegraphics[width=0.8\linewidth]{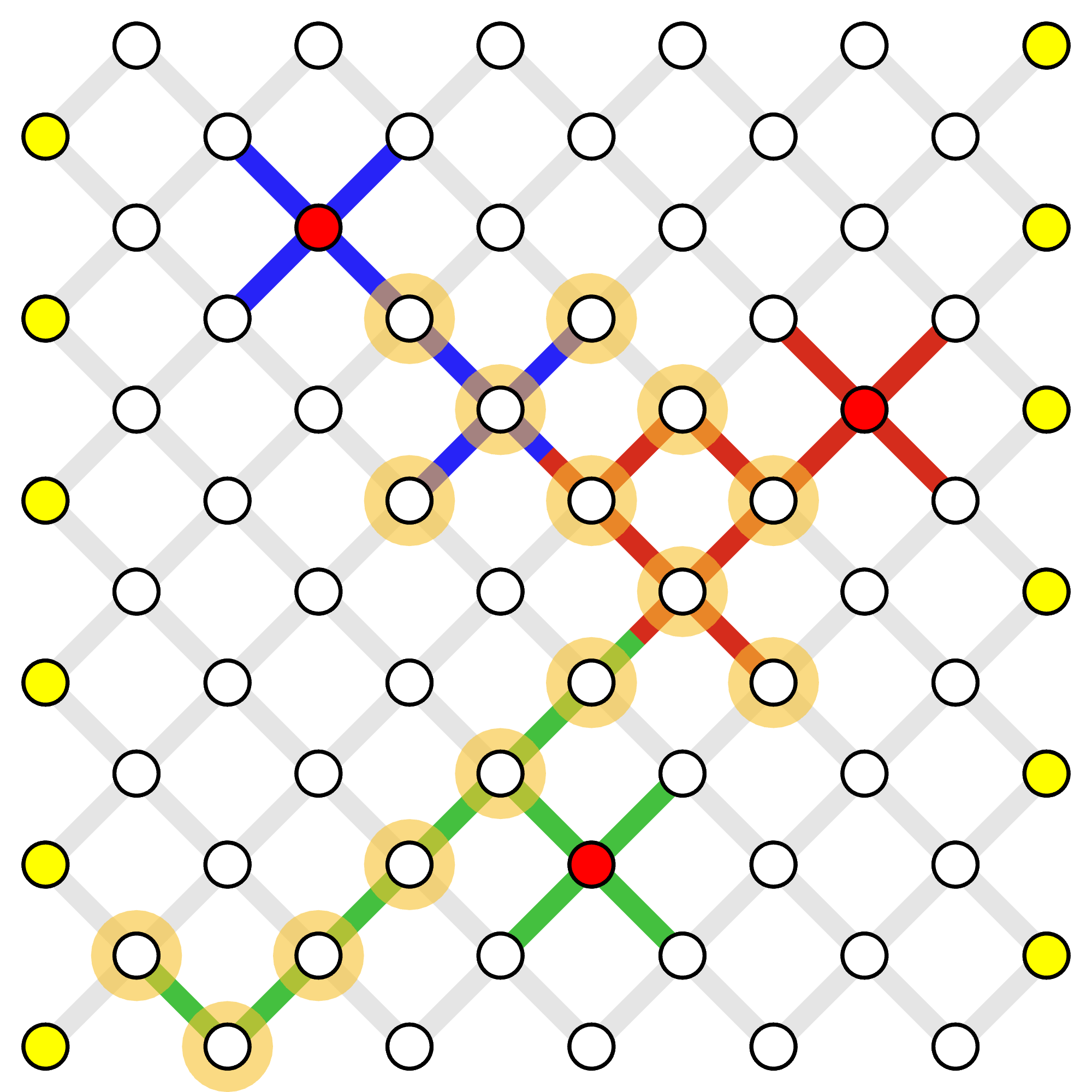}
        \caption{Invalid \ref{def:pseudo-covers}}
        \label{fig:pseudo-cover-island-initial}
    \end{subfigure}
    \begin{subfigure}{.49\linewidth}
        \centering
        \includegraphics[width=0.8\linewidth]{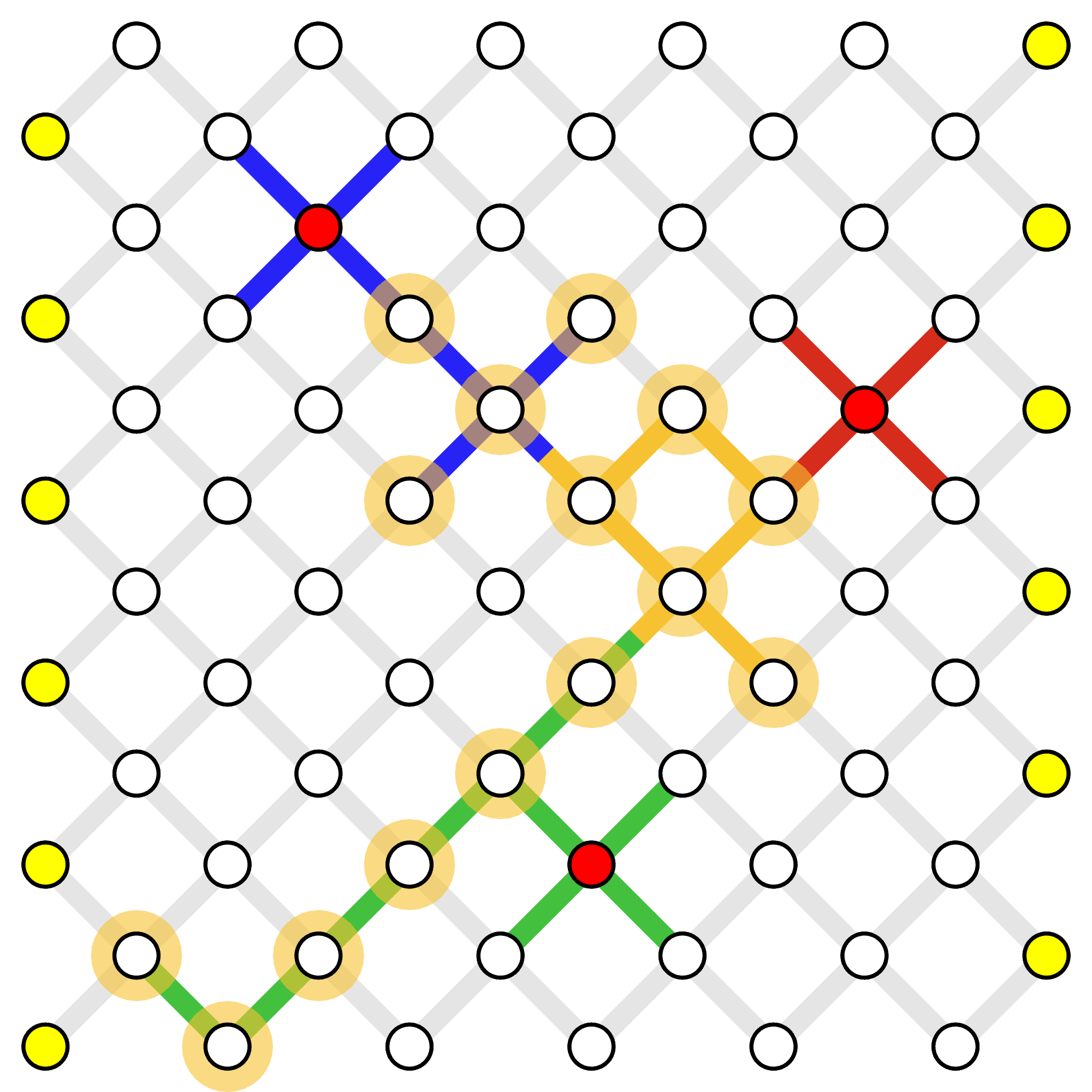}
        \caption{$\Delta y_R = -1$, $\overline{\text{Cover}}(R)$}
        \label{fig:pseudo-cover-island-intrude}
    \end{subfigure}
    \begin{subfigure}{.49\linewidth}
        \centering
        \includegraphics[width=0.8\linewidth]{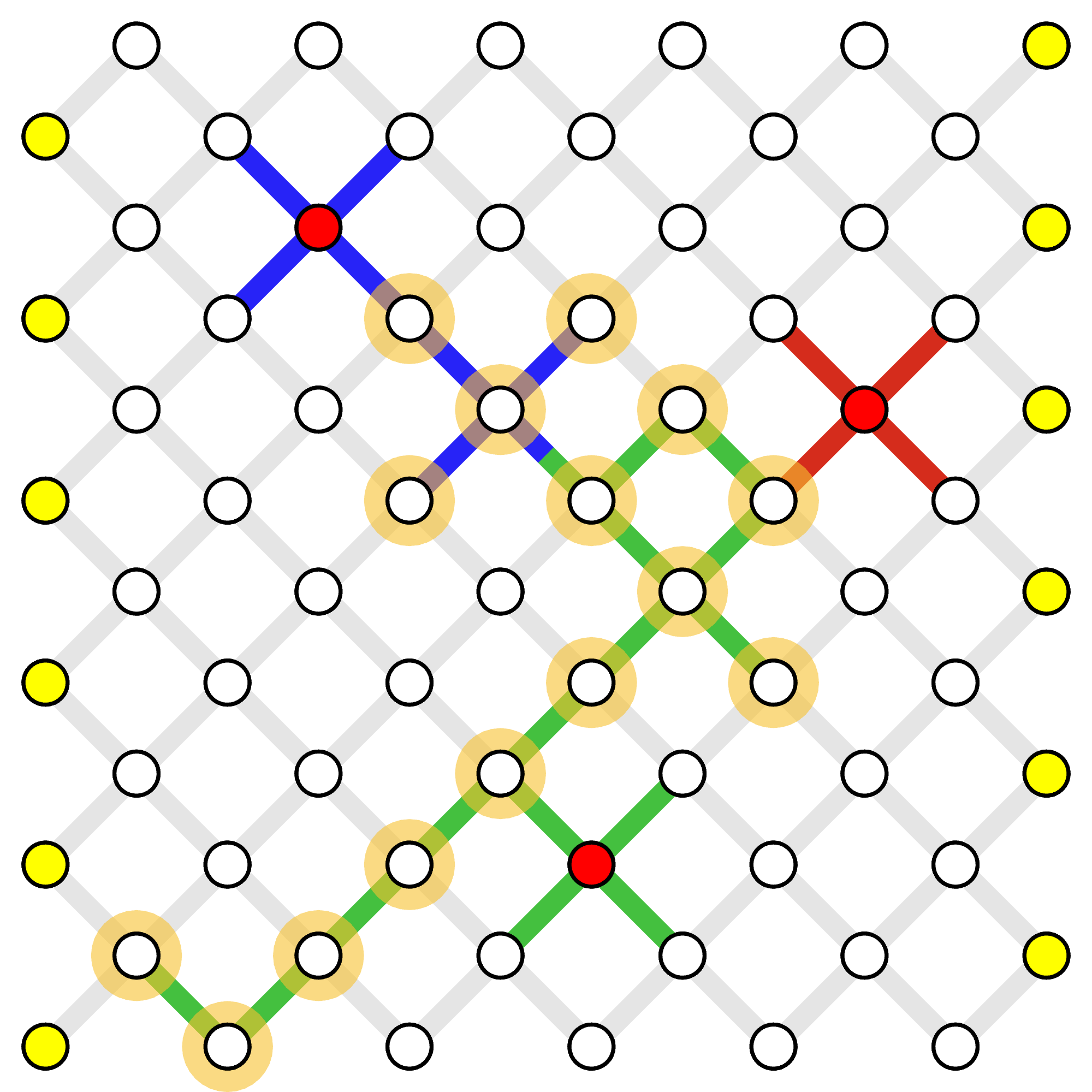}
        \caption{$\Delta y_G = +1$, $\overline{\text{Cover}}(G)$}
        \label{fig:pseudo-cover-island-extrude}
    \end{subfigure}
    \caption{Example of maintaining Pseudo-Covers to detect at least one obstacle. (1) The \ref{def:island} is marked in yellow circles, connected by zero edges in yellow. When a vertex in the Island is inside a Pseudo-Cover, we mark half of the incident edges with the color of the node. (2) Suppose the nodes are R,G,B for red, green and blue respectively. Suppose $\Delta y_R = -1$, $\Delta y_G = \Delta y_B = +1$. Although there is an obstacle between $G$ and $B$ given $\text{Cover}(G) \cap \text{Cover}(B) \neq \varnothing$, their initial yet invalid Pseudo-Covers $\overline{\text{Cover}}(G)$ and $\overline{\text{Cover}}(B)$ do not touch on any decoding graph edge, isolated by $\overline{\text{Cover}}(R)$ as shown in the figure, thus forbidding an obstacle to be detected. In order to let $\overline{\text{Cover}}(G)$ and $\overline{\text{Cover}}(B)$ touch, the algorithm constructs valid Pseudo-Covers by (3) removing boundary vertices from $\overline{\text{Cover}}(R)$ given $\Delta y_R = -1$ and (4) absorbing as many boundary vertices as possible into $\overline{\text{Cover}}(G)$ given $\Delta y_G = +1$. After that, $\overline{\text{Cover}}(G)$ and $\overline{\text{Cover}}(B)$ touch on one decoding graph edge and thus detects an obstacle.}
    \label{fig:pseudo-cover-island}
\end{figure}

\vspace{1ex}\definitionlabel{def:island}{Island} We define the \emph{island} of a vertex $v$ as the set of vertices that have zero distance to $v$, noted as
\[ \text{Island}(v) = \{ u | \text{Dist}(u, v) = 0, u \in V_M \} \]

\vspace{1ex}Obviously, $v \in \text{Island}(v)$.
Also, $|\text{Island}(v)| > 1$ iff $\exists e = (u, v) \in E_M,\ w_e = 0$.
That is, an island is non-trivial only when there are zero edges.
An example is shown in \cref{fig:pseudo-cover-island-island}, where the vertices in yellow circles constitute an island, connected by yellow zero edges.

\vspace{1ex}\definitionlabel{def:occupancy}{Occupancy} We define the \emph{Occupancy} of a vertex $v$ as the set of nodes whose Cover includes $v$, noted as
\[ \text{Occupancy}(v) = \{ S | v \in \text{Cover}(S) \land S~\text{is node} \} \]
A member of the Occupancy is called an Occupier. 

\vspace{1ex}\definelemma{occupier-boundary}{Occupier Boundary} If $v\in V_M$ has more than one occupiers, it must be on the boundary of the Covers of all its occupiers. 
\begin{proof}
Assume $S_1, S_2\in \text{Occupancy}(v)$.
We have $v\in \text{Cover}(S_1)\cap \text{Cover}(S_2)$.
We prove by contradiction, assuming $v$ is not on the boundary of $\text{Cover}(S_1)$.

Since $v$ is strictly inside $\text{Cover}(S_1)$, there exists a defect vertex $u_1 \in S_1$ such that $v$ is strictly inside the circle $C(u_1)$.
That is, $\text{Dist}(u_1, v) < \sum_{A \in \mathcal{A}(u_1)} y_A$.
Since $v \in \text{Cover}(S_2)$, there exists a defect vertex $u_2 \in S_2$ such that $v \in C(u_2)$.
That is, $\text{Dist}(u_2, v) \leqslant \sum_{A \in \mathcal{A}(u_2)} y_A$.
We have
\begin{align*}
    \text{Dist}(u_1, u_2) &\leqslant \text{Dist}(u_1, v) + \text{Dist}(u_2, v)\\
    &< \qns\sum_{A \in \mathcal{A}(u_1)}\qns y_A + \qns\sum_{A \in \mathcal{A}(u_2)}\qns y_A = \sum_{S \in \mathcal{O}^* | (u_1, u_2) \in \delta(S)}\dqns y_S
\end{align*}
This violates the dual constraint (\ref{eq:dual-lp-constraint-a}) of $e = (u_1, u_2) \in E$, contradiction.
Thus, the theorem must be true.
\end{proof}

We can extend the notion of Occupancy to an Island: the Occupancy of Island($v$) is the same as Occupancy($v$). This is because
\begin{align*}
    v \in \text{Cover}(S) &\Longleftrightarrow \text{Island}(v) \subseteq \text{Cover}(S)
\end{align*}

As shown in example \cref{fig:pseudo-cover-island-island}, the Occupancy of the island is $\varnothing$, while in \cref{fig:pseudo-cover-island-initial} the Occupancy of the island is $\{ R,G,B \}$.

\vspace{1ex}\definelemma{min-max-cover-overlap}{Pseudo-Cover Touching} Consider a vertex $v \in V_M$, if there exists two node $S_1, S_2 \in \text{Occupancy}(v)$ with $\Delta y_{S_1}+\Delta y_{S_2}>0$, there exists two nodes $S_3, S_4 \in \text{Occupancy}(v)$ with $\Delta y_{S_3}+\Delta y_{S_4}>0$ whose Pseudo Covers border each other.
That is,
\begin{gather*}
\exists S_1,S_2 \in \text{Occupancy}(v), \Delta y_{S_1} +\Delta y_{S_2} > 0\\%
\Longrightarrow\qquad \forall S_3 \in \text{Occupancy}(v), \Delta y_{S_3} > 0,\\%
\exists S_4 \in \text{Occupancy}(v) \setminus \{S_3\}, \Delta y_{S_4} \geqslant 0, e = (a, b) \in E_M, \\%
a \in \overline{\text{Cover}}(S_3), b \in \overline{\text{Cover}}(S_4),\\
e \subseteq \overline{\text{Cover}}(S_3) \cup \overline{\text{Cover}}(S_4)%
\end{gather*}

\begin{proof}
Because $v$ has at least two occupiers ($S_1$ and $S_2$), $v$ and all members of Island$(v)$ must be on the boundaries of the Covers of all its occupiers, per \reflemma{occupier-boundary}.

According to how a Pseudo Cover is derived, $\forall S \in \text{Occupancy}(v)$ with $\Delta y_S < 0$, we have $\overline{\text{Cover}}(S) \cap \text{Island}(v) = \varnothing$, because  all zero edges and their vertices must be removed from Cover$(S)$ to create $\overline{\text{Cover}}(S)$.

For every $S_3 \in \text{Occupancy}(v)$ with $\Delta y_{S_3}>0$, there are only two cases. 
\begin{itemize}
    \item 
 In the first case, $\text{Island}(v)$ does not overlap with any other Pseudo-Covers beyond $\overline{\text{Cover}}(S_3)$. We have $\text{Island}(v) \subseteq\overline{\text{Cover}}(S_3)$ given the island is a connected graph and how Pseudo-Covers are derived.
Since there exists another node $S_4 \in \text{Occupancy}(v) \setminus \{ S_3 \}$ with $\Delta y_{S_4} \geqslant 0$ and Island$(v)\cap\overline{\text{Cover}}(S_4)=\varnothing$, there must exist a decoding graph edge $e=(a,b)\in\text{Cover}(S_4)$ where $a\in \text{Island}(v)\subseteq \overline{\text{Cover}}(S_3)$  and $b\in \overline{\text{Cover}}(S_4)$. That is, $a\in \text{Island}(v)$ must be removed from $\text{Cover}(S_4)$ to form $\overline{\text{Cover}}(S_4)$. 
Because $b\notin\text{Island}(v)$, $w_e$ must be greater than zero. As a result, 
$e\subseteq \overline{\text{Cover}}(S_4)$. Therefore, $e \subseteq \overline{\text{Cover}}(S_4) \subseteq \overline{\text{Cover}}(S_3) \cup \overline{\text{Cover}}(S_4)$, and $\Delta y_{S_3}+\Delta y_{S_4}>0$.

\item $\text{Island}(v)$ overlaps with a Pseudo-Cover beyond that of $S_3$. 
There must exists a zero edge $e=(a,b)\in E_M$ that $a,b\in\text{Island}(v)$ and $a$ and $b$ are covered by Pseudo-Covers of $S_3$ and another nodes $S_4$, respectively. 
We have $\Delta y_{S_4}\geqslant 0$ because $\overline{\text{Cover}}(S) \cap \text{Island}(v) = \varnothing$ for any $\Delta y_S < 0$.
We have $e=\varnothing \subseteq \overline{\text{Cover}}(S_3) \cup \overline{\text{Cover}}(S_4)$, and $\Delta y_{S_3}+\Delta y_{S_4}>0$.

\end{itemize}

\end{proof}

\vspace{1ex}\definelemma{edge-fully-cover}{Edge Fully Cover} If a decoding graph edge $e = (u, v) \in E_M$ is covered by the union of two different Pseudo-Covers $S_1, S_2$ and $u, v$ are inside $\overline{\text{Cover}}(S_1), \overline{\text{Cover}}(S_2)$ respectively, then the Covers of $S_1$ and $S_2$ overlap. That is,
\begin{gather*}
e \subseteq \overline{\text{Cover}}(S_1) \cup \overline{\text{Cover}}(S_2), u \in \overline{\text{Cover}}(S_1), v \in \overline{\text{Cover}}(S_2) \\
\Longrightarrow \text{Cover}(S_1) \cap \text{Cover}(S_2) \neq \varnothing
\end{gather*}
\begin{proof}
We prove there exists a point in $\text{Cover}(S_1) \cap \text{Cover}(S_2)$.
We have $u \in \overline{\text{Cover}}(S_1) \subseteq \text{Cover}(S_1)$ and $v \in \overline{\text{Cover}}(S_2) \subseteq \text{Cover}(S_2)$.

Clearly, if $u \in \text{Cover}(S_2)$ or $v \in \text{Cover}(S_1)$, the vertex $u$ or $v$ belongs to $\text{Cover}(S_1) \cap \text{Cover}(S_2)$.

If not, then $w_e > 0$ per definition of \ref{def:island} and \ref{def:occupancy}.
Since $v \notin \text{Cover}(S_1)$, $\text{Cover}(S_1) \cap (e \cup \{ u, v \})$ is a closed edge segment from $u$ to $p_1 \in (u, v)$.
Similarly, $\text{Cover}(S_2) \cap (e \cup \{ u, v \})$ is a closed edge segment from $v$ to $p_2 \in (u, v)$.
We have $e \subseteq \overline{\text{Cover}}(S_1) \cup \overline{\text{Cover}}(S_2) \subseteq \text{Cover}(S_1) \cup \text{Cover}(S_2)$.
Thus, we have $\text{Dist}(u, p1) + \text{Dist}(v, p2) \geqslant w_e$, because otherwise there exists a point on segment edge $p \in (p_1, p_2) \subseteq e$ where $p \notin \text{Cover}(S_1) \cup \text{Cover}(S_2)$.
Also, $\text{Dist}(u, p1) + \text{Dist}(v, p2) \leqslant w_e$ given \ref{theorem:cover-finite-overlap-with-proof}, because otherwise $\text{Cover}(S_1)$ and $\text{Cover}(S_2)$ overlap on segment edge $(p_1, p_2)$ with infinite points.
Thus, we have $\text{Dist}(u, p1) + \text{Dist}(v, p2) = w_e$, i.e., point $p_1 = p_2$ belongs to both $\text{Cover}(S_1)$ and $\text{Cover}(S_2)$.

\end{proof}

\vspace{2ex}

\theoremobstacledetectionoptimized{theorem:obstacle-detection-optimized-with-proof}

\begin{proof}
\nosection{Sufficiency.}
We prove that when $S_1 = S_3$, $S_2 = S_4$, there exists such a tight edge $e$.
Given \ref{theorem:obstacle-detection-with-proof} and $\Delta y_{S_1} + \Delta y_{S_2} > 0$, we only need to prove $\text{Cover}(S_3) \cap \text{Cover}(S_4) \neq \varnothing$.
Although their Pseudo-Covers do not overlap on vertices, i.e., $\overline{\text{Cover}}(S_3) \cap \overline{\text{Cover}}(S_4) \cap V_M = \varnothing$, we have $\text{Cover}(S_3) \cap \text{Cover}(S_4) \neq \varnothing$ given \reflemma{edge-fully-cover}.

\nosection{Necessity.}
Given \ref{theorem:obstacle-detection-with-proof}, we have $\text{Cover}(S_1) \cap \text{Cover}(S_2) \neq \varnothing$ and $\Delta y_{S_1} + \Delta y_{S_2} > 0$.
Since $\Delta y_S \in \{ 0, \pm1\}$, without loss of generality, we have $\Delta y_{S_1} = +1$ and $\Delta y_{S_2} \in \{ 0, +1 \}$.
Also, there exists a point $p \in \text{Cover}(S_1) \cap \text{Cover}(S_2)$.

If such a point belongs to an edge $p \in e' = (v_3, v_4) \in E_M$, we have $w_e > 0$ and $\text{Dist}(v_3, p) > 0$, $\text{Dist}(v_4, p) > 0$.
Given \ref{theorem:cover-finite-overlap-with-proof}, $v_3$ and $v_4$ must belong to different Covers.
Without loss of generality, we assume $v_3 \in \text{Cover}(S_1)$ and $v_4 \in \text{Cover}(S_2)$.
Thus, segment edge $(v_3, p) \subseteq \text{Cover}(S_1)$ and $(v_4, p) \subseteq \text{Cover}(S_2)$.
Given edge segments contain no vertex and $\text{Cover}(S) \setminus \overline{\text{Cover}}(S) \subseteq V_M$, we have $(v_3, p) \subseteq \overline{\text{Cover}}(S_1)$, $(v_4, p) \subseteq \overline{\text{Cover}}(S_2)$ and $p \in \overline{\text{Cover}}(S_1) \cap \overline{\text{Cover}}(S_2)$.
Thus, 
\[ e' = (v_3, p) \cup \{ p \} \cup (p, v_4) \subseteq \overline{\text{Cover}}(S_1) \cup \overline{\text{Cover}}(S_2) \]
Since $v_3,p \in \text{Cover}(S_1)$, $v_4 \notin \text{Cover}(S_1)$, $v_3$ is not on the boundary of $\text{Cover}(S_1)$.
Given $\text{Cover}(S) \setminus \overline{\text{Cover}}(S)$ only consists of boundary vertices per definition of \ref{def:pseudo-cover}, we have $v_3 \in \overline{\text{Cover}}(S_1)$.
Similarly, $v_4 \in \overline{\text{Cover}}(S_2)$.
That is, there exist $S_3 = S_1$ and $S_4 = S_2$ satisfying the conditions.

If such a point is a vertex $v$, we have $S_1, S_2 \in \text{Occupancy}(v)$.
Now we can simply invoke \reflemma{min-max-cover-overlap} to complete the proof.
Note that in this case, $S_3$ and $S_4$ are not necessarily $S_1$ and $S_2$.

\end{proof}

\section{\seqalgo Example}

\begin{figure*}[!htb]
    \renewcommand*\thesubfigure{(\arabic{subfigure})}  
    \centering
    \begin{subfigure}{.116\textwidth}
        \centering
        \color{lightgray}\includegraphics[width=0.81\linewidth]{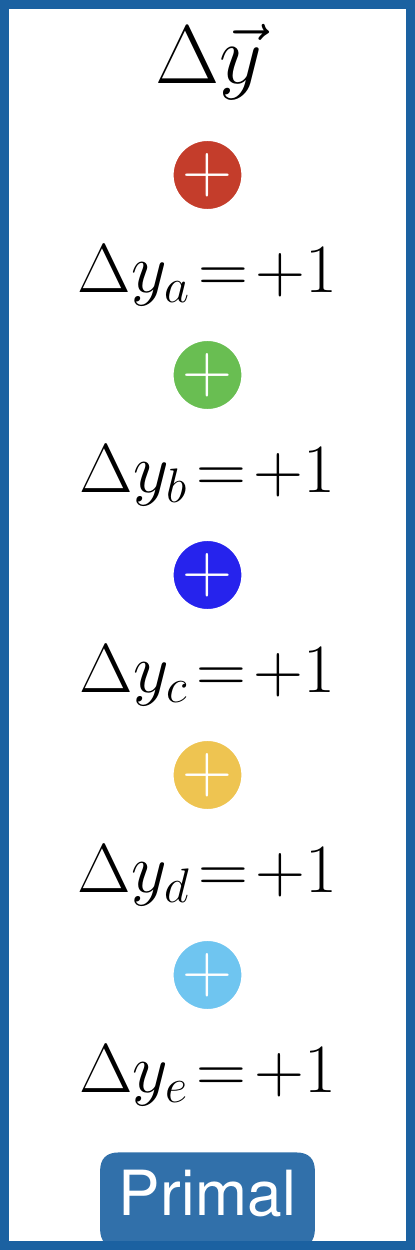}
        \caption{Initial}
    \end{subfigure}
    \hfill
    \begin{subfigure}{.2821\textwidth}
        \centering
        \color{lightgray}\includegraphics[width=\linewidth]{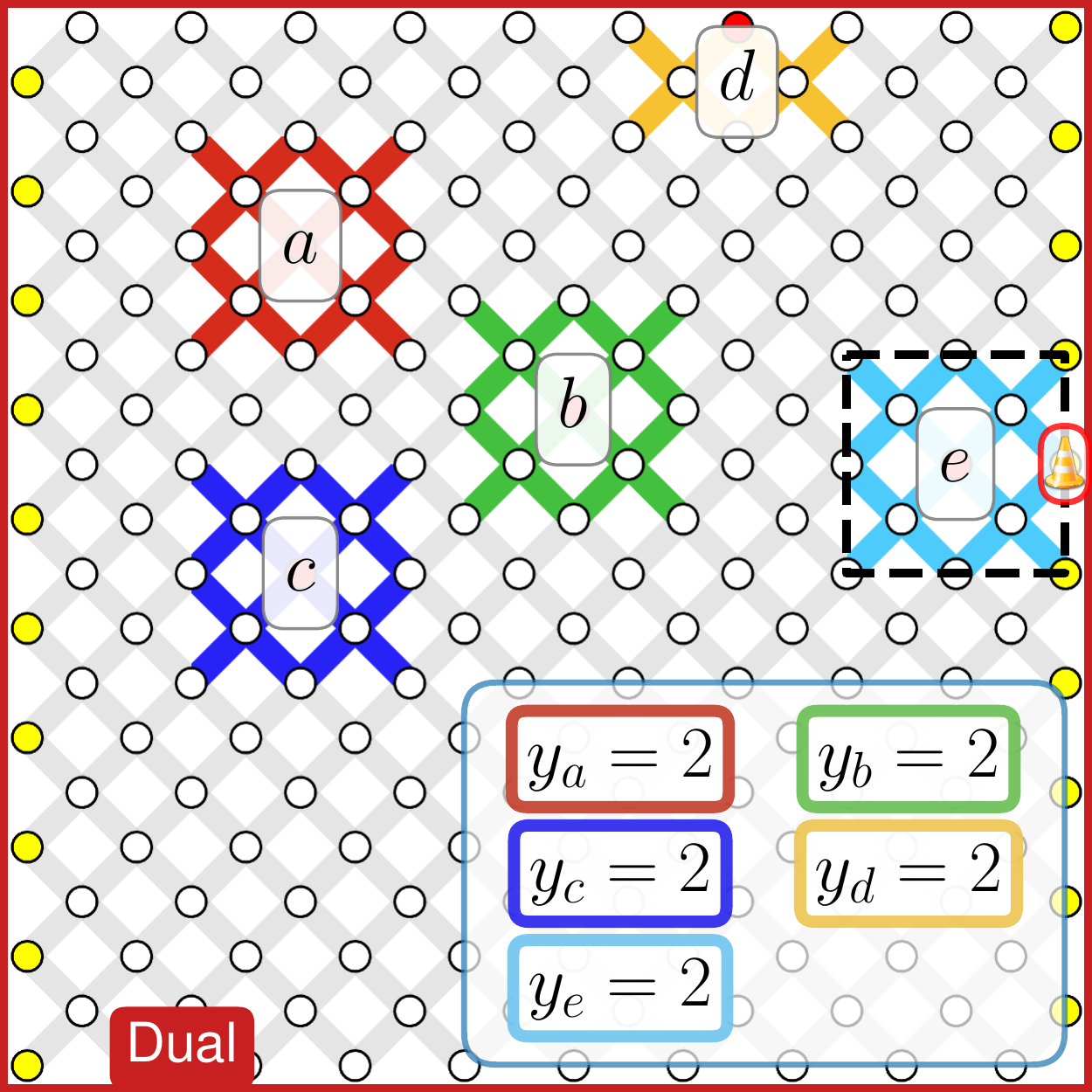}
        \caption{Obstacle $(e, \text{virtual})$}
        \label{fig:cover-steps-touch-virtual}
    \end{subfigure}
    \hfill
    \begin{subfigure}{.116\textwidth}
        \centering
        \color{lightgray}\includegraphics[width=0.81\linewidth]{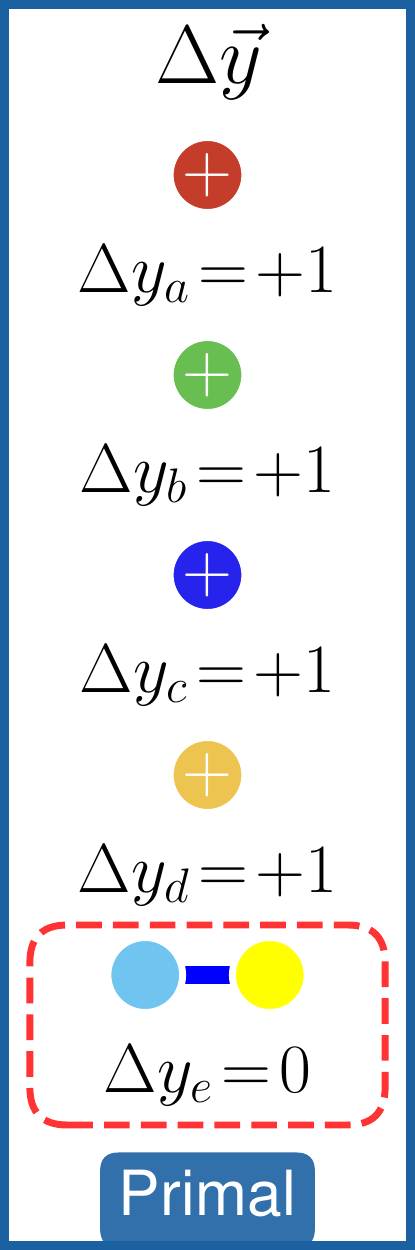}
        \caption{Match}
    \end{subfigure}
    \hfill
    \begin{subfigure}{.2821\textwidth}
        \centering
        \color{lightgray}\includegraphics[width=\linewidth]{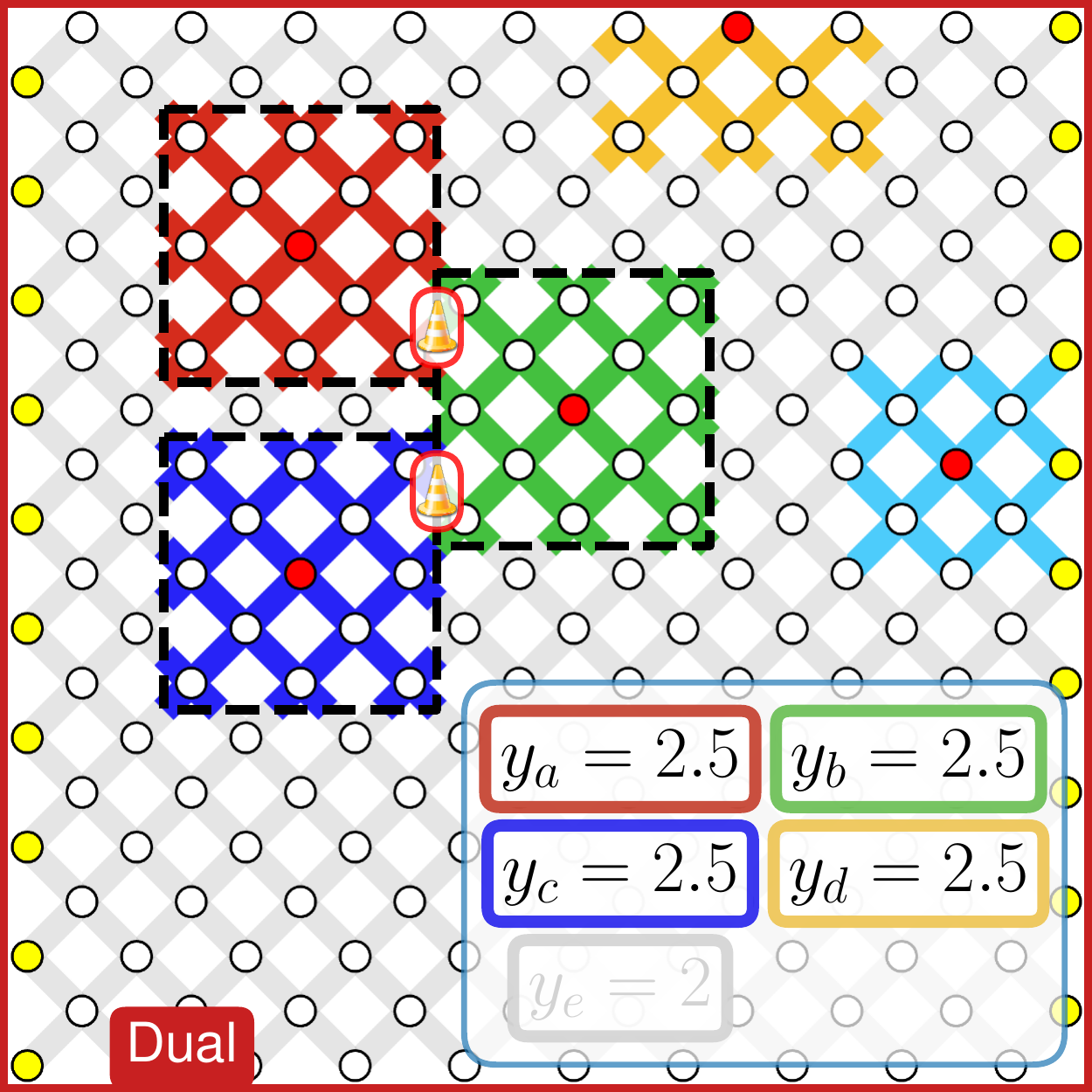}
        \caption{Obstacle $(a, b)$ and $(b, c)$}
    \end{subfigure}
    \hfill
    \begin{subfigure}{.116\textwidth}
        \centering
        \color{lightgray}\includegraphics[width=0.81\linewidth]{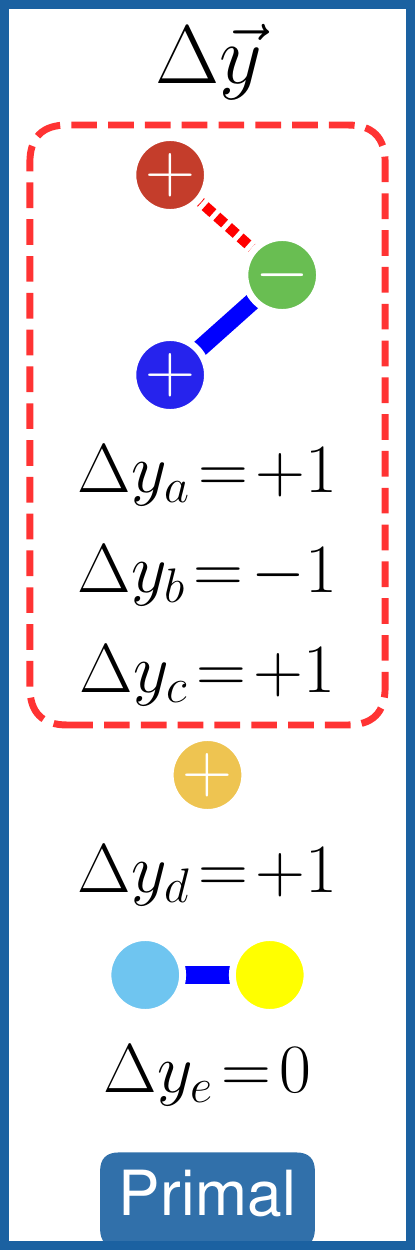}
        \caption{Alternating}
    \end{subfigure}
    \hfill
    \vspace{1ex}
    \begin{subfigure}{.2821\textwidth}
        \centering
        \color{lightgray}\includegraphics[width=\linewidth]{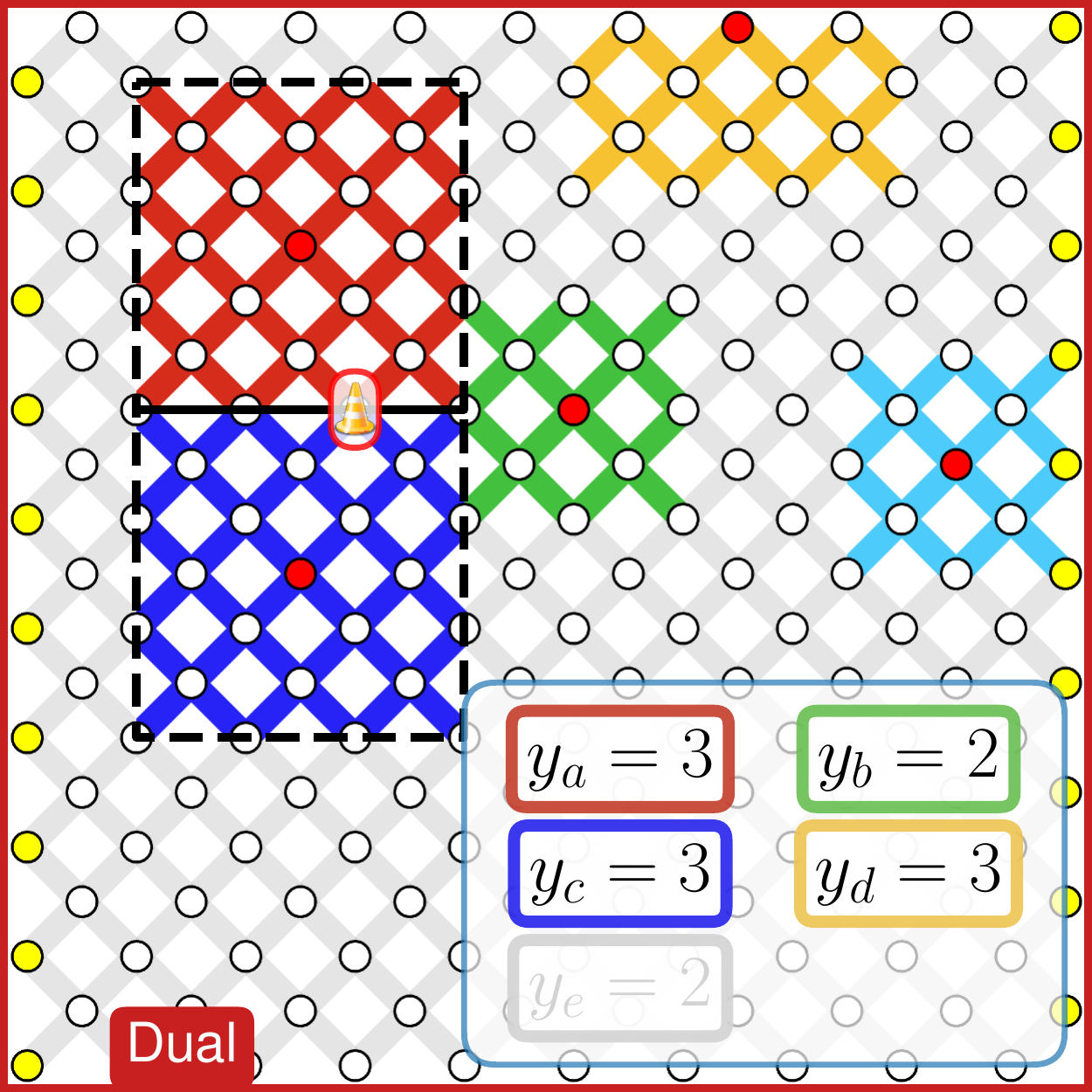}
        \caption{Obstacle $(a, c)$}
    \end{subfigure}
    \hfill
    \begin{subfigure}{.116\textwidth}
        \centering
        \color{lightgray}\includegraphics[width=0.81\linewidth]{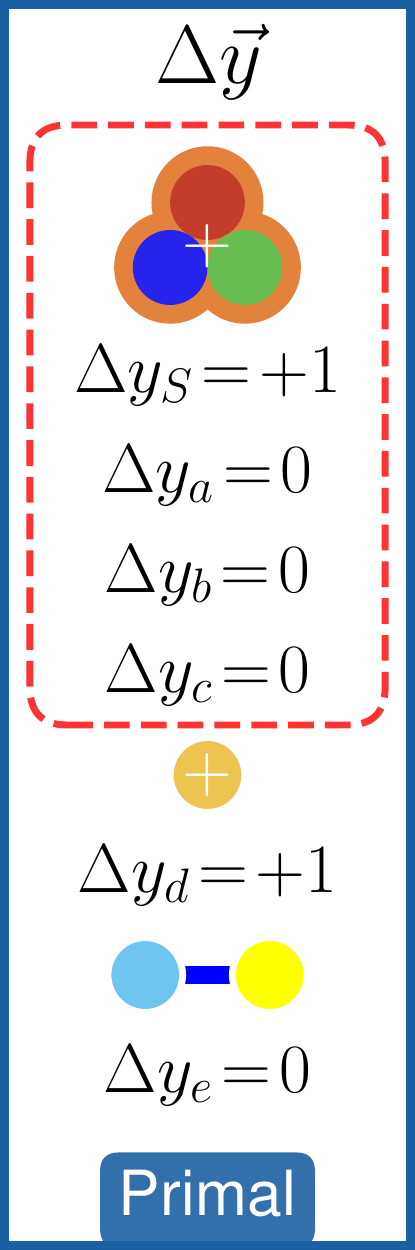}
        \caption{Blossom}
    \end{subfigure}
    \hfill
    \begin{subfigure}{.2821\textwidth}
        \centering
        \color{lightgray}\includegraphics[width=\linewidth]{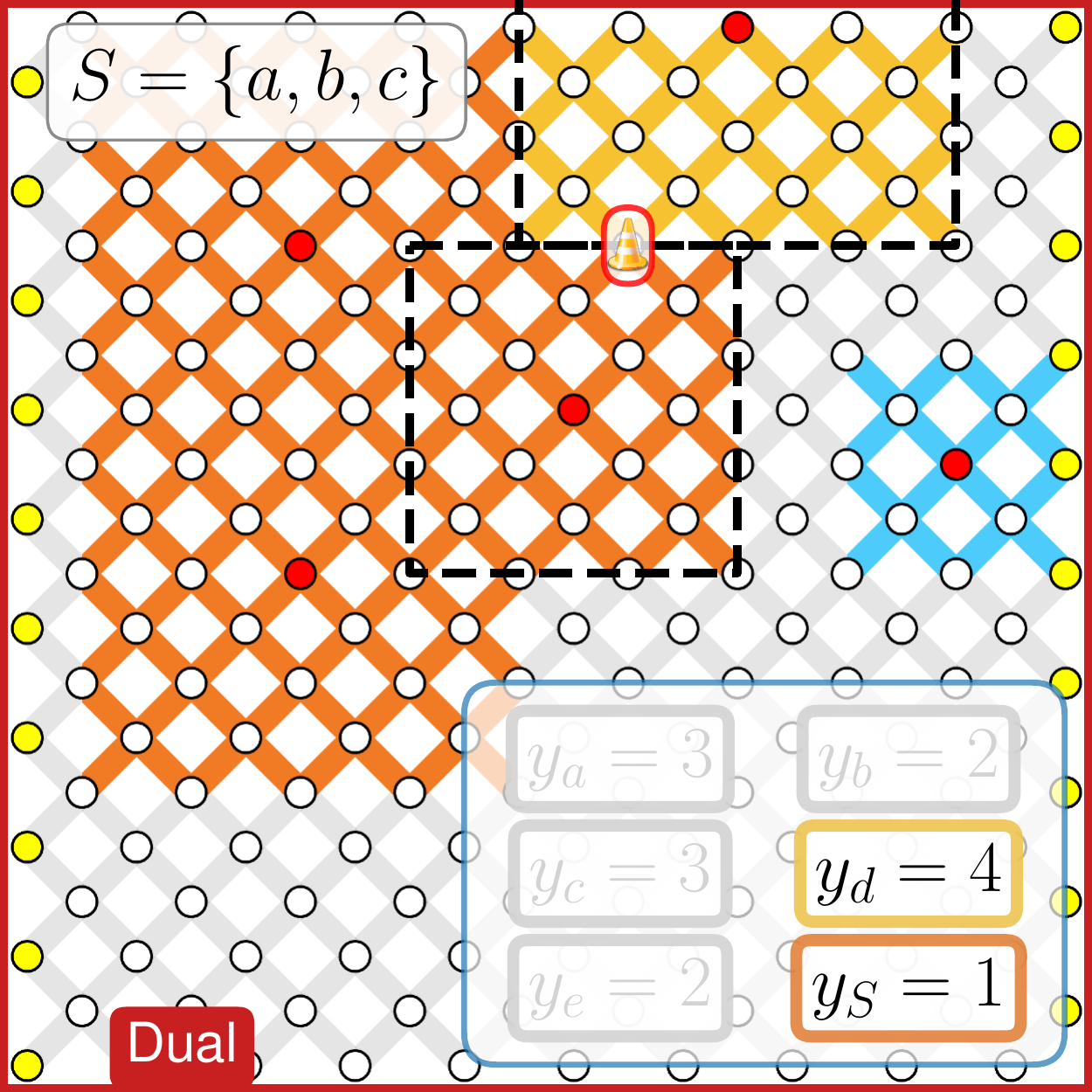}
        \caption{Obstacle $(b, d)$}
        \label{fig:cover-steps-blossom-grow}
    \end{subfigure}
    \hfill
    \begin{subfigure}{.116\textwidth}
        \centering
        \color{lightgray}\includegraphics[width=0.81\linewidth]{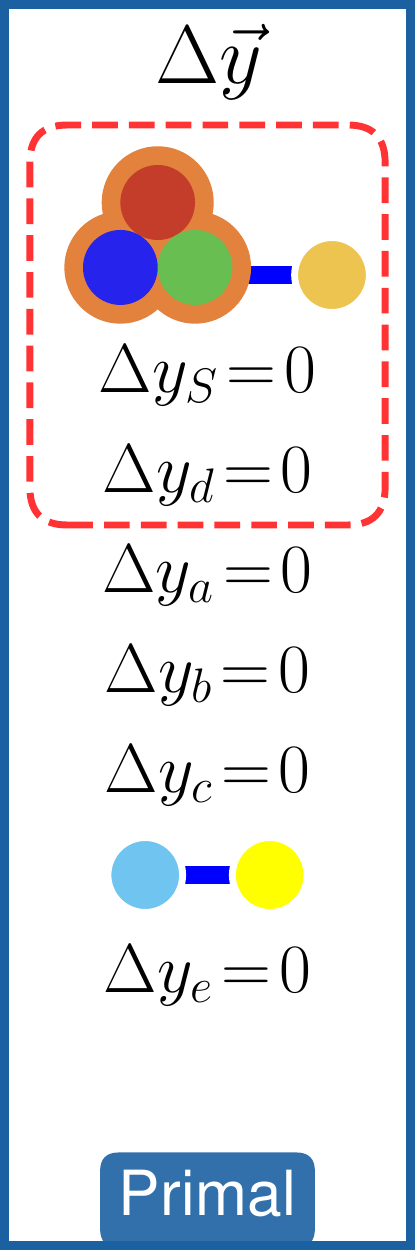}
        \caption{Matched}
    \end{subfigure}
    \hfill
    \begin{subfigure}{.116\textwidth}
        \centering
        \color{lightgray}\includegraphics[width=0.81\linewidth]{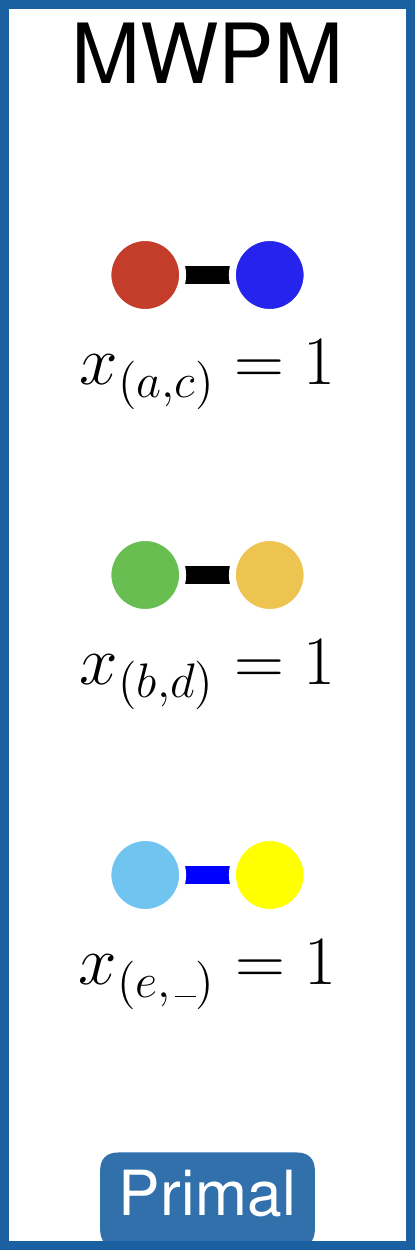}
        \caption{MWPM}
    \end{subfigure}
    \caption{Blossom on Decoding Graph using \seqalgo as example. The primal phase works on the matchings and outputs directions $\Delta \vec{y}$. The dual phase works on the \ref{def:covers} of nodes (in different colors) and outputs \ref{def:obstacles} (with \includegraphics[height=1em,keepaspectratio]{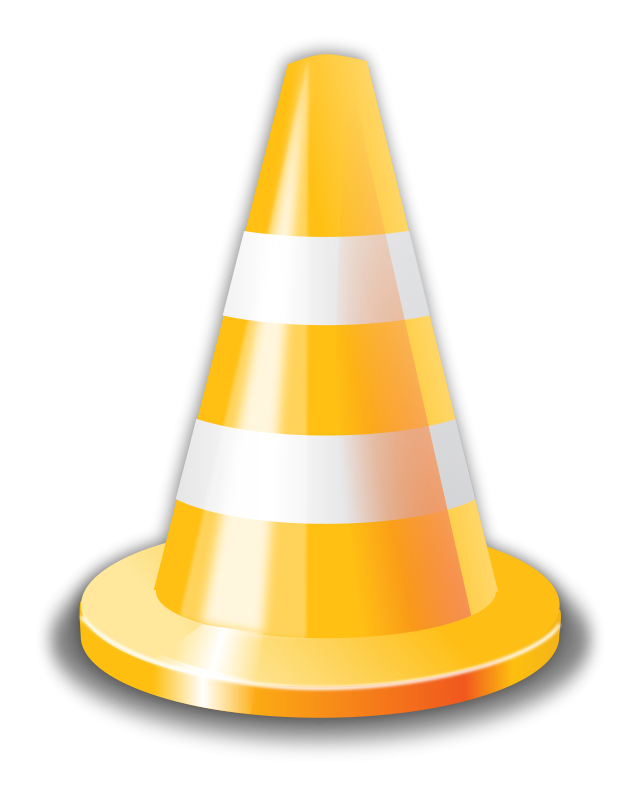} icon). When an obstacle is detected between vertices $u$ and $v$, the black dashed lines show the \ref{def:circles} $C(u)$ and $C(v)$ as part of their individual Covers. (1) The algorithm starts with direction $\Delta y_v = +1,\ \forall v \in V$ and initial dual variables $y_S = 0,\ \forall S \in \mathcal{O}^*$. (2) Dual phase grows $2\Delta \vec{y}$ and finds an obstacle between vertex $e$ and a virtual vertex on the right. (3) Primal phase overcomes the obstacle by matching $e$ with the virtual vertex, and set $\Delta y_e = 0$. (4) Dual phase grows $\frac{1}{2}\Delta \vec{y}$ and finds two obstacles at tight edges $(a, b)$ and $(b, c)$. (5) Primal phase constructs an alternating tree with alternating grow and shrink $\Delta y_a,\Delta y_b,\Delta y_c = +1,-1,+1$ to overcome the obstacles. (6) Dual phase grows $\frac{1}{2}\Delta \vec{y}$ and finds an obstacle between $(a, c)$. (7) Primal phase constructs a blossom $S = \{a,b,c\}$ with $\Delta y_S = +1$. (8) Dual phase grows $\Delta \vec{y}$ and finds an obstacle between $(b, d)$. It's found by first detecting $\text{Cover}(S) \cap \text{Cover}(\{d\}) \neq \varnothing$ and then detecting $C(b \compactsetminus d) \cap C(d \compactsetminus b) \neq \varnothing$. (9) Primal phase matches $S$ to $b$. All nodes are matched. (10) Primal phase expands the blossoms and outputs an MWPM.}
    \label{fig:parity-example}
\end{figure*}

An example of \seqalgo is shown in \cref{fig:parity-example}, demonstrating the whole procedure from receiving the defect vertices to calculating the MWPM.

\fi

\end{document}